\documentclass[10pt]{article}

\usepackage{amsmath}    
\usepackage{amssymb}    
\usepackage{graphicx}
\usepackage{cite} 

\usepackage[usenames]{color}
\usepackage{cancel}
\usepackage{hyperref}
\usepackage[normalem]{ulem}

\usepackage[labelfont=bf,labelsep=period,justification=raggedright]{caption}

\makeatletter
\renewcommand{\@biblabel}[1]{\quad#1.}
\makeatother

\date{}

\begin{document}

\begin{flushleft}
{\Large
\textbf{Incentive and stability in the Rock-Paper-Scissors game: an experimental investigation\footnote{This version is for the presentations in 2014 Economic Science Association (ESA)  European meeting in Prague, Czech Republic and North American meeting, Fort Lauderdale, US. Suggestions and comments are strongly welcome.}}
}
\\
Zhijian Wang$^{1}$,
Bin Xu$^{2,1,3,\ast}$
\\
\bf{1} Experimental Social Science Laboratory, Zhejiang University, Hangzhou 310058, China
\\
\bf{2} Public Administration College, Zhejiang Gongshang University, Hangzhou 310018, China
\\

\bf{3} State Key Laboratory of Theoretical Physics, Institute of Theoretical Physics,
Chinese Academy of Sciences, Beijing 100190, China
\\
$\ast$ E-mail: xubin211@zju.edu.cn
\end{flushleft}

\section*{Abstract}

In a two-person Rock-Paper-Scissors (RPS) game, if we set a loss worth nothing and a tie worth 1, and the payoff of winning (the incentive $a$) as a variable, this game is called as generalized RPS game. The generalized RPS game is a representative mathematical model to illustrate the game dynamics, appearing widely in textbook. However, how actual motions in these games depend on the incentive has never been report quantitatively. Using the data from 7 games with different incentives, including 84 groups of 6 subjects playing the game in 300-round, with random-pair tournaments and local information recorded, we find that, both on social and individual level, the actual motions are changing continuously with the incentive. More expressively, some representative findings are, (1) in social collective strategy transit views, the forward transition vector field is more and more centripetal as the stability of the system increasing; (2) In the individual behavior of strategy transit view, there exists a phase transformation as the stability of the systems increasing, and the phase transformation point being near the standard RPS; (3) Conditional response behaviors are structurally changing accompanied by the controlled incentive. As a whole, the best response behavior increases and the win-stay lose-shift (WSLS) behavior declines with the incentive. Further, the outcome of win, tie, and lose influence the best response behavior and WSLS behavior. Both as the best response behavior, the win-stay behavior declines with the incentive while the lose-left-shift behavior increase with the incentive. And both as the WSLS behavior, the lose-left-shift behavior increase with the incentive, but the lose-right-shift behaviors declines with the incentive. We hope to learn which one in tens of learning models can interpret the empirical observation above.


\section*{Introduction}

Recently, collective cycle phenomenon has been found in Rock-Paper-Scissors (RPS) games no matter it is unstable, neutral or stable~\cite{wang2014conditional}. According to the classical game theory~\cite{VonNeumann1944}, the players who have the complete rationality will choose each strategy in the probability of a third choice. If it is the truth, the outcome of a play in long run will satisfy the detailed balance condition, i.e., each elementary process should be equilibrated by its reverse process~\cite{Boltzmann1964,Young2008,Blume2007}. The existence of cyclic motion seems to indicate that the human behaviour deviates systematically fully rational behaviour and has a hidden pattern in it. Wang et al~\cite{wang2014conditional} further confirmed the relationship between macro cycle phenomena and micro behaviour pattern by a phenomenological model. The result shows that the frequency of cycle of each population can be explained quite well by the average individuals' conditional response behaviors in that population, no matter the RPS game is unstable, neutral or stable.

A straightforward question addressed here is, what is the difference between different RPS games? The evolutionary game theory, for example, the replicator dynamics~\cite{Taylor1978}, not only predicts the cyclicly collective evolution in the state space, but also predicts the difference among different RPS games. The generalized Rock-Paper-Scissors (RPS) game is the prototypic model
to study the stability of a game process~\cite{Maynard1982evolution,HofbauerSigmund1998,Sandholm2011,hofbauer2009stable,Nowak2012}.
From the evolutionary game theory point of view, it is a single-population symmetric game. If the payoff for losing is 0 and tie is 1, then the payoff for win can be denoted by $a$. According to the replicator dynamics~\cite{Taylor1978}, the stability is dependant on the incentive $a$ despite of the common unique mixed-strategy Nash Equilibrium (NE) ($\frac{1}{3},\frac{1}{3},\frac{1}{3}$)~\cite{hahn2012evolutionary,hofbauer2009stable}. A RPS game will be unstable if $a<2$, neutral if $a=2$ or stable if $a>2$. Persistent cycling motions are predicted to occur in unstable and neutral games, but will not in stable games~\cite{Nowak2012}, where the mixed-strategy Nash equilibrium is equivalent to a stable fixed point~\cite{hofbauer2009stable,hahn2012evolutionary,tang2001anticipatory,Shapley1964,norman2010cycles,crawford1974learning,crawford1989learning,echenique2004mixed}.
On the other hand, stability is one of the central concepts in game theory~\cite{VonNeumann1944,Maynard1982evolution,harsanyi1988general}, which
has wide application in economics and also in biology~\cite{nowak2006evolutionary,Bowles2004,Plott2004globalScarf,Friedman1998Rev}.
Both evolutionary dynamics and experimental economics methods have been
employed to quantitatively describe the stability of a game process~\cite{tang2001anticipatory,friedman2010tasp,2013arXiv1301.3238X,Huyck1999,Nowak2012,Friedman2014}.

However, how actual motions in these games depend on the incentive of winning $a$ has never been report quantitatively. In the present investigation, we aim at answering this question. We manipulate the incentive parameter $a$ from 1, 1.1, 2, 4, 9, 100, to $\infty$ to yield, in the sense of the replicator dynamics, unstable, neutral, stable and extremely stable RPS game, respectively to study how the actual motions change with the incentive $a$.

\section{Experiment}
\begin{figure}[!ht]
\begin{center}
 \includegraphics[width=0.86in]{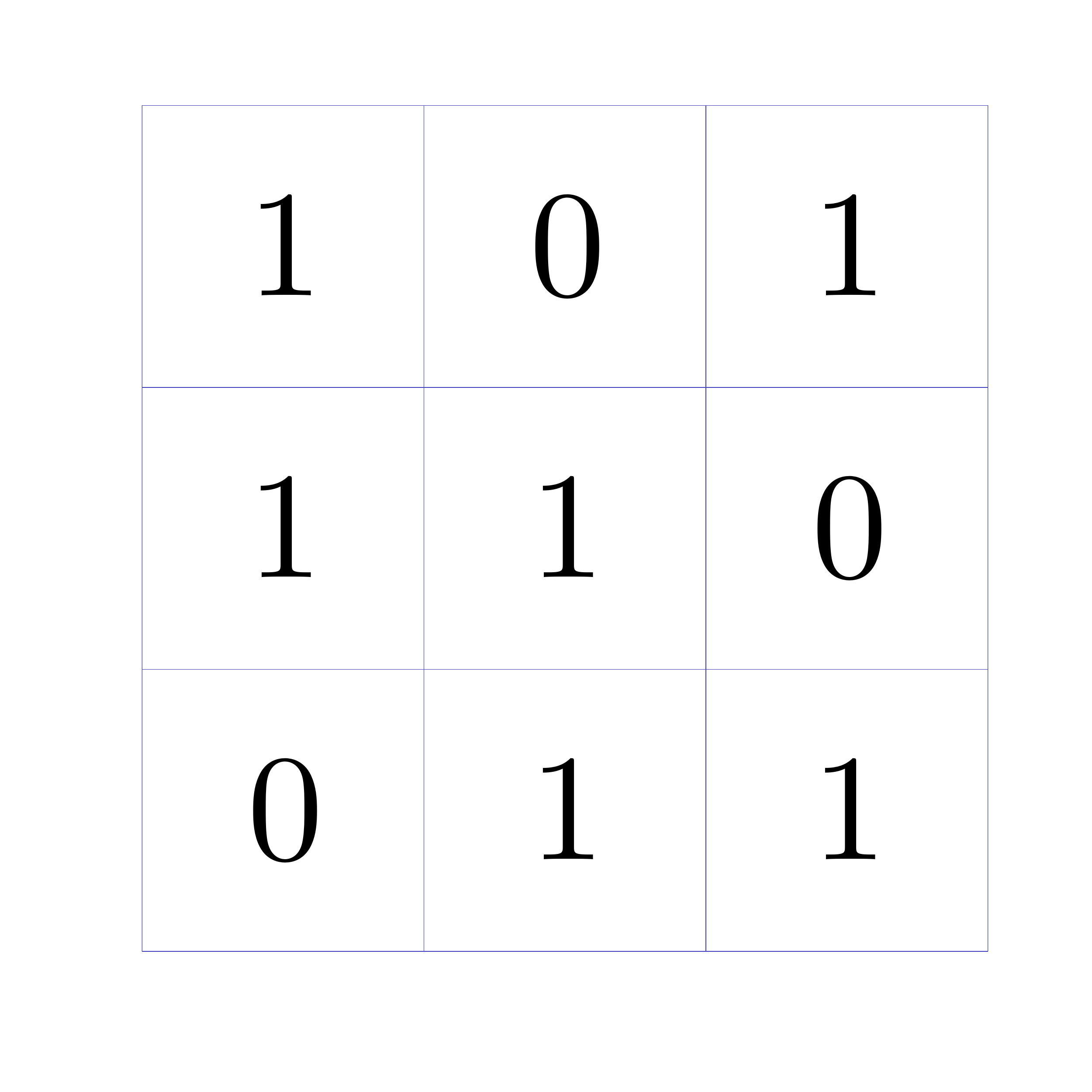}
  \includegraphics[width=0.86in]{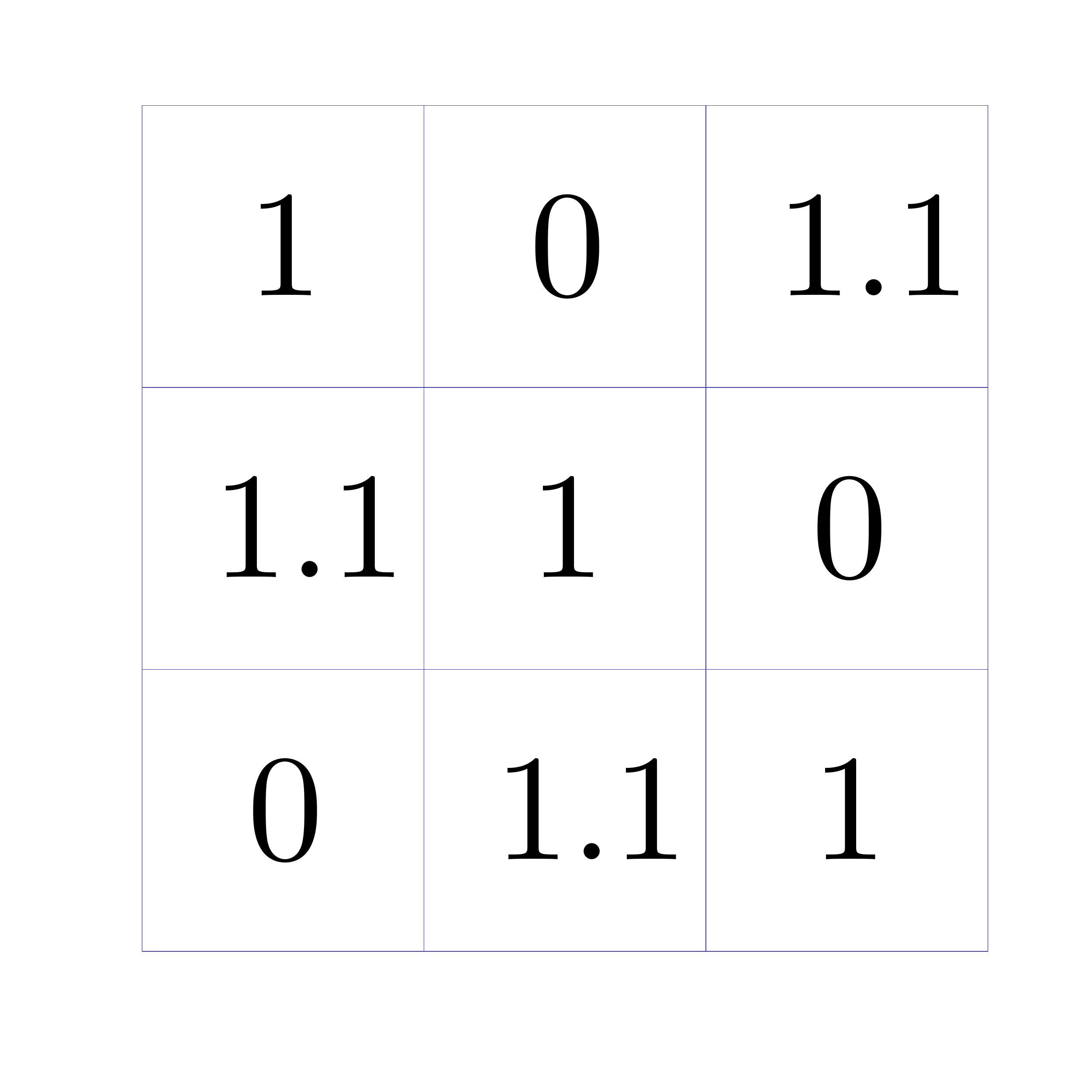}
   \includegraphics[width=0.86in]{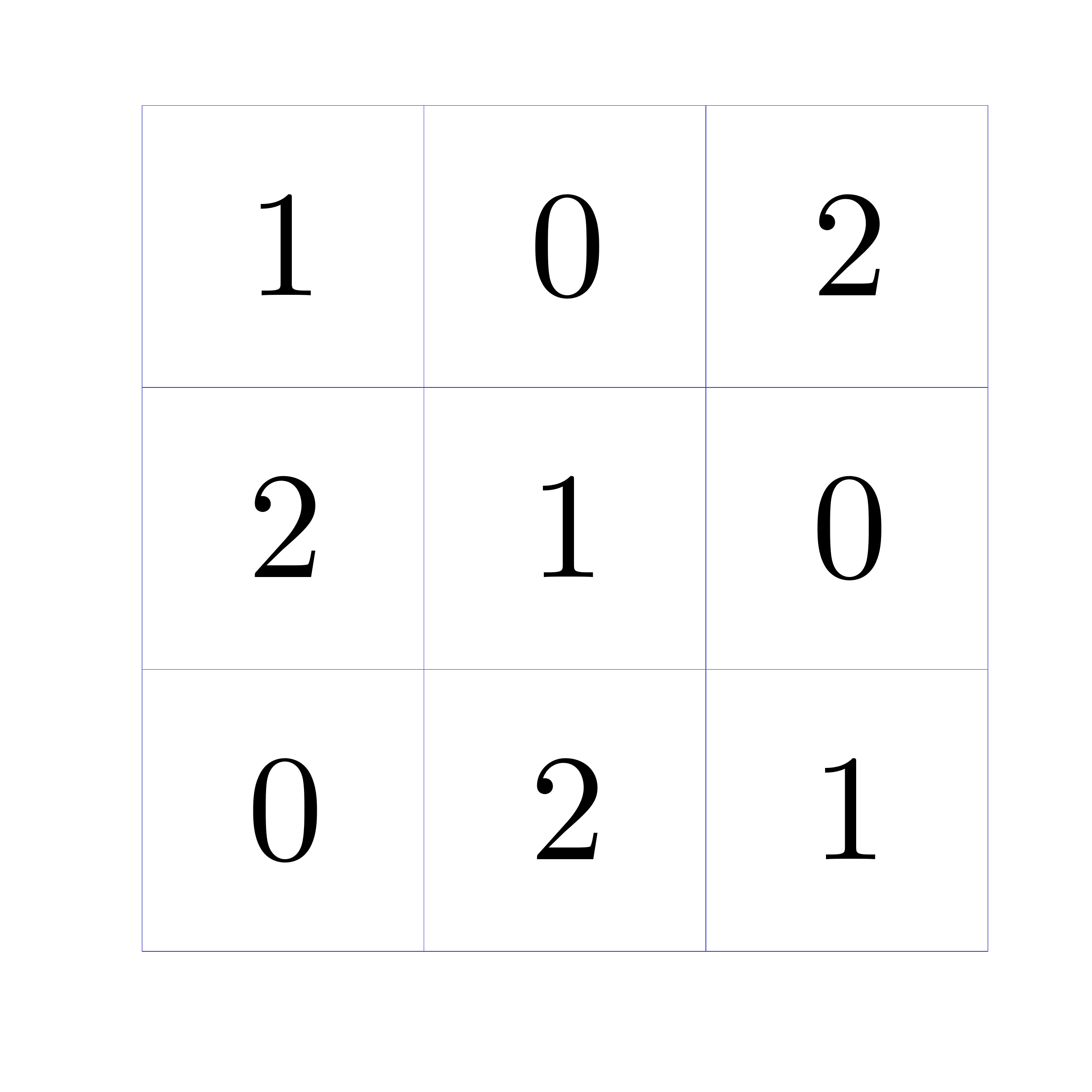}
    \includegraphics[width=0.86in]{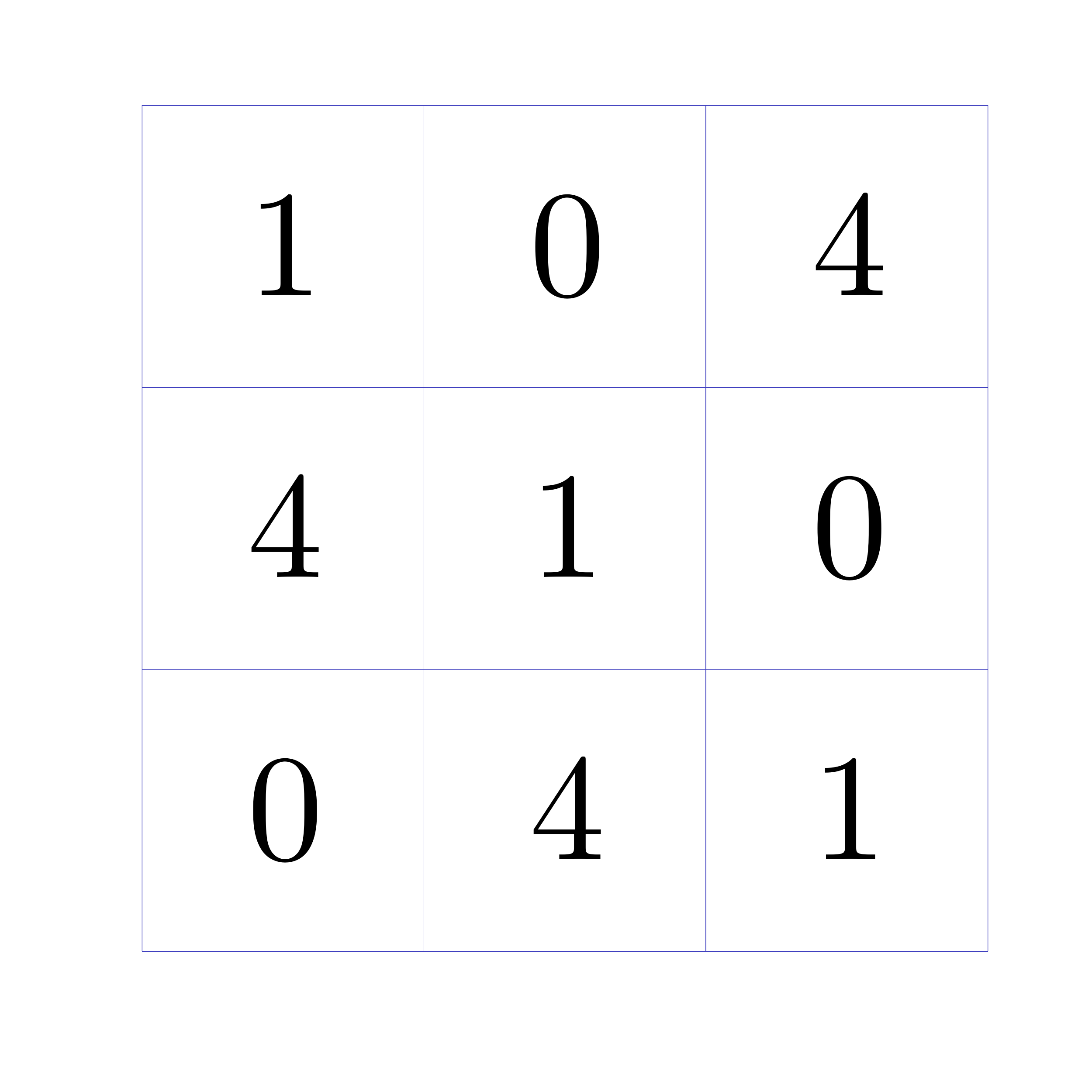}
     \includegraphics[width=0.86in]{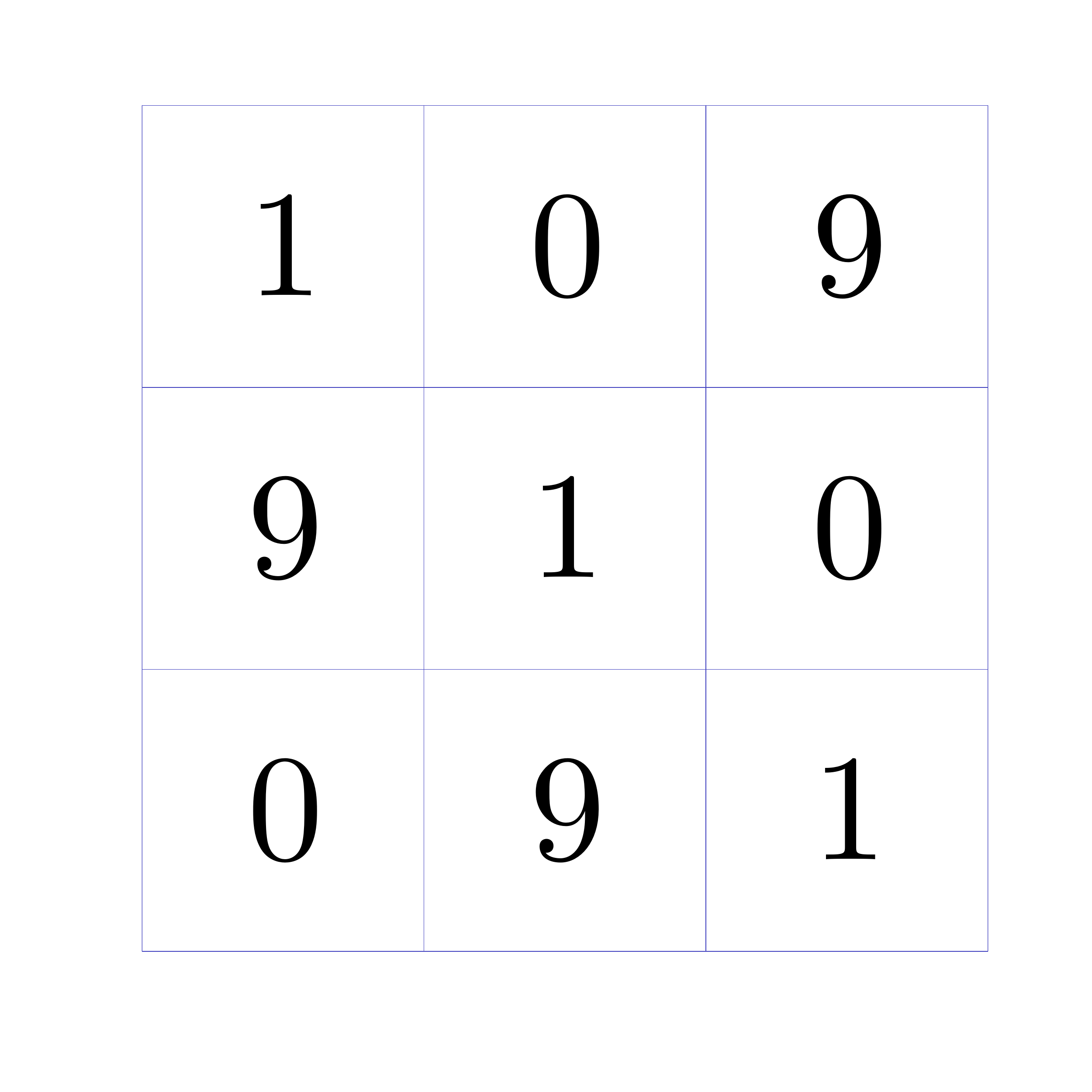}
      \includegraphics[width=0.86in]{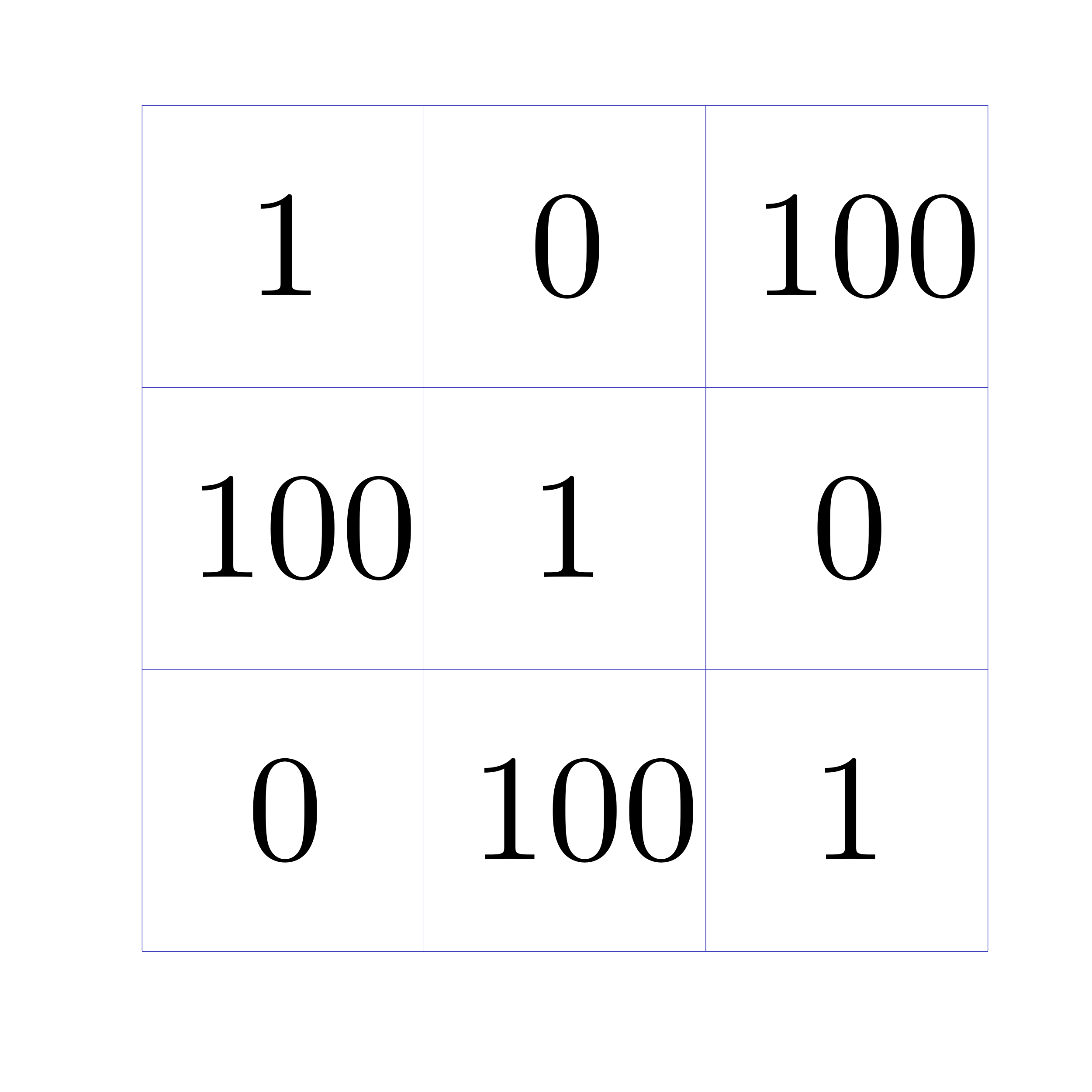}
       \includegraphics[width=0.86in]{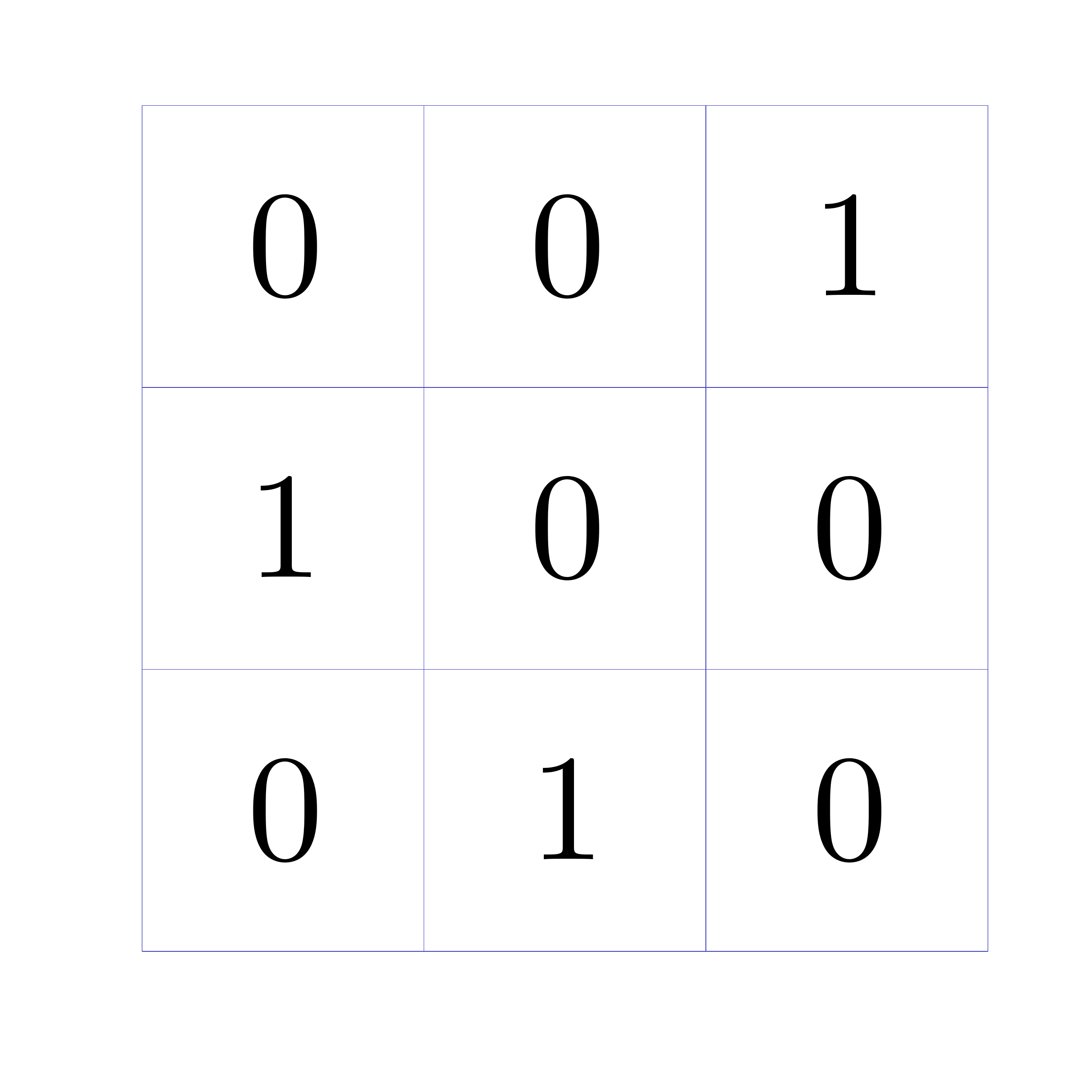}
 \includegraphics[width=0.86in]{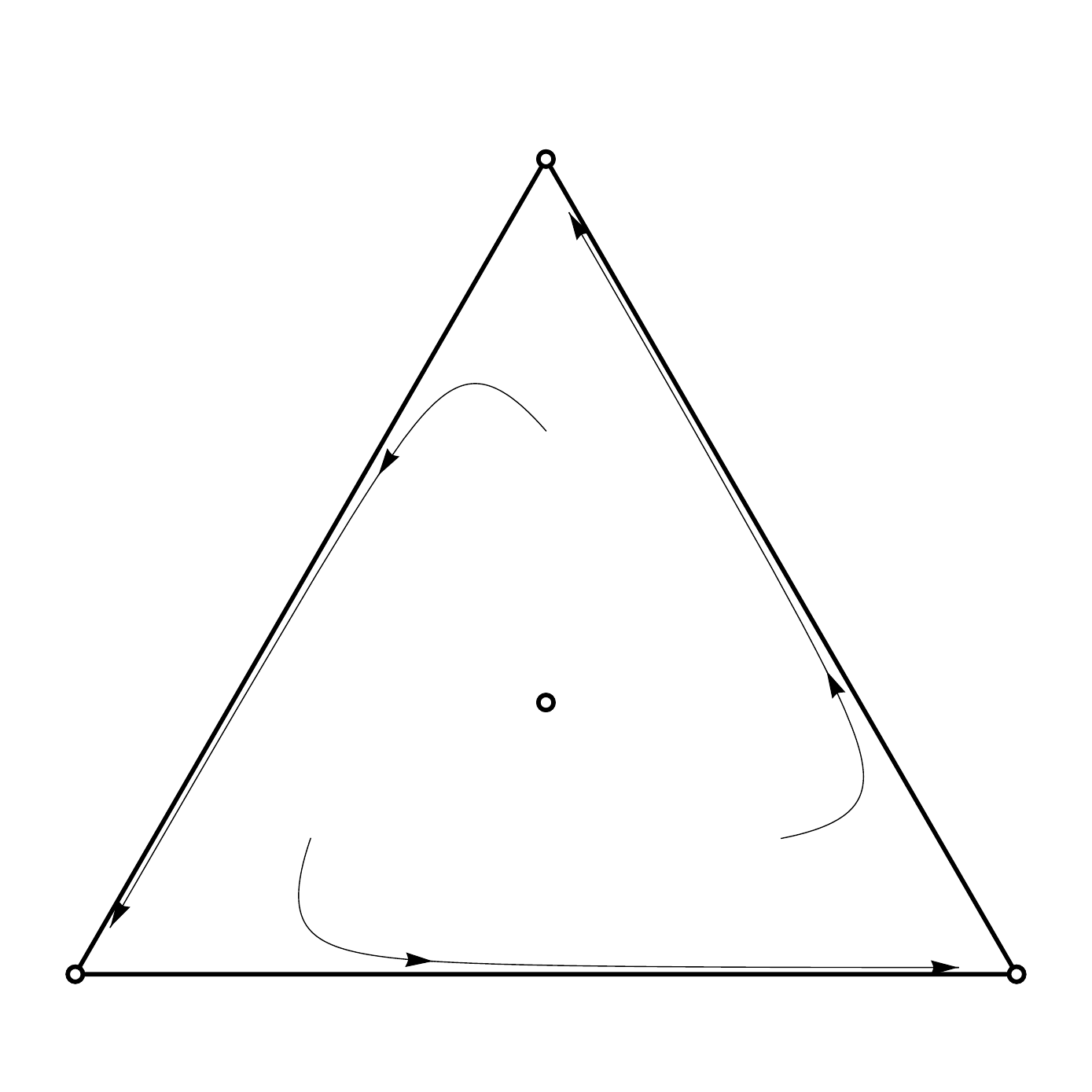}
  \includegraphics[width=0.86in]{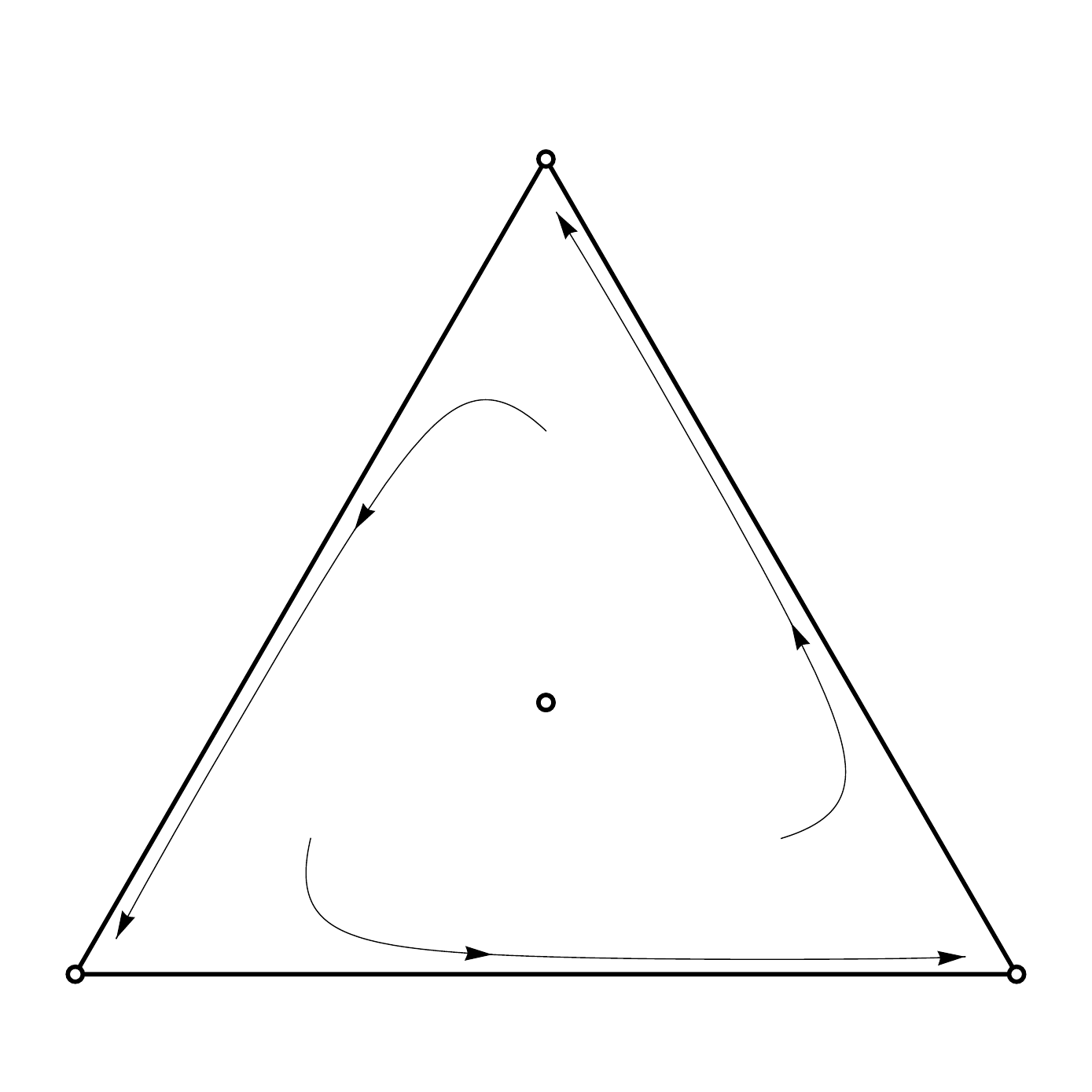}
   \includegraphics[width=0.86in]{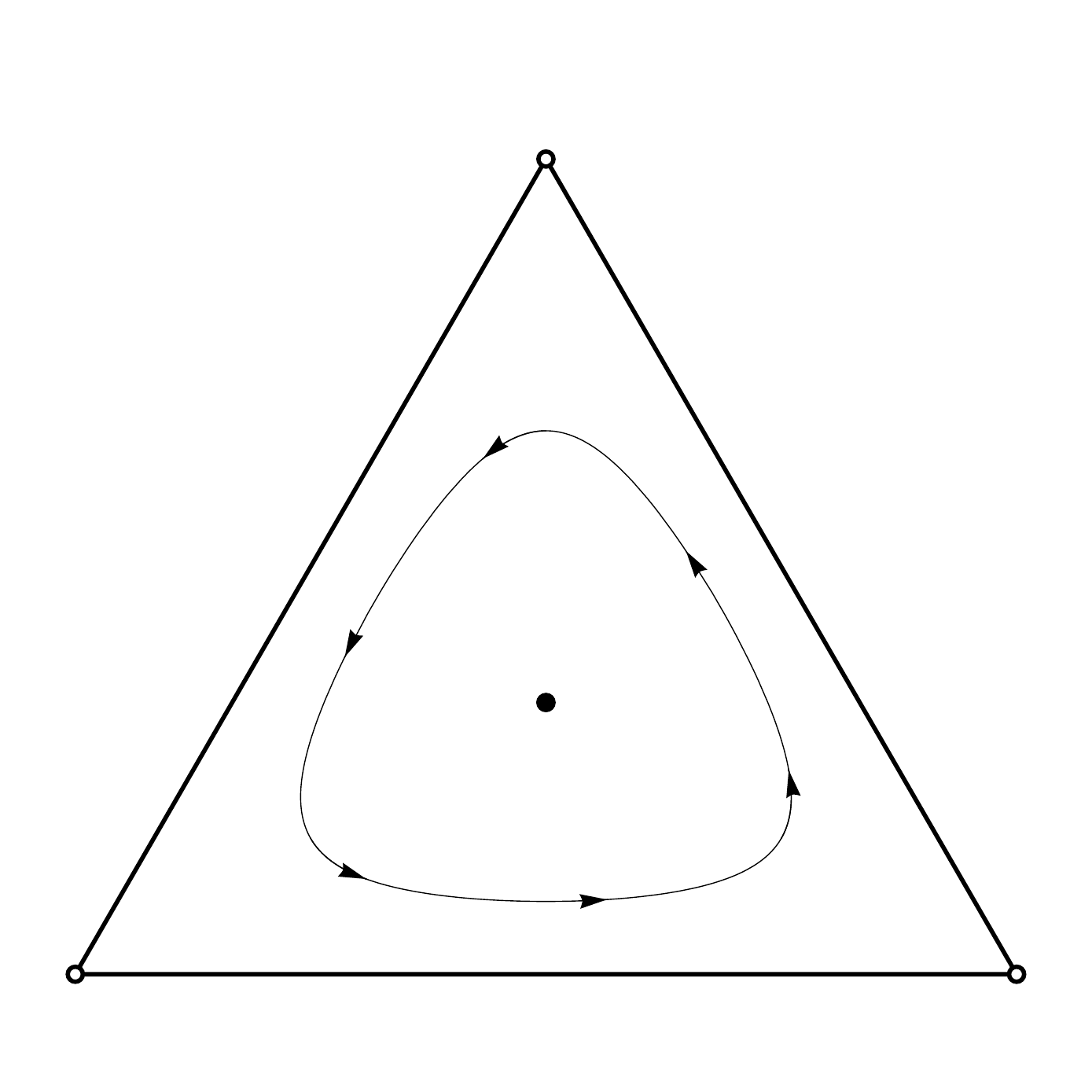}
    \includegraphics[width=0.86in]{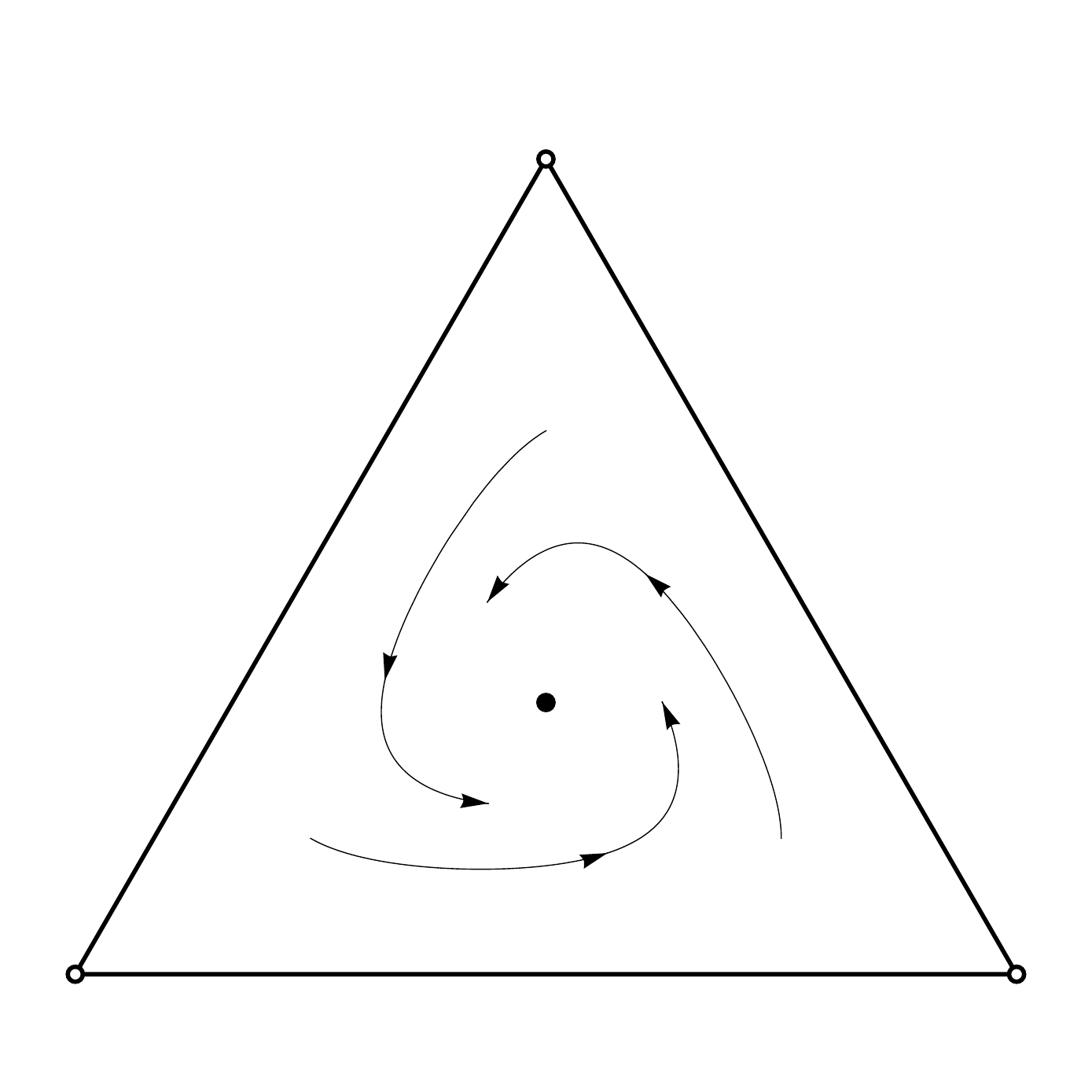}
     \includegraphics[width=0.86in]{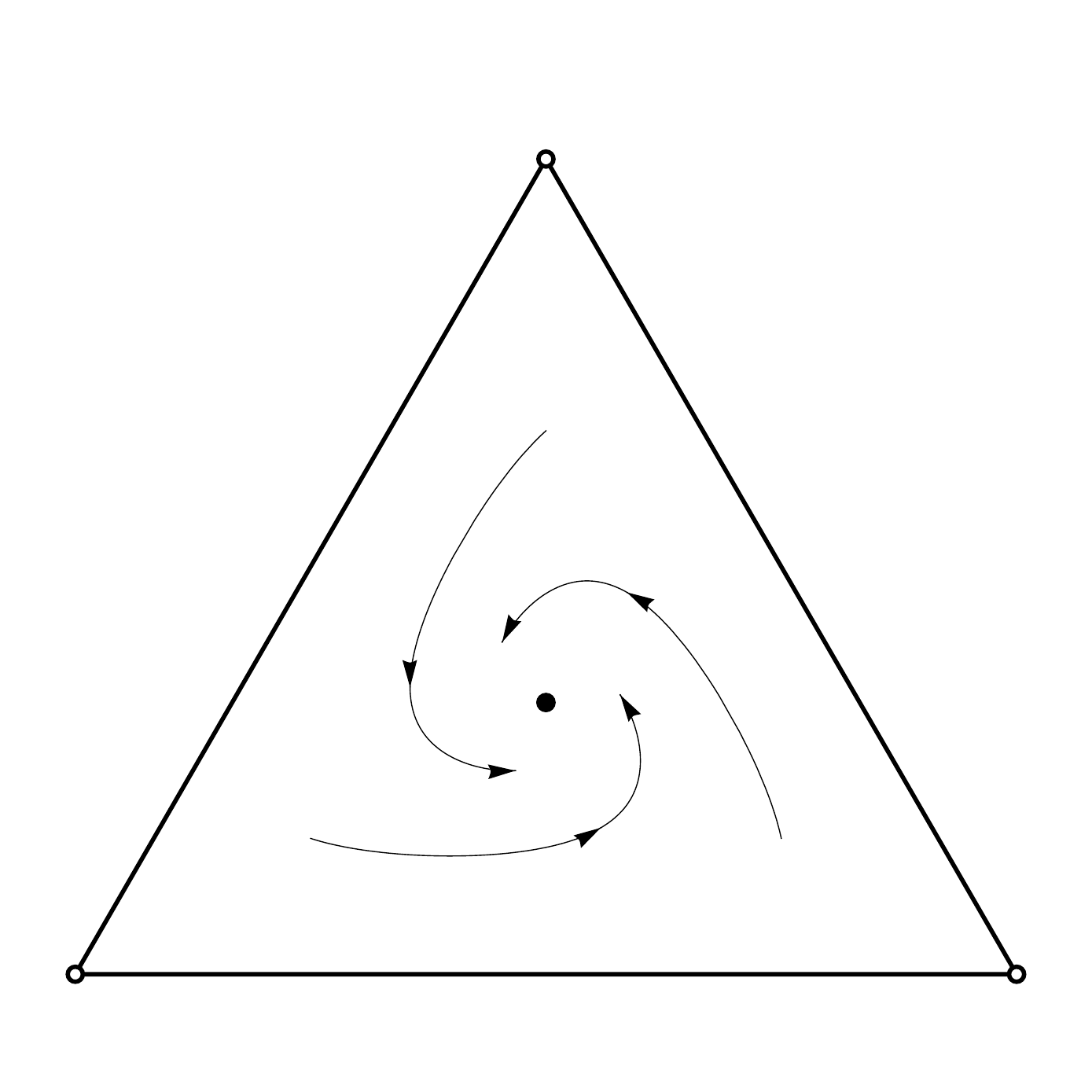}
      \includegraphics[width=0.86in]{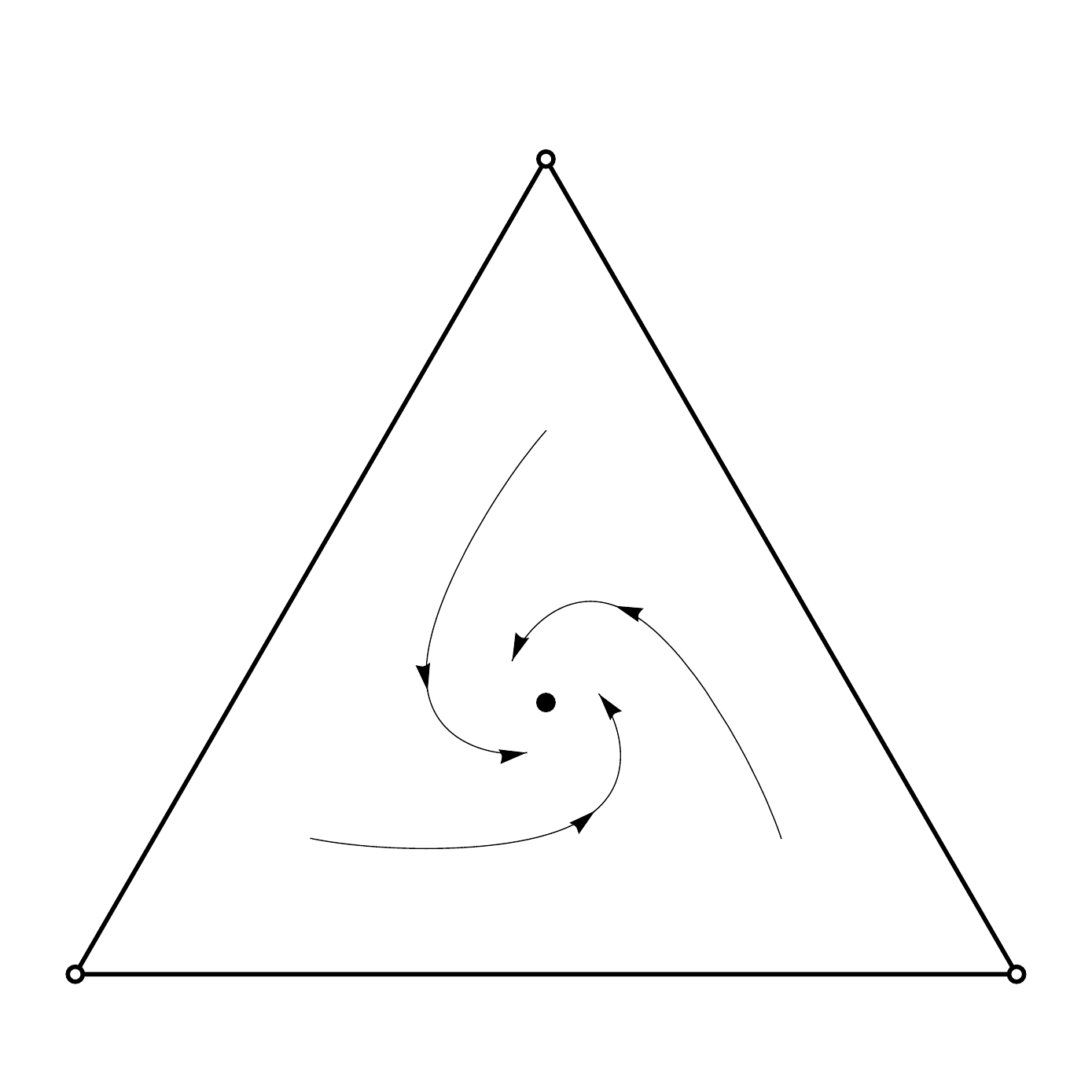}
       \includegraphics[width=0.86in]{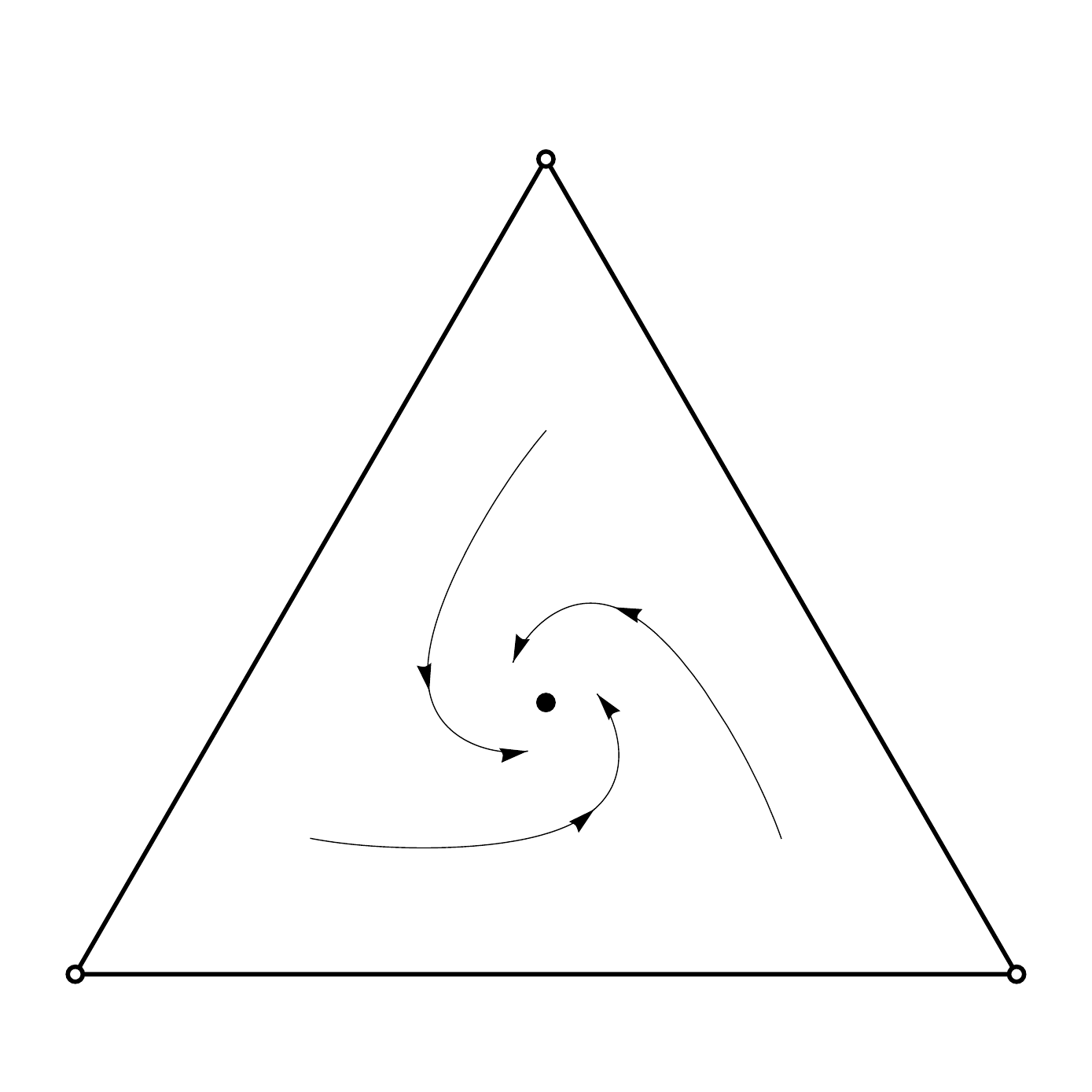}
  \end{center}
  \caption{
    \textbf{Experimental design.} First row, the seven payoff matrices employed in the experiment. Second row, the replicator dynamics pattern corresponding to each payoff matrix.
}
   \label{fig:payoffmatrices}
\end{figure}

The experiment is almost the same with the experiment in Ref.~\cite{wang2014conditional}, except two special Rock Paper Scissors games which are very important for the analyzing of difference among different types of RPS games. Specifically, the payoff matrices employed in this investigate are as Fig.~\ref{fig:payoffmatrices}. The first one and the last one are the two special RPS games added here, and the remain five payoff matrices are the same with those of in Ref.~\cite{wang2014conditional}. And the experimental setting is same. So for more details of experimental setting, see Ref.~\cite{wang2014conditional}. As an incentive, the students subjects were paid in local currency (RMB) in proportion to the number points of their victories. The mean earnings a subject got is about 50 yuan RMB, including 5 RMB show-up fee. A total of 504 undergraduate and graduate students from different disciplines at Zhejiang University were recruited in the experiment with each student participate into the experiment only once.

\label{sec:results}
All payoff matrices used here have a common property that they all have mixed strategy Nash equilibrium (MSNE) ($\frac{1}{3},\frac{1}{3},\frac{1}{3}$), though the first one has extra three pure Nash equilibria, ($0,0,1$), ($0,1,0$), and ($1,0,0$). The payoff of winning is equal to the payoff of drawing in this special RPS game. We use this special game as the extremely unstable case. On contrary, the payoff matrices in the right side is extremely stable, where the payoff of tie is set to be 0 and the payoff of winning is set to be 1. This means the relative payoff of winning comparing with tie is $\infty$, while the tie can be 1, for convenience, we call the last one as $a=\infty$. From the evolutionary dynamics of view, for example, the replicator dynamics~\cite{Taylor1978}, the RPS game is unstable if $a<2$, neutral if $a=2$, and stable if $a>2$, see Fig.~\ref{fig:payoffmatrices}, second row.

\section{Results}
\subsection{Spire social jump out pattern}

First of all, let's have a overview on the three type of the laboratory Rock Paper Scissors games.

\begin{figure}[!ht]
\begin{center}
 \includegraphics[width=2in]{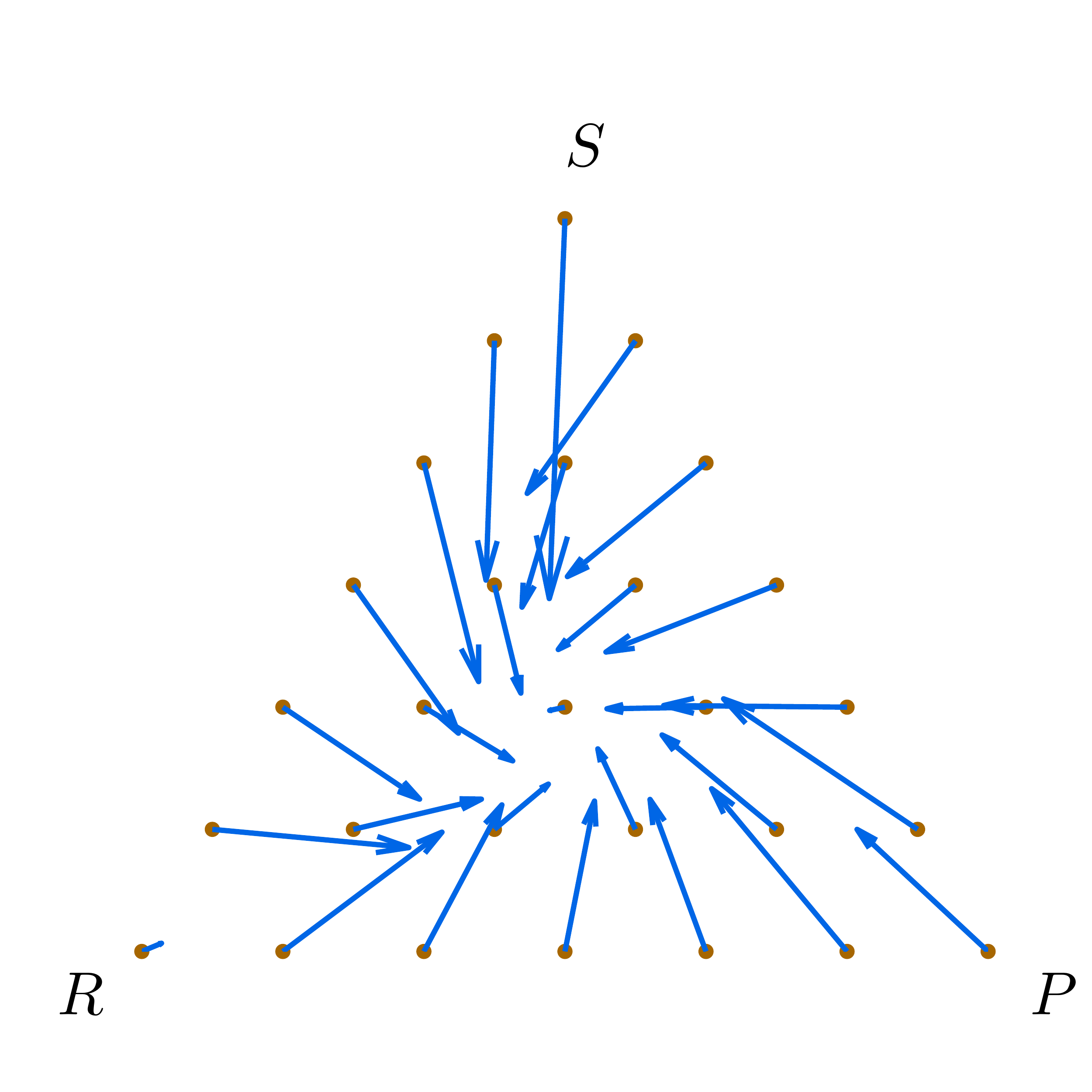}
    \includegraphics[width=2in]{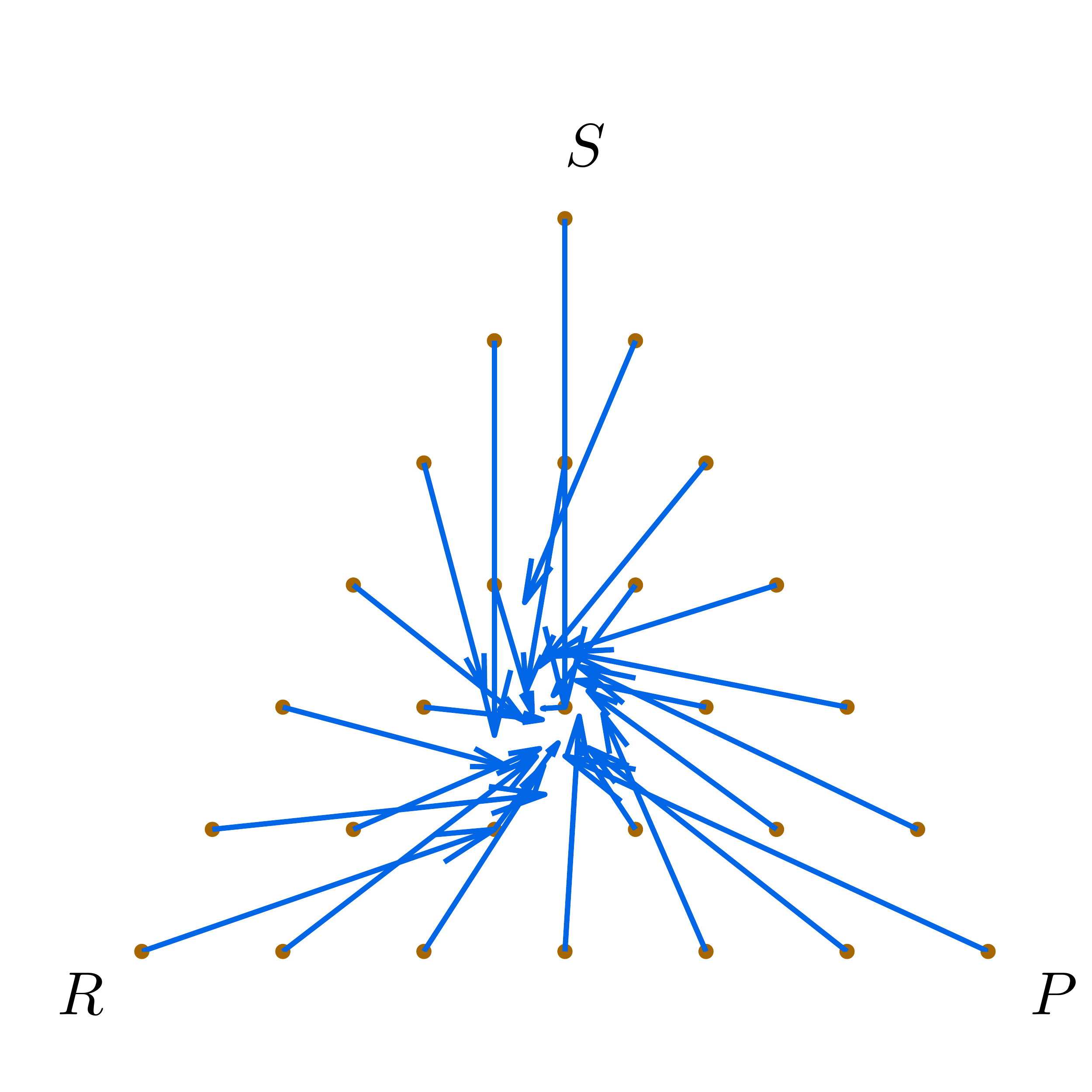}
    \includegraphics[width=2in]{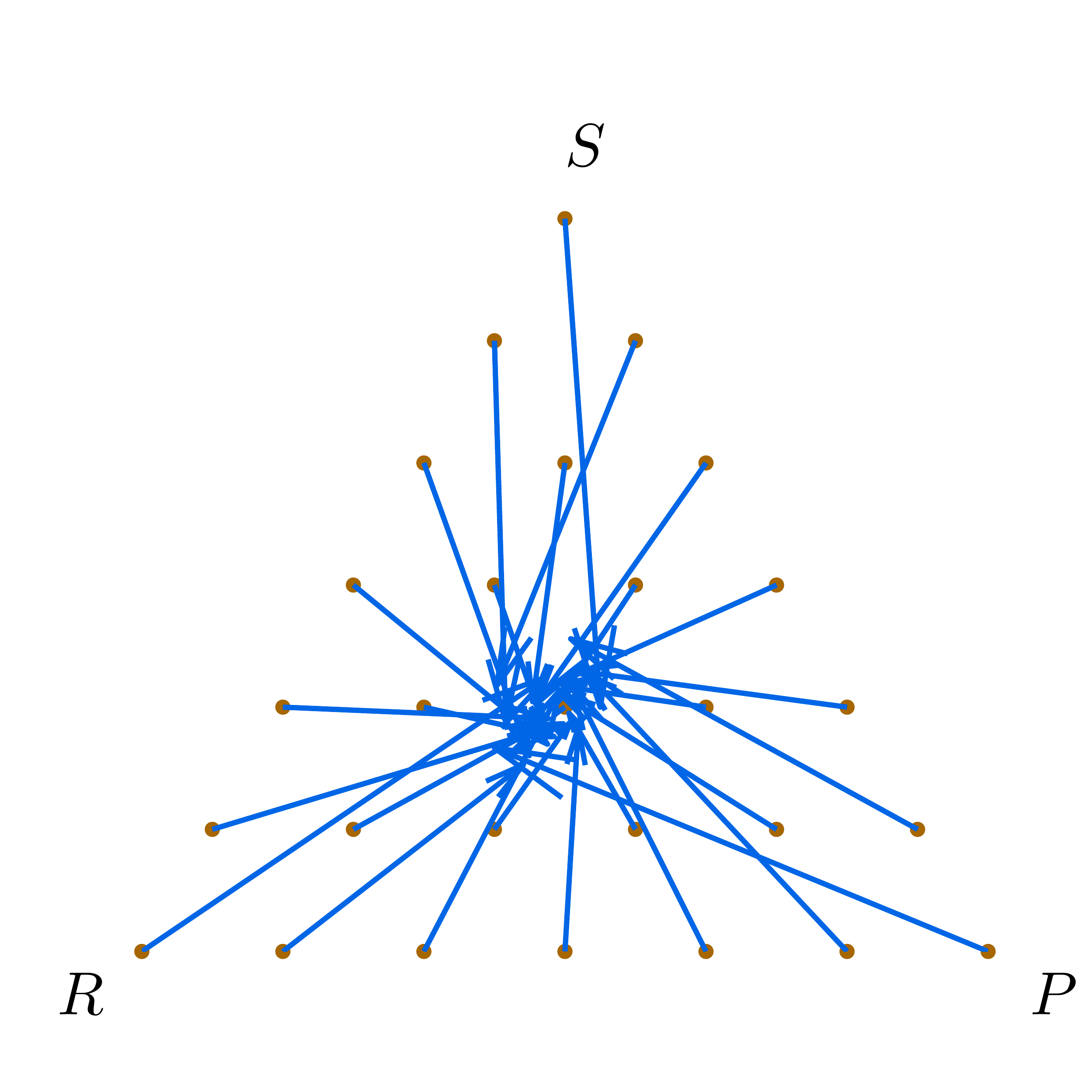}
  \end{center}
  \caption{
    \textbf{Spire pattern of social jump out in unstable, neutral, and stable RPS games}. From left to right, $a=1.1, 2, 9$, respectively. For all of the patterns, see Figure~\ref{fig:payoffandall} in appendix)
}
   \label{fig:Spire}
\end{figure}

From the perspective of overview, the outcome of a game can be described as population's state which is the combination of strategies all the players used, and a social state is a coarse-grained description of the population's state~\cite{Sandholm2011}. In one population RPS game, the state can be represented by ($r, p,s$), where $r$ $(p,~s)$ is the proportion of strategy $R(P,~S)$ in the population, and $r+p+s=1$. The total number of different social states for a population of size $N$ is $\frac{(N+1) (N+2)}{2}$. In the studied case of $N=6$ this number is $28$, and all of them fall into the state space, $\triangle RPS$, see Figure~\ref{fig:Spire}. The Nash equilibrium point is identical to the social state $(\frac{1}{3}, \frac{1}{3}, \frac{1}{3})$. 

\label{sec:results}

In the repeated RPS game, the individuals may change their strategies, meanwhile the population mix may jump from one social state to another. The social state at round $t$ is denoted as $x_{t}$ := $(r_{t},p_{t}, s_{t})$; and in the next round is denoted as $x_{t+1}$:=$(r_{t+1},p_{t+1}, s_{t+1})$. During an experimental session of 300 rounds, a state might be passed many times. For a given state $x$, there is a jump out vector which is called as a social forward transition vector indicating the transition from $x_t$ to $x_{t+1}$~\cite{xu2012test}, the average terminal state is $\bar{x}_{t+1}$, and the average jump out vector is $\bar{T}_{x}$. We calculated all of the jump out vectors of each state $x$ for each game, and graph them in Figure~\ref{fig:Spire} and Figure~\ref{fig:payoffandall} (see appendix).

Obviously, as a whole, the jump out vector form a counter-clockwise spire pattern in all of the three types of RPS games. In addition, these patterns look different among different games. The average terminal states at unstable RPS game is far way from the Mixed Strategy Nash Equilibrium (MSNE) while the average terminal states at stable RPS game is near the MSNE, and the neutral RPS game is in the middle.

\subsection{Ascending social nearness centrality}

We use the distance from centre to each terminal state of average jump out vector to quantitatively compare the nearness centrality over different parameter $a$.

First, similar to Ref.~\cite{Nowak2012,Friedman2014}, for a given terminal state $(\bar{x}_{t+1})$, we define the distance refer to Nash equilibrium ($\frac{1}{3}, \frac{1}{3}, \frac{1}{3}$) as
\begin{equation}\label{eq:distance1}
  Dis_x = \sqrt{[\bar{r}_{t+1} - \frac{1}{3}]^2 + [\bar{p}_{t+1} - \frac{1}{3}]^2 + [\bar{s}_{t+1} - \frac{1}{3}]^2}.
\end{equation}
The interval of $Dis_x$, from the center of state space $(\frac{1}{3},\frac{1}{3},\frac{1}{3})$ to the vertex of the state space $(0,0,1)$, $(0,1,0)$ or $(1,0,0)$, is $[0, \sqrt{\frac{2}{3}}]$.

\begin{figure}[!ht]
\begin{center}
                \includegraphics[width=4in]{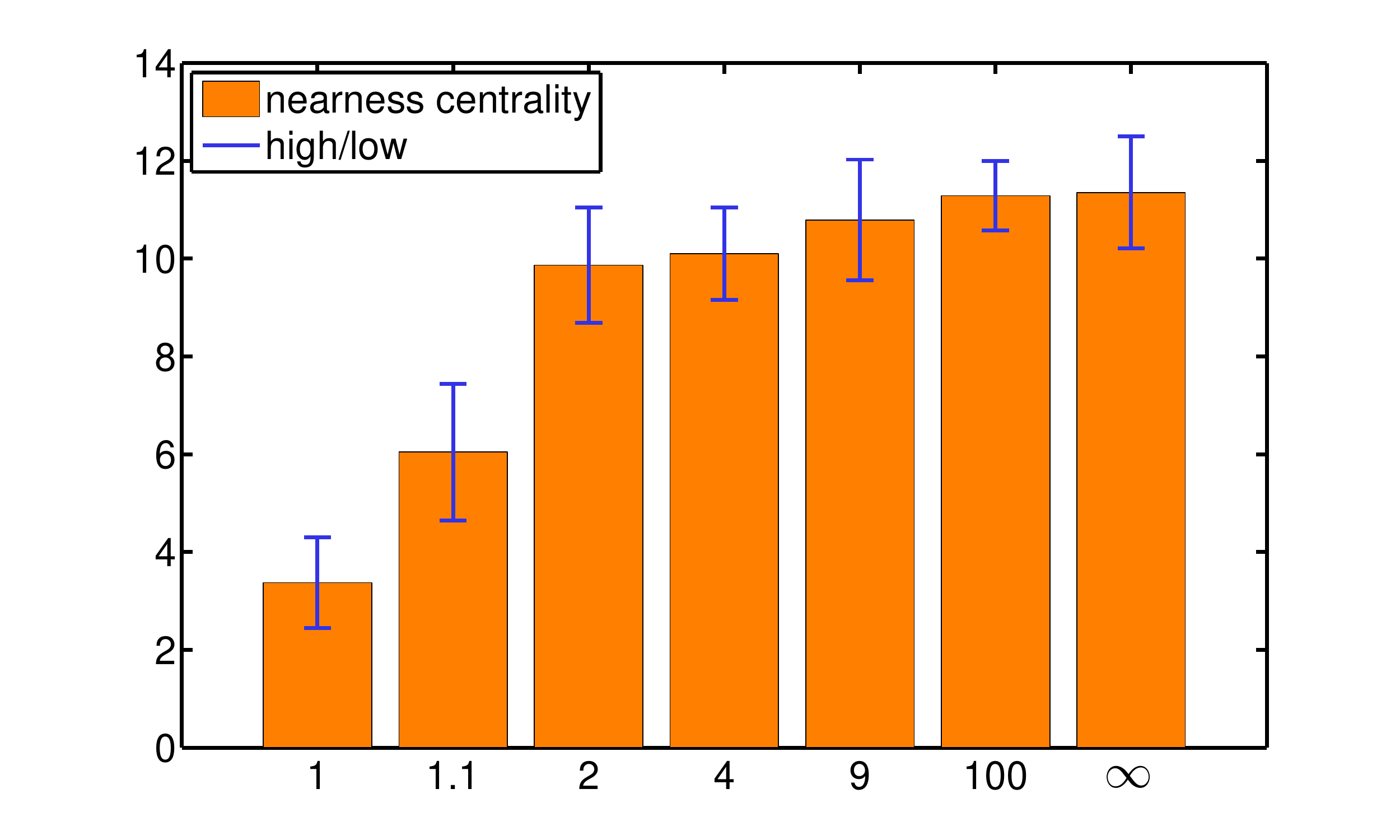}
         \end{center}
  \caption{
    \textbf{Nearness centrality from unstable to deeply stable RPS game.} The numerics at the horizontal axis is the winning payoff $a$, the numerics at the vertical axis is the number of NCI.  
}
   \label{fig:nearness}
\end{figure}

For convenience, we introduce a single Global Distance Index (GDI):
\begin{equation}\label{eq:distance2}
  Dis_\triangle=\sum_{x=1}^{28} Dis_x p_x,
\end{equation}
where $p_x$ is the proportion of frequency of jump out vector of the state $x$. And now we introduce nearness centrality index (NCI) as a single index to describe the nearness centrality.
\begin{equation}\label{eq:CCI}
  NCI=\frac{1}{Dis_\triangle}
\end{equation}
Table~\ref{tab:nearness} exhibits the results of closeness centrality of each parameter, and Fig.~\ref{fig:nearness} provides a graph. The nearness centrality tends to increasing with the increasing of incentive parameter $a$. (Spearman's $rho=0.7600$, Prob $> |t|$ =0.0000, n=84) (Spearman's $rho = 1.0000$, Prob $> |t|$ =0.0000, n=7) As if the population in larger incentive $a$ get larger centripetal force. 

In addition, the nearness centrality is related to the instability. Comparing with neutral RPS game, the nearness centrality in unstable RPS game is significant smaller (Two-sample Wilcoxon rank-sum (Mann-Whitney) test, $n_1$=$n_2$=12, Ho: $NCI_{a=1}=NCI_{a=2}$, z=-4.157, p=0.0000, Ho: $NCI_{a=1.1}=NCI_{a=2}$, z =-3.695, p=0.0002). On contrary, comparing with neutral RPS game, the nearness centrality in the stable RPS game is larger, but only the deeply stable RPS games are significantly larger. (Two-sample Wilcoxon rank-sum (Mann-Whitney) test, $n_1$=$n_2$=12, Ho: $NCI_{a=4}=NCI_{a=2}$, z=0.520, p=0.6033, Ho: $NCI_{a=9}=NCI_{a=2}$, z=1.328, p=0.6033, Ho: $NCI_{a=100}=NCI_{a=2}$, z=2.598, p=0.0094,
Ho: $NCI_{a=100}=NCI_{a=2}$, z=2.309, p=0.0209).

\begin{table}
\center
\small
\caption{
\bf{Nearness Centrality}}
\begin{tabular}{c|crrrrr}
\hline
a	&obs.	&Mean	&Std.Err.	&$[95\%	Conf.$&$Interval]$	\\
\hline
1	&12	&3.371	&0.422	&2.533	&4.210	\\
1.1	&12	&6.043	&0.635	&4.779	&7.307	\\
2	&12	&9.865	&0.536	&8.800	&10.931	\\
4	&12	&10.098	&0.433	&9.244	&10.952	\\
9	&12	&10.788	&0.562	&9.670	&11.905	\\
100	&12	&11.286	&0.322	&10.644	&11.927	\\
$\infty$	&12	&11.350	&0.520	&10.316	&12.384	\\
\hline
\end{tabular}
\begin{flushleft}
\end{flushleft}
\label{tab:nearness}
\end{table}

\subsection{Enhanced individual activity}

\begin{figure}[!ht]
\begin{center}
  \includegraphics[width=4in]{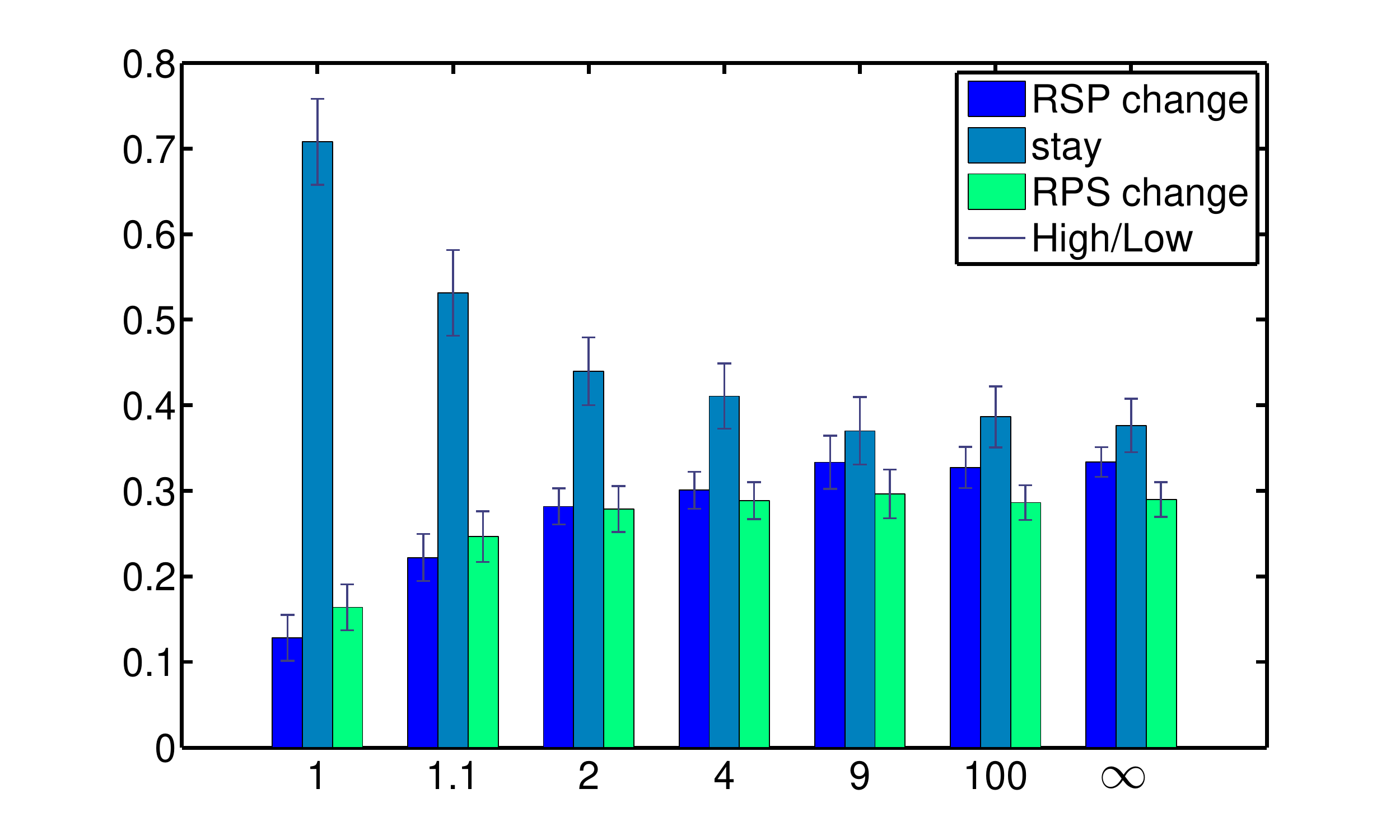}
      \end{center}
  \caption{
    \textbf{Three type of transit.} From left to right, $a=1,1.1, 2, 4, 9, 100, \infty$, respectively. For each subgraph, the left stick is the proportion of ``RSP" shift, $T_{RSP}$, the right stick is the proportion of ``RPS" shift, $T_{RPS}$, and the middle is the proportion of ``stay". (The err bar represents the $95\%$ confidence interval, obs=72.)
}
   \label{fig:unconditionalshift}
\end{figure}

After a round of game, a player can remain her/his the current strategy, called as ``stay", or change to another strategy. The ratio of change is high the activity is high. So we use the ratio of change as the  indicator of activity. Table~\ref{tab:transit} (column $5^{th}$) exhibits the average proportion stay as well as the shift with the direction of $R\rightarrow S\rightarrow P$ ($T_{RSP}$) (column $7^{th}$) and the shift with the direction of $R\rightarrow P\rightarrow S$ ($T_{RPS}$) (column $7^{th}$) and Figure~\ref{fig:unconditionalshift} illustrates them. 

In all of the treatments, the most frequently used strategy is ``stay" strategy, and ``stay" ratio is significant larger than RSP shift ratio except treatment $a=9$, and is significant larger than RPS shift in all treatments, see Table~\ref{tab:transit} (column $9^{th}$, $11^{th}$). More important, the individuals are more likely to change their strategies when $a$ goes up, the proportion of ``stay" tends to decline as $a$ goes up (see Fig.~\ref{fig:unconditionalshift}) (Spearman's rho $=$ -0.422, obs $=$504, $p=0.000$). The activity is enhanced as $a$ goes up. This tendency seems to be a suitable explain to that result of nearness centrality (see Fig.~\ref{fig:nearness}), i.e., the more active the individual is the more near center the population jumps towards to. Using group level centrality as dependent variable and group level shift ratio as independent variable, linear regression result shows these two variables are correlative (Coef.$=18.410$, $p=0.000$, two-tailed, s.e.$=$1.416, obs=84, $R^2=0.674$).


\begin{table}
\center
\small
\caption{
\bf{The three types of transit}}
\begin{tabular}{|c|c|cc|cc|cc|cc|cc|}
\hline
a	&Obs	&$T_{RSP}$	&s.e.	&$stay$	&s.e.	&$T_{RPS}$	&s.e.	&$T_{RSP}$-stay &$p$&$T_{RPS}$-stay &$p$ \\
\hline								
1	        &72	&0.128	&0.013	&0.708	&0.025	&0.164	&0.013	&-0.580  &0.000  &-0.544  &0.000\\
1.1	        &72	&0.222	&0.014	&0.531	&0.025	&0.247	&0.015	&-0.309  &0.000  &-0.285  &0.000\\
2	        &72	&0.282	&0.011	&0.440	&0.020	&0.279	&0.013	&-0.158  &0.000  &-0.161  &0.000\\
4	        &72	&0.301	&0.011	&0.411	&0.019	&0.288	&0.011	&-0.110  &0.000  &-0.122  &0.000\\
9	        &72	&0.333	&0.016	&0.370	&0.020	&0.297	&0.014	&-0.037  &0.131  &-0.074  &0.010\\
100	        &72	&0.327	&0.012	&0.386	&0.018	&0.286	&0.010	&-0.059  &0.022  &-0.100  &0.000\\
$\infty$	&72	&0.334	&0.009	&0.376	&0.016	&0.290	&0.010	&-0.043  &0.036  &-0.087  &0.001\\

   \hline
\end{tabular}
\begin{flushleft}
\end{flushleft}
\label{tab:transit}
\end{table}

\subsection{Phase change of individual shift direction}

\begin{figure}[!ht]
\begin{center}
           \includegraphics[width=4in]{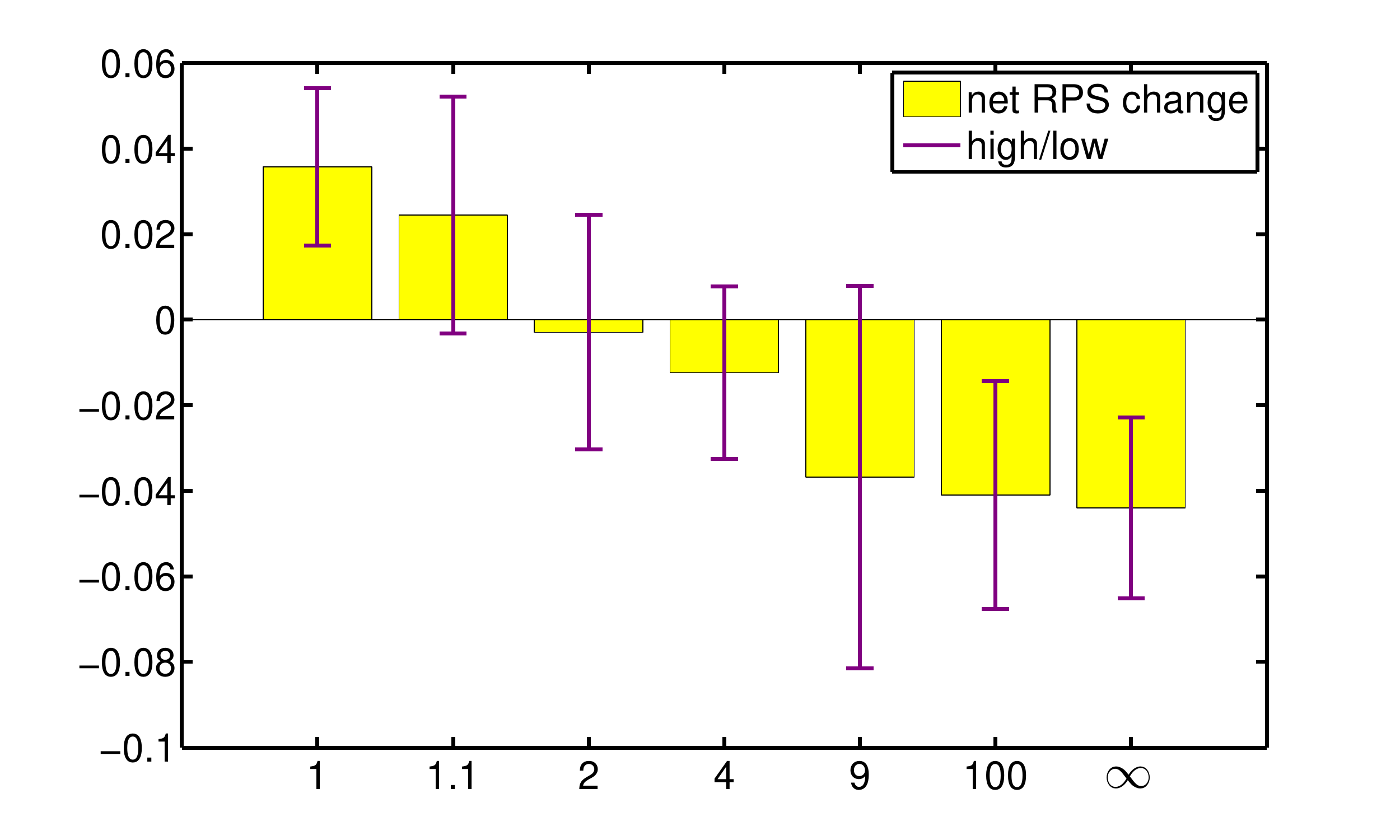}
      \end{center}
  \caption{
    \textbf{Net RPS change.} From left to right, $a=1,1.1, 2, 4, 9, 100, \infty$, respectively.
}
   \label{fig:netRPS}
\end{figure}

Since the social spire pattern is in the direction of counter-clockwise, i.e., $R\rightarrow P\rightarrow S$ direction. So we are especially interested in the net ``RPS" shift, which is amount number of ``RPS" shift after deducting the ``RSP" shift. Let $n_{RPS}$ be the number of `RPS" shift, $n_{RSP}$ be the `RSP" shift, and $n_{stay}$ be the number of ``stay". Then the net RPS shift proportion denoted as ``$nT_{RPS}$" is equal to $\frac{n_{RPS}-n_{RSP}}{n_{RPS}+n_{RSP}+n_{stay}}$. Fig.~\ref{fig:netRPS} illustrates the proportion of net ``RPS" shift.

It's amazing that the net RPS shift behavior declines as $a$ goes up (Spearman's rho $=$-0.2959, $Prob > |t| =0.0000$, obs=504) (Spearman's rho $=-0.5133$, $Prob > |t| =0.0000$, obs=84). The proportion of net RPS shift changes from positive ($z_{a=1} = 3.595$, $p=0.0001$, $z_{a=1.1} = 1.939$, $p=0.0525$, Wilcoxon signed-rank test, two-tailed, n=72) to negative ($z_{a=4} = -1.827$, $p=0.0667$, $z_{a=9} = -1.695$, $p=0.0901$, $z_{a=100} = -2.949$, $p=0.0032$, $z_{a=\infty} = -4.226$, $p=0.0000$, Wilcoxon signed-rank test, two-tailed, n=72). The transformation point is corresponding to neutral RPS game ($z_{a=2} = -1.501$, $p=0.1333$, Wilcoxon signed-rank test, two-tailed, n=72).

Intuitively, someone may expect that the individuals in unstable RPS game are more likely to move counter-clockwise, ``RPS" shift, and in stable RPS game are more likely to move clockwise, namely, ``RSP" shift, and this will induce population ``RPS" spire in the unstable game and ``RSP" spire in the stable game. However, as we have shown in above, the population jump out vectors in all of the games form counter-clockwise spire patterns.

Therefore, we haven't explained why jump out patterns in all of the games are counter-clockwise. Besides, why people are more likely to move counter-clockwise in unstable game and move clockwise in stable game?

\subsection{Conditional response behaviors}
As Ref.~\cite{wang2014conditional}, we are also interested in the conditional response behaviors. However, here we are especially interested in the change of conditional response behavior with winning payoff $a$.

\textbf{Lose condition}

\begin{figure}[!ht]
\begin{center}
     \includegraphics[width=4in]{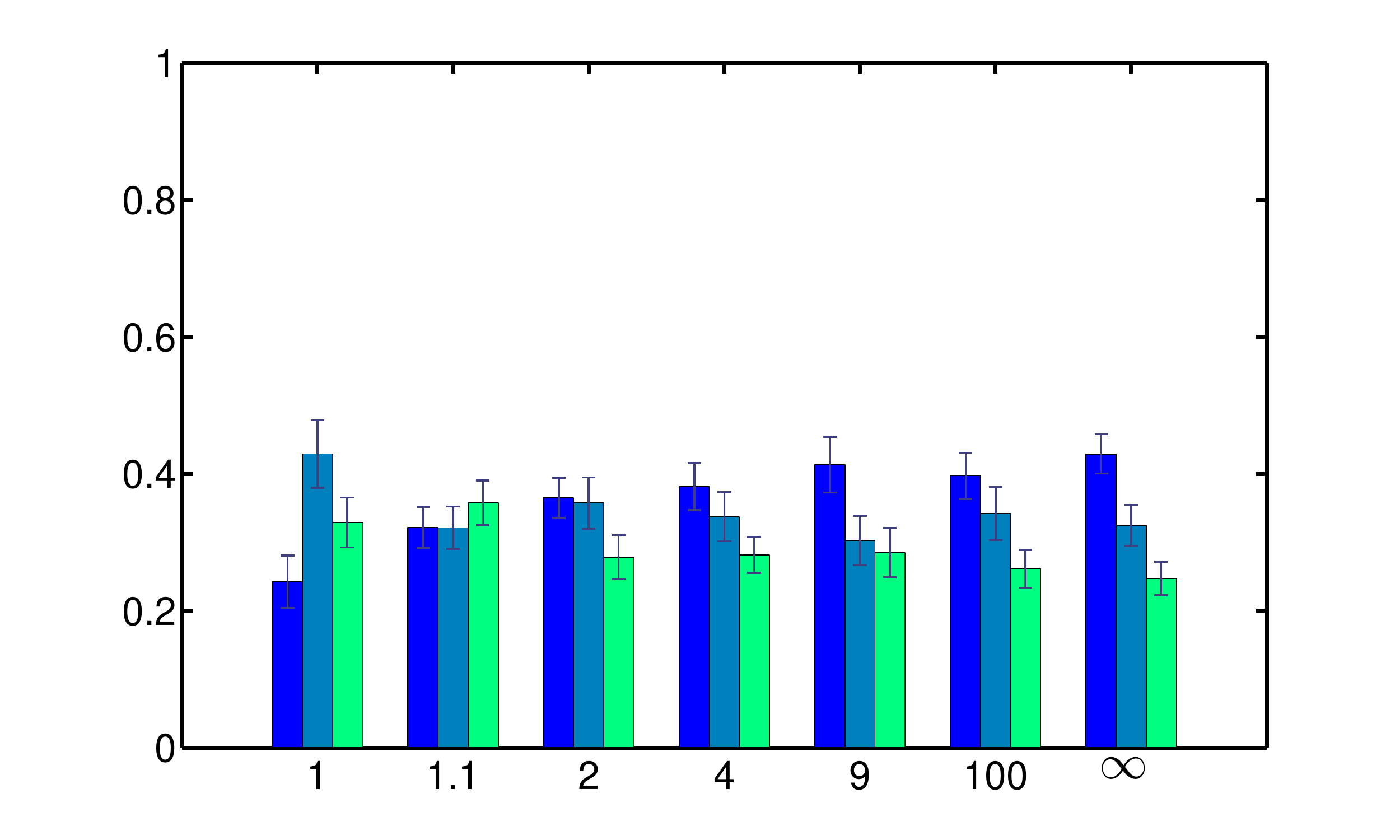}
            \end{center}
  \caption{
    \textbf{Lose Condition}. For each group of three, the first stick is$L_-$, the second is $L_0$, and the third is $L_+$.}
   \label{fig:losecondition}
\end{figure}

First, we report the outcome in the condition of lose. Fig.~\ref{fig:losecondition} illustrate the three individual shift when people lose. We find that, when people lose, they tend to increase $L_-$ behavior with the winning payoff $a$ (Spearman's rho=0.355, Prob $> |t|=0.000$), while decrease $L_0$ (Spearman's rho=-0.115, Prob $> |t|=0.010$)) and $L_+$ (Spearman's rho=-0.225, Prob $> |t|=0.000$) behaviors with winning payoff $a$.

\textbf{Tie condition}

\begin{figure}[!ht]
\begin{center}
          \includegraphics[width=4in]{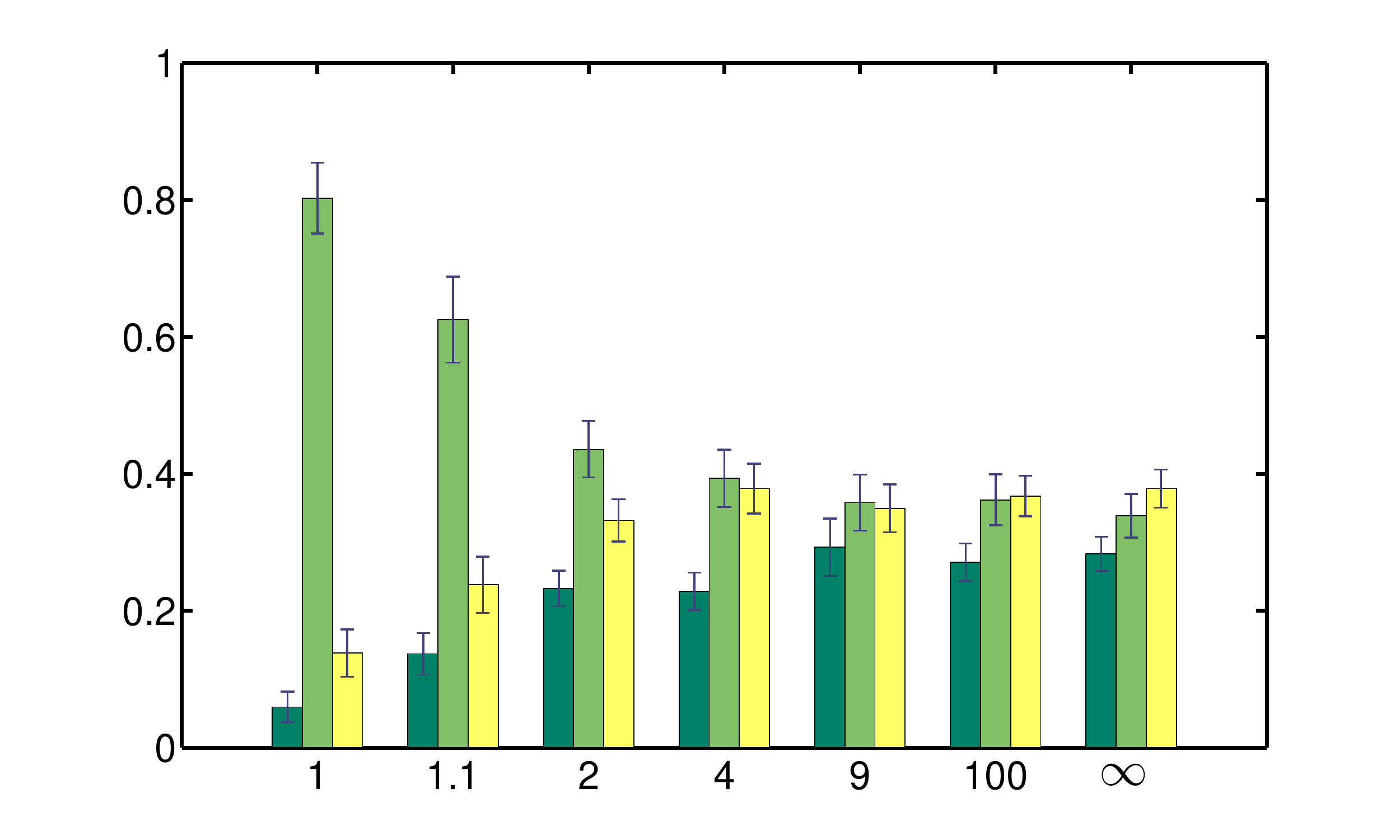}
           \end{center}
  \caption{
    \textbf{Tie Condition}. For each of the group of three, the first stick is $T_-$, the second is $T_0$, and the third is $T_+$.}
   \label{fig:tiecondition}
\end{figure}

Second, we report the outcome in the condition of tie. Fig.~\ref{fig:tiecondition} illustrate the three individual shift when people tie. We find that, when people tie, they tend to decrease $T_0$ behavior with the winning payoff $a$ (Spearman's rho=-0.531 Prob $> |t|=0.000$), while increase $T_-$ (Spearman's rho=0.507, Prob $> |t|=0.000$)) and $T_+$ (Spearman's rho=0.432, Prob $> |t|=0.000$) behaviors with winning payoff $a$.

\textbf{Win condition}

\begin{figure}[!ht]
\begin{center}
          \includegraphics[width=4in]{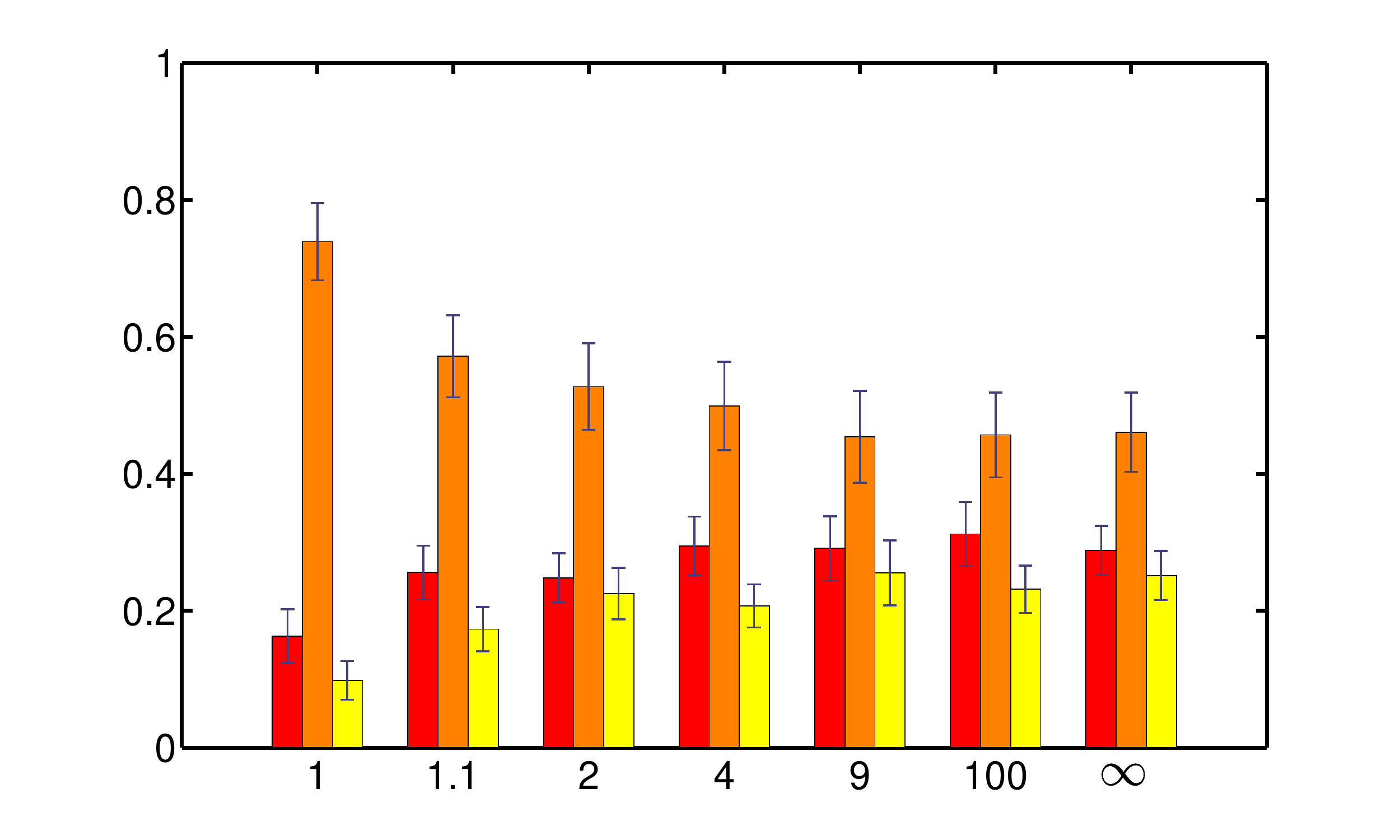}
           \end{center}
  \caption{
    \textbf{Win Condition}. For each of the group of three, the first stick is $W_-$, the second is $W_0$, and the third is $W_+$.}
   \label{fig:wincondition}
\end{figure}

Third, we report the outcome in the condition of win. Fig.~\ref{fig:wincondition} illustrate the three individual shift when people win. We find that, when people win, they tend to decrease $W_0$ behavior with the winning payoff $a$ (Spearman's rho=-0.292 Prob $> |t|=0.000$), while increase $W_-$ (Spearman's rho=0.213, Prob $> |t|=0.000$)) and $W_+$ (Spearman's rho=0.297, Prob $> |t|=0.000$) behaviors with winning payoff $a$.

\section{An explanation}

\subsection{Different payoffs corresponding to probability of win of state}
In the studied case, there 28 social states in state space. Since the proportions of three strategies are different, the probabilities of win, tie or lose are different. And therefore, as parameter $a$ changes, the payoff in different state will change.

Given a state $x_{(r,p,s)}$, the probabilities of winning, tie and losing are:

\begin{eqnarray}\label{eq:probabilitywintielose}
    p^{W} |x_{(r,p,s)}= \sum_{j\in\{R,P,S\}}\frac{n(k^w)}{N-1}p(j)|x_{(r,p,s)}\nonumber\\
    p^{T} |x_{(r,p,s)}= \sum_{j\in\{R,P,S\}}\frac{n(k^t)-1}{N-1}p(j)|x_{(r,p,s)} \\
    p^{L} |x_{(r,p,s)}= \sum_{j\in\{R,P,S\}}\frac{n(k^l)}{N-1}p(j)|x_{(r,p,s)}, \nonumber
\end{eqnarray}
where $k^w\not=j$ is the dominant strategy against $j$, $n(k^w)$ is the number of this strategy in the state,  $k^t=j$ is the same strategy as $j$, and $n(k^t)$ is the number of this strategy in the state, $k^l\not=j$ is the dominated strategy against $j$, $n(k^l)$ is the number of this strategy in the state, $p(j)$ is the proportion of strategy $j$, and $N$ is the total number of people in a population. And the three probabilities satisfy:
\begin{equation}\label{eq:indprobablityst}
p^{W} |x_{(r,p,s)}+ p^{T} |x_{(p,s)}+p^{L} |x_{(r,p,s)}=1.
\end{equation}
For a given state, one win always means the other lose. So, we have,
\begin{equation}\label{eq:stateprobablityst2}
p^{W} |x_{(r,p,s)}=p^{L}|x_{(r,p,s)}.
\end{equation}

\begin{figure}
\begin{center}
\includegraphics[width=0.45\linewidth]{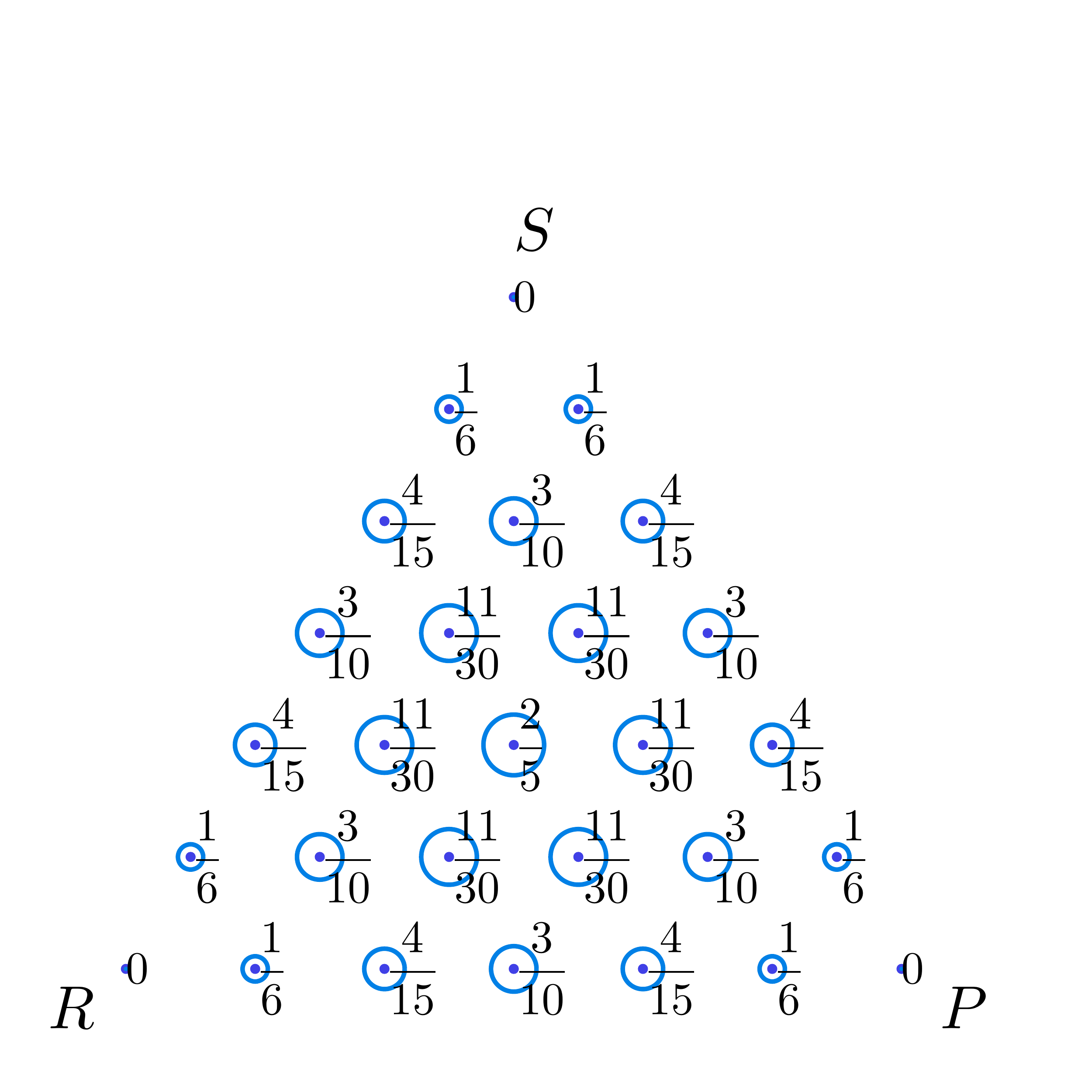}
 \includegraphics[width=0.45\linewidth]{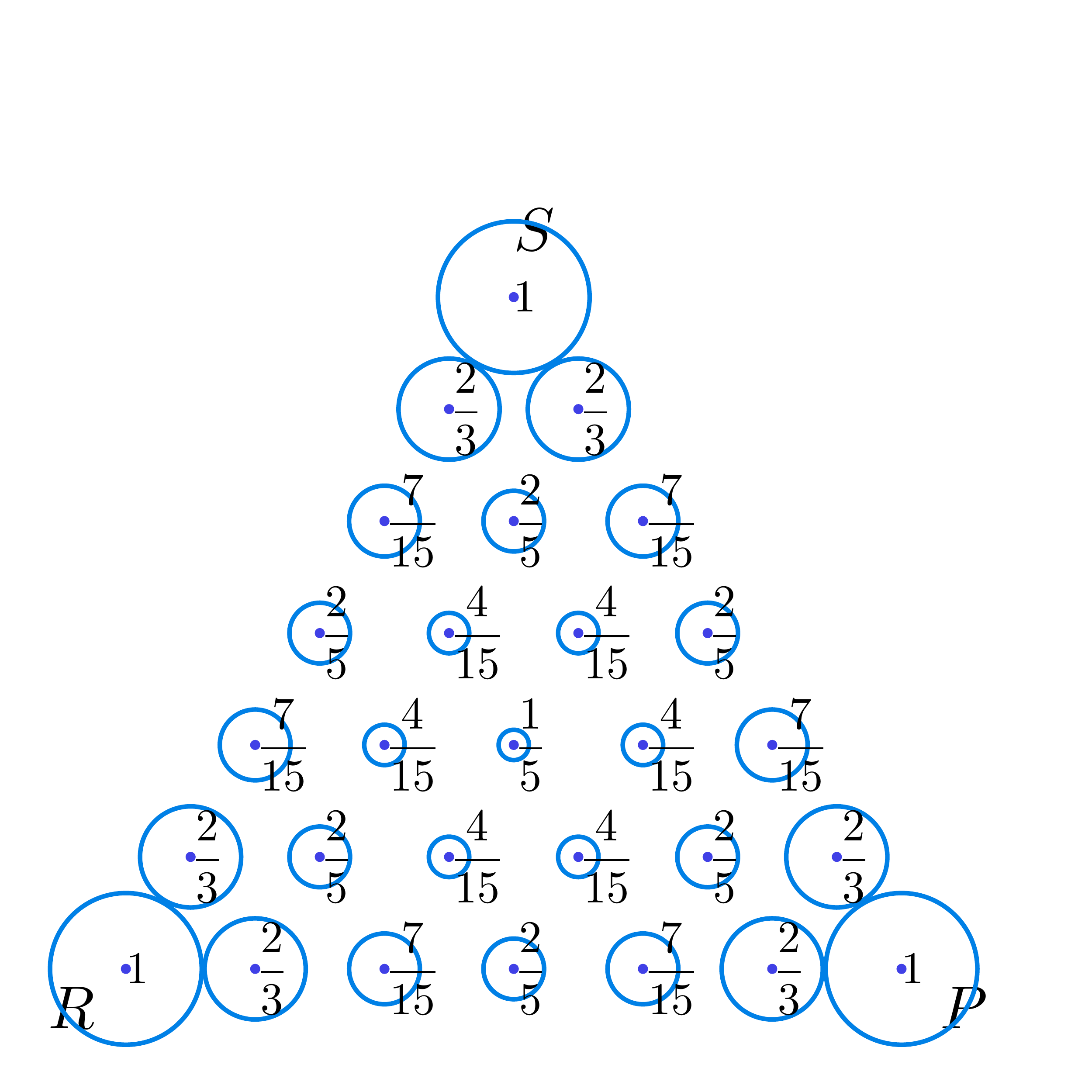}
\end{center}
\caption{Proportion of win, tie and lose in the studied case, population of six. Left, proportion of win or lose for each state, right, proportion of tie for each state.
}
  \label{fig:Proportionwintielose}
\end{figure}

Figure~\ref{fig:Proportionwintielose} illustrate the probabilities of win or lose (Figure~\ref{fig:Proportionwintielose}, left) and tie (Figure~\ref{fig:Proportionwintielose}, right) at each state by Eq.~\ref{eq:probabilitywintielose}. The probability for wining or losing in the center is larger than that of in the periphery, on contrary, the probability of tie in the center is smaller than in the periphery of the state space. The lowest probability of tie is 0.2 which appears at the center point $(\frac{1}{3},\frac{1}{3},\frac{1}{3})$, while the probability of win or lose is 0.4. There are six grades of amplitudes, from center to periphery, corresponding to these amplitudes, $p^{T}|x_{(p,s)}$ is $\frac{1}{5}, \frac{4}{15}, \frac{2}{5},\frac{7}{15},\frac{2}{3},1$ from small to large successively, and the probability $p^{W}|x_{(p,s)}$  ($p^{L}|x_{(p,s)}$), on contrary, from center to periphery is $\frac{2}{5}, \frac{11}{30}, \frac{3}{10},\frac{4}{15},\frac{1}{6},0$, from large to small successively.

Theoretically, if $a$=2, the game is a constant sum game in which each two person get the constant sum of payoffs 2 no matter tie or win-loss. Therefore, every state yields same sum of payoffs in neutral RPS game. On contrary, the sum is nonconstant in both unstable RPS game and stable RPS games. The sum of payoffs is $2$ for both tying and the sum of payoffs for one win and one loss is $a$. Therefore, the average payoff of one win and one lose is larger than the payoff of one tie if $a>2$, the average payoff of one win and one lose is smaller than the payoff of one tie if $a<2$, and the average payoff of one win one lose is equal to the payoff of one tie if $a=2$. So, the total payoffs of a state depends on the probabilities of win/lose and tie. If a state has more opportunity to tie, the state will yield larger sum of payoffs in unstable game, and on contrary, the state in stable game will yield smaller sum of payoffs. In unstable RPS game, the states near the margin yield larger sum of payoffs, on contrary, in stable RPS game, the states near the center yield larger sum of payoffs. The larger the winning payoff is, the higher the relative payoff in  the center is.

\subsection{Decentralized individual responses and collective phenomenon}
So far, we seem to explain why the population more likely to run towards the center when $a$ goes up. This is because that the payoffs in the center of the state space are relative higher in stable RPS games.  However, how can a population know what state they arrived and then decide to jump out to a state together? Actually, every player in the experiment only gets the information about her/his and her/his opponent's strategies, nobody know what social state the population arrived. Further, every round, the pairs of players are randomly generated, and no information about the opponent's historical choice would be provided. It is impossible to command the population stay or shift. The spire jump out pattern come out to be the result of ``invisible hand".

After one round of play, an individual will face three possible outcomes, i.e., win (W), tie (T) and lose (L), and has three choices stay (0), turn to the dominated strategy beaten by the current strategy, we call left-shift ($-$), or turn to the dominate strategy beating the current strategy, we call it right-shift ($+$). Therefore, there are total 9 conditional probabilities, $L_{-}$, $L_{0}$, $L_{+}$, $T_{-}$, $T_{0}$, $T_{+}$, $W_{-}$, $W_{0}$, $W_{+}$~\cite{wang2014conditional}.

Suppose the conditional probabilities are constant and common among a population under a parameter. Then we apply these empirical conditional probabilities to every social state to get the theoretical jump out vector.

Given a state (r, p, s), the average next round state ($r_2, p_2, s_2$) is

\begin{eqnarray}\label{eq:conditional}
  r_2=p_R^L L_0+p_P^L L_-+p_S^L L_+ +p_R^T T_0+p_P^T T_- +p_S^T T_+ +p_R^W W_0+p_P^W W_-+p_S^W W_+, \nonumber\\
   p_2=p_R^L L_+ +p_P^L L_0+p_S^L L_- +p_R^T T_+ +p_P^T T_0 +p_S^T T_- +p_R^W W_++p_P^W W_0+p_S^W W_-, \\
   s_2=p_R^L L_- +p_P^L L_+ +p_S^L L_0 +p_R^T T_- +p_P^T T_+ +p_S^T T_0 +p_R^W W_- +p_P^W W_+ +p_S^W W_0, \nonumber
\end{eqnarray}
where, $p_R^L =r \frac{pN}{N-1}$, $p_P^L =p \frac{sN}{N-1}$, $p_S^L =s \frac{rN}{N-1}$, $p_R^T =r \frac{rN-1}{N-1}$, $p_P^T =p \frac{pN-1}{N-1}$, $p_S^T =s \frac{sN-1}{N-1}$, $p_R^W =r \frac{sN}{N-1}$, $  p_P^W =p \frac{rN}{N-1}$, $p_S^W =s \frac{pN}{N-1}$.

\begin{figure}[!ht]
\begin{center}
        \includegraphics[width=2in]{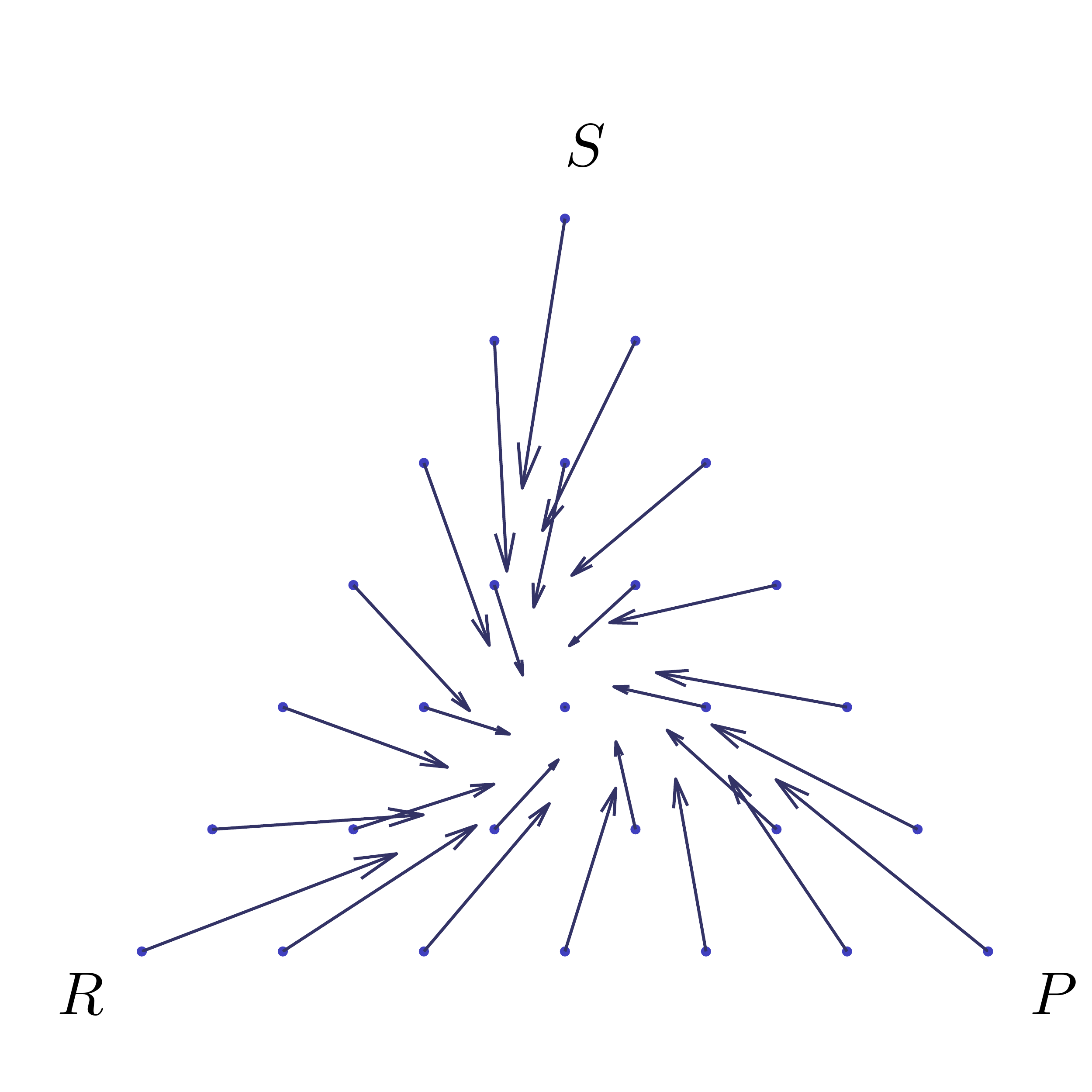}
    \includegraphics[width=2in]{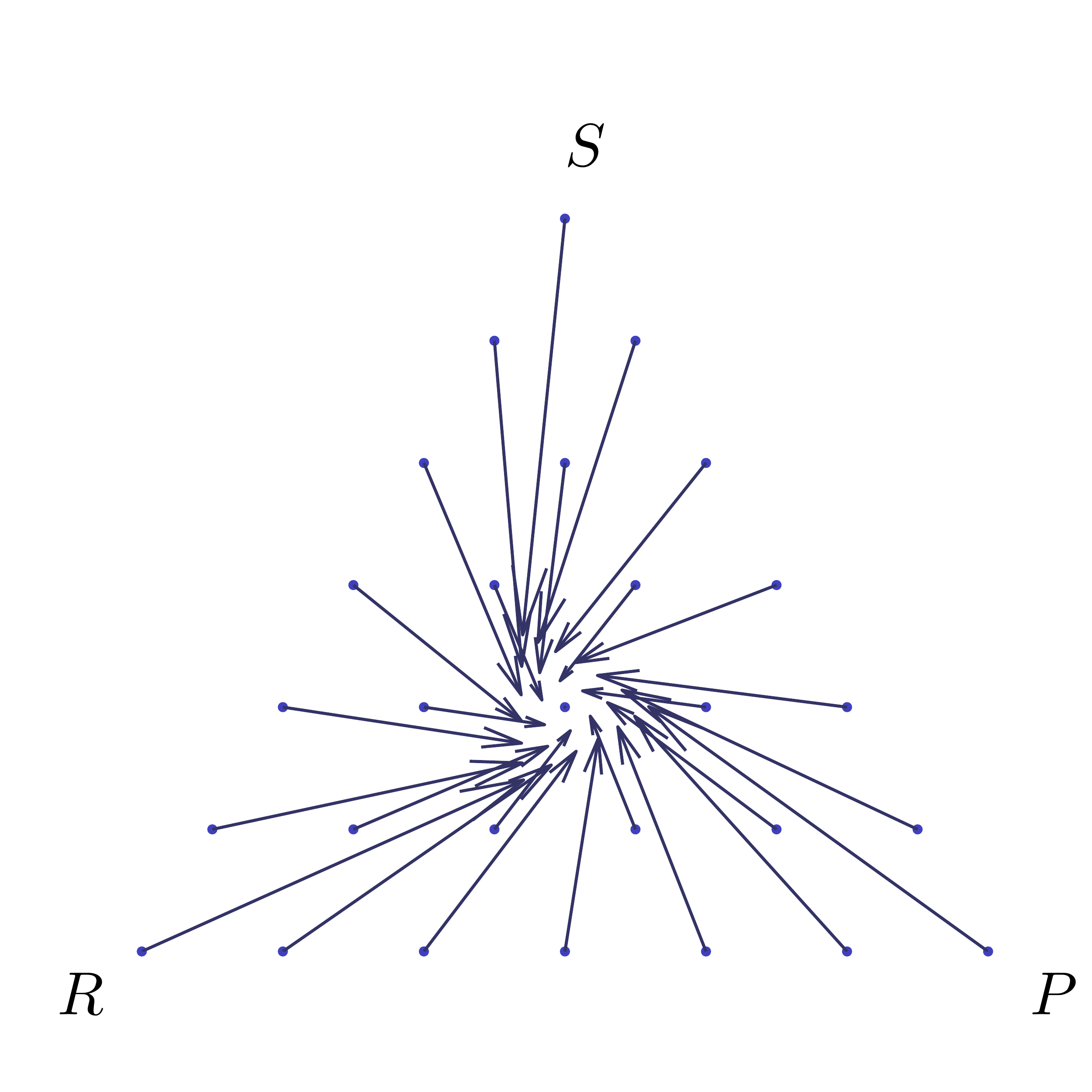}
       \includegraphics[width=2in]{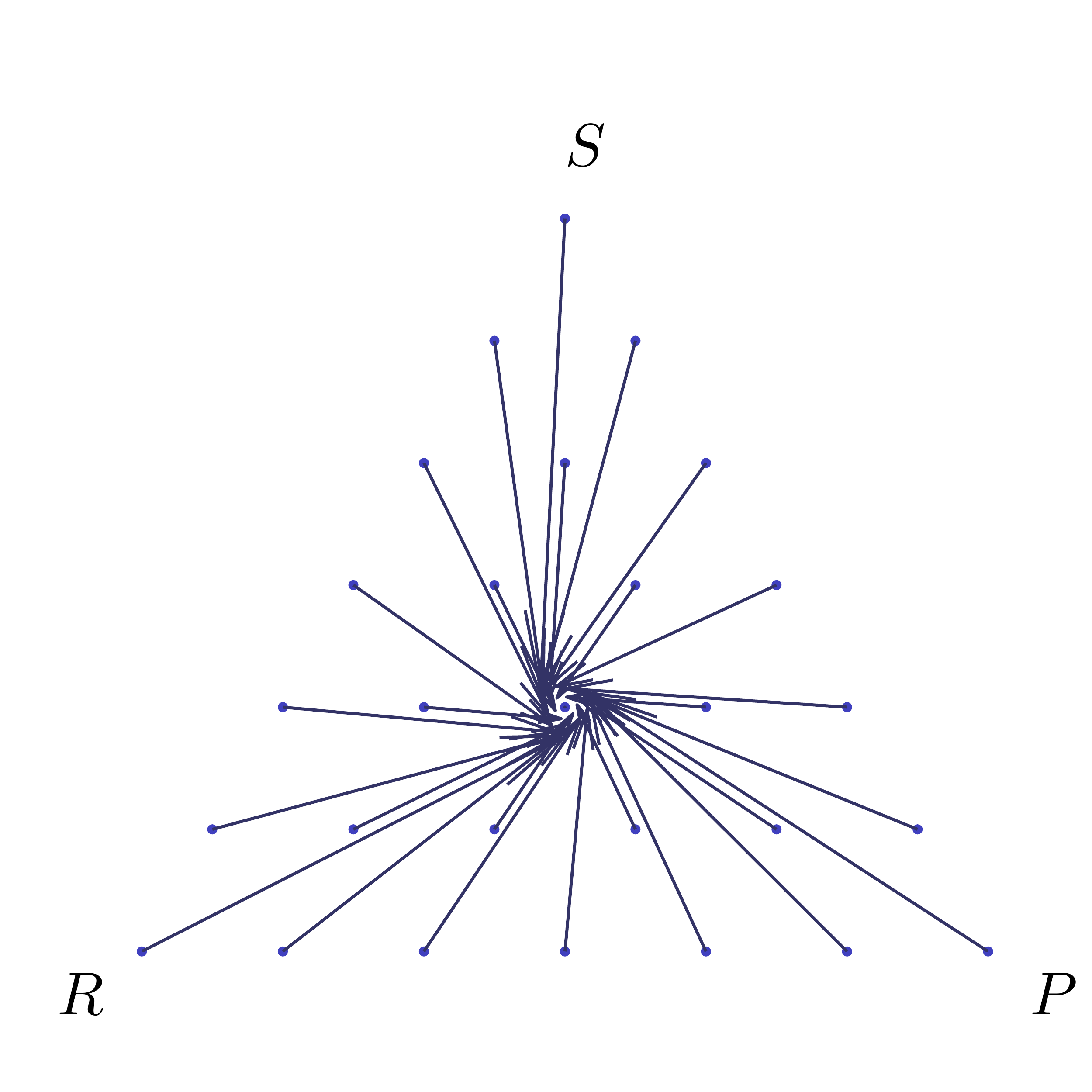}
      \end{center}
  \caption{
    \textbf{Decentralized conditional responses form  collective spire patterns}. Left, $a=1.1$, middle, $a=2$, right, $a=9$.}
   \label{fig:conditional}
\end{figure}

Using the empirical conditional responses parameters $L_{-}$, $L_{0}$, $L_{+}$, $T_{-}$, $T_{0}$, $T_{+}$, $W_{-}$, $W_{0}$, $W_{+}$ from experiments, we calculate the theoretical jump out vector for each state in each game according to Eq.~\ref{eq:conditional} and graph them in Figure~\ref{fig:conditional} and Figure~\ref{fig:payoffandall}, see Appendix.

Figure~\ref{fig:conditional} looks very like Figure~\ref{fig:Spire}, the real results from real experimental data. This indicator that the collective spire jump out pattern can be explained by the decentralized individuals' conditional response behaviors. That means, without centralized command, only given the local limit information about the previous both parties' strategies, the aggregated individuals' behaviors form a clear collective pattern.

\subsection{The elemental driving force of revolve}
So far, we have explained the collective revolving jump out pattern will form even if there is no centralized command and information. However, what kind of individual behavior drives the jump out vector revolve?

As mentioned above, there are total 9 conditional probabilities, $L_{-}$, $L_{0}$, $L_{+}$, $T_{-}$, $T_{0}$, $T_{+}$, $W_{-}$, $W_{0}$, $W_{+}$~\cite{wang2014conditional}. Among them, three kind of individual behaviors contribution nothing to move, i.e., $L_{0}$, $T_{0}$, $W_{0}$. The remain six individual behaviors move either counter-clockwise or clockwise. So we decompose behaviors into six elements, $L_{-}$, $L_{+}$, $T_{-}$, $T_{+}$, $W_{-}$, $W_{+}$ to analyze the driving force of revolve.

\begin{figure}[!ht]
\begin{center}
       \includegraphics[width=1in]{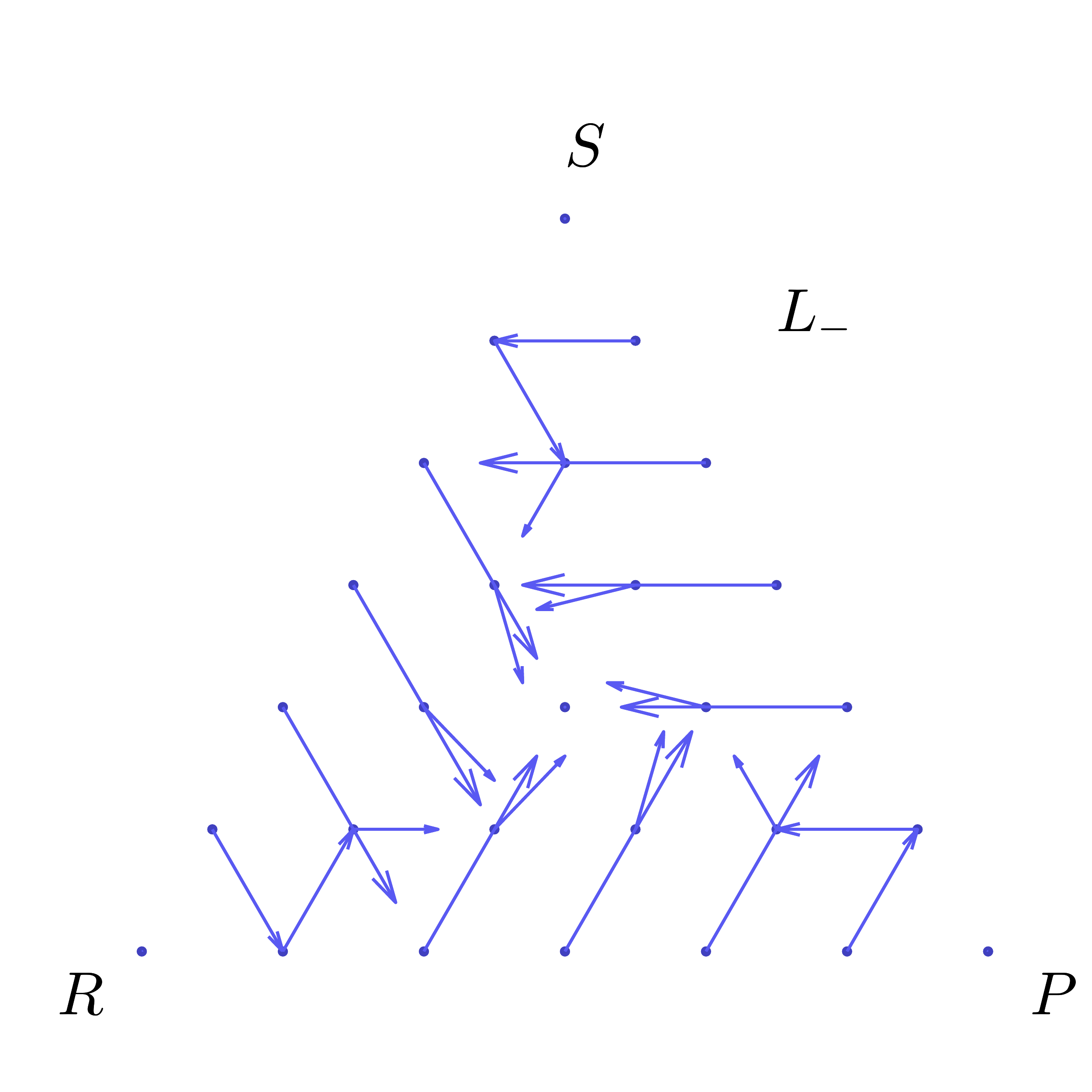}
       \includegraphics[width=1in]{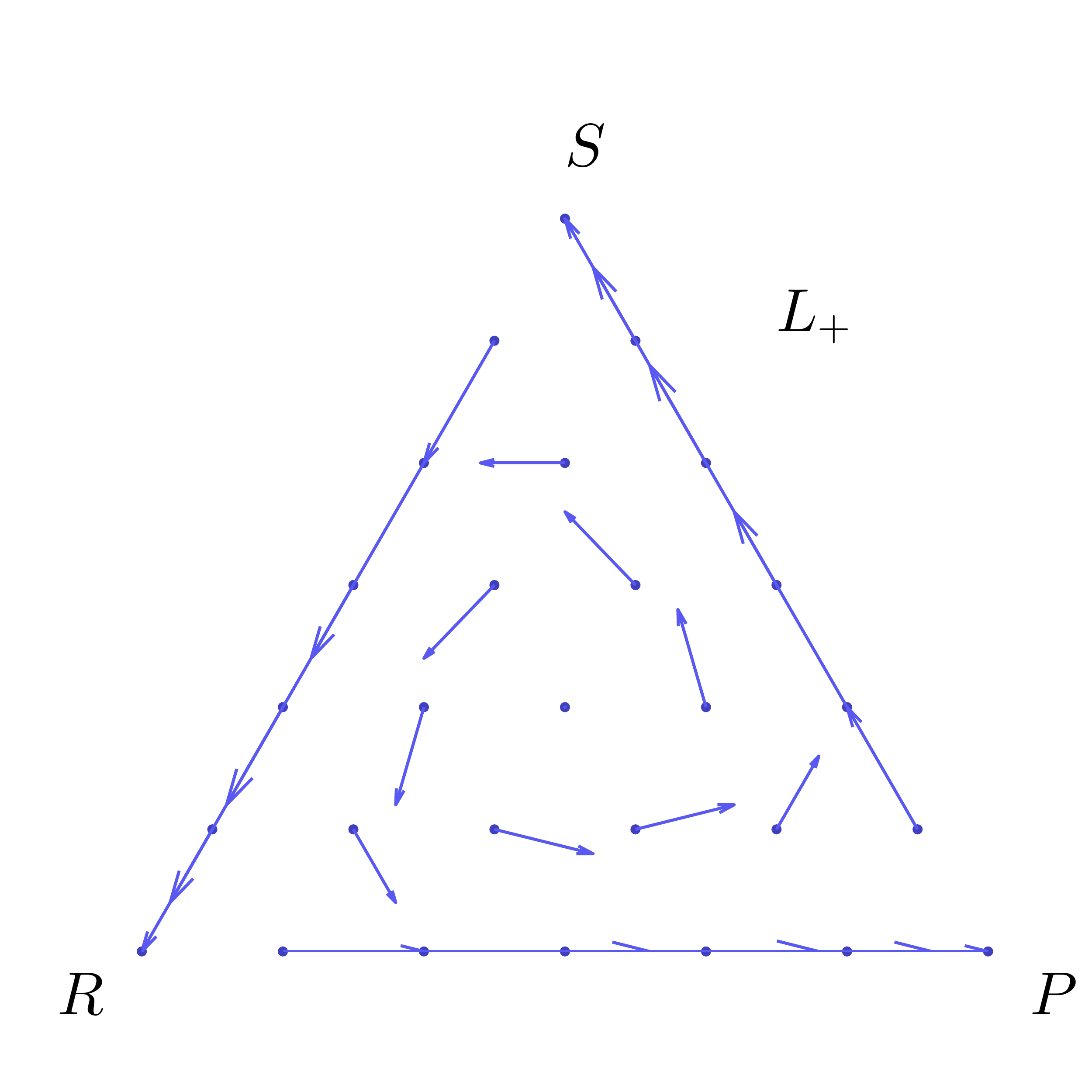}
       \includegraphics[width=1in]{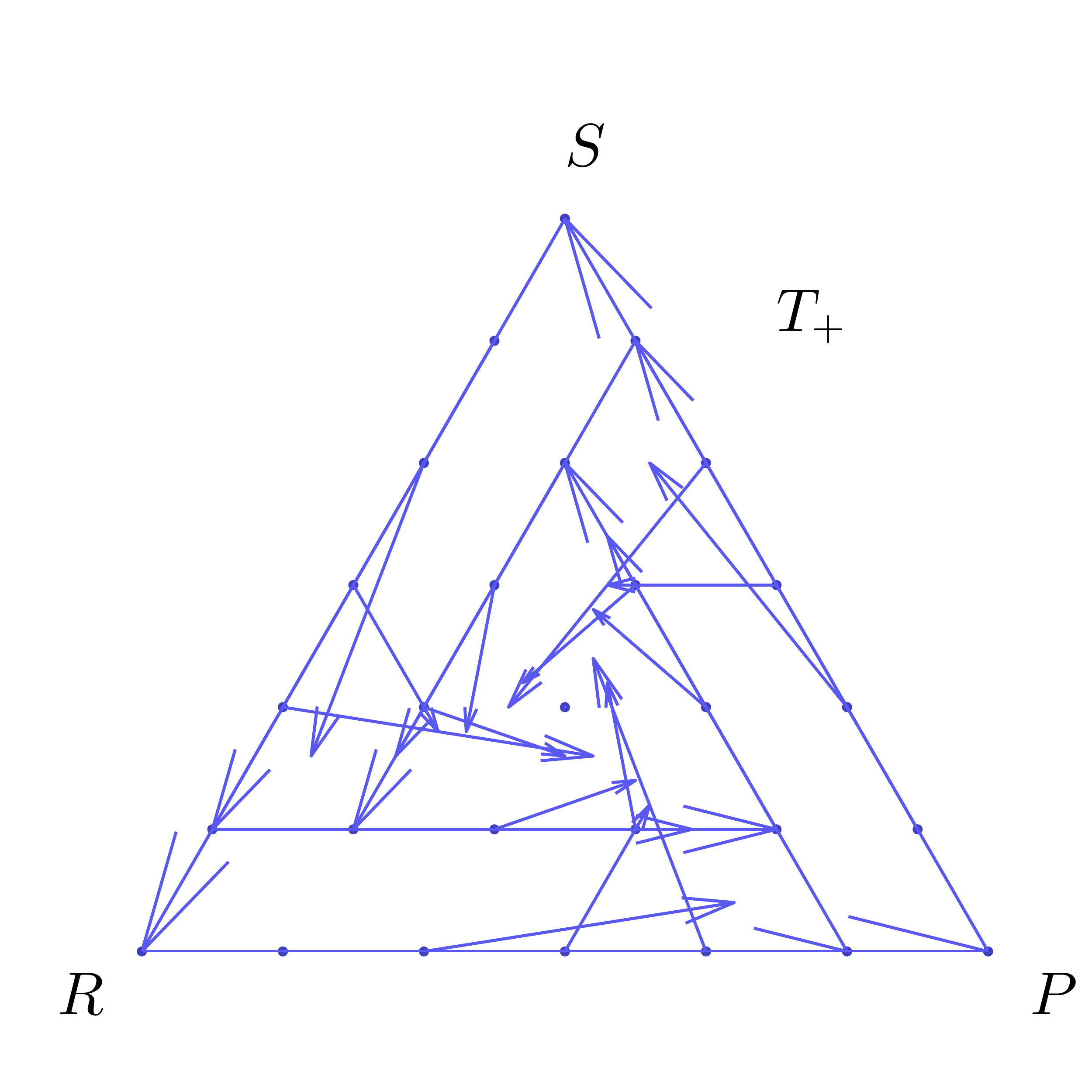}
       \includegraphics[width=1in]{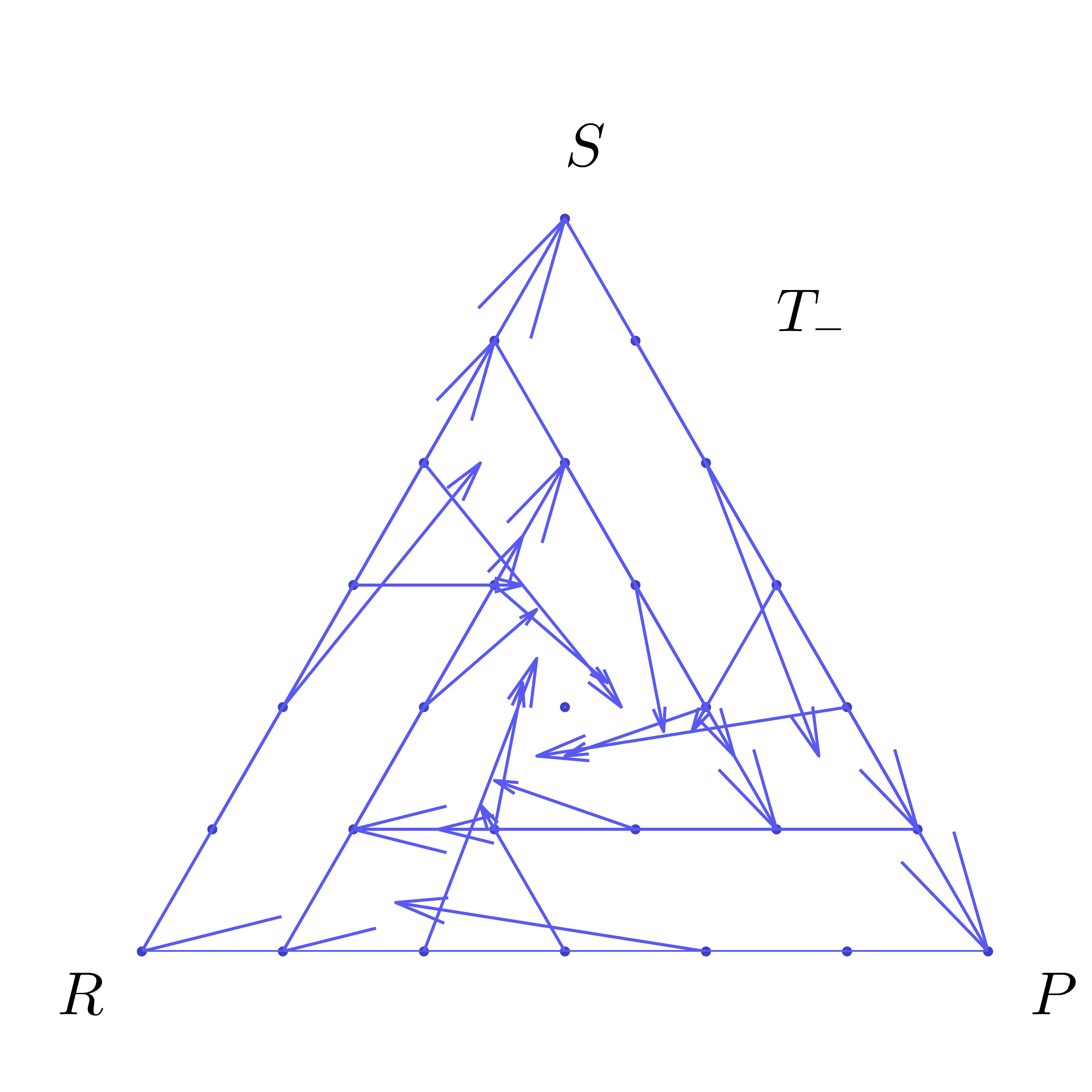}
       \includegraphics[width=1in]{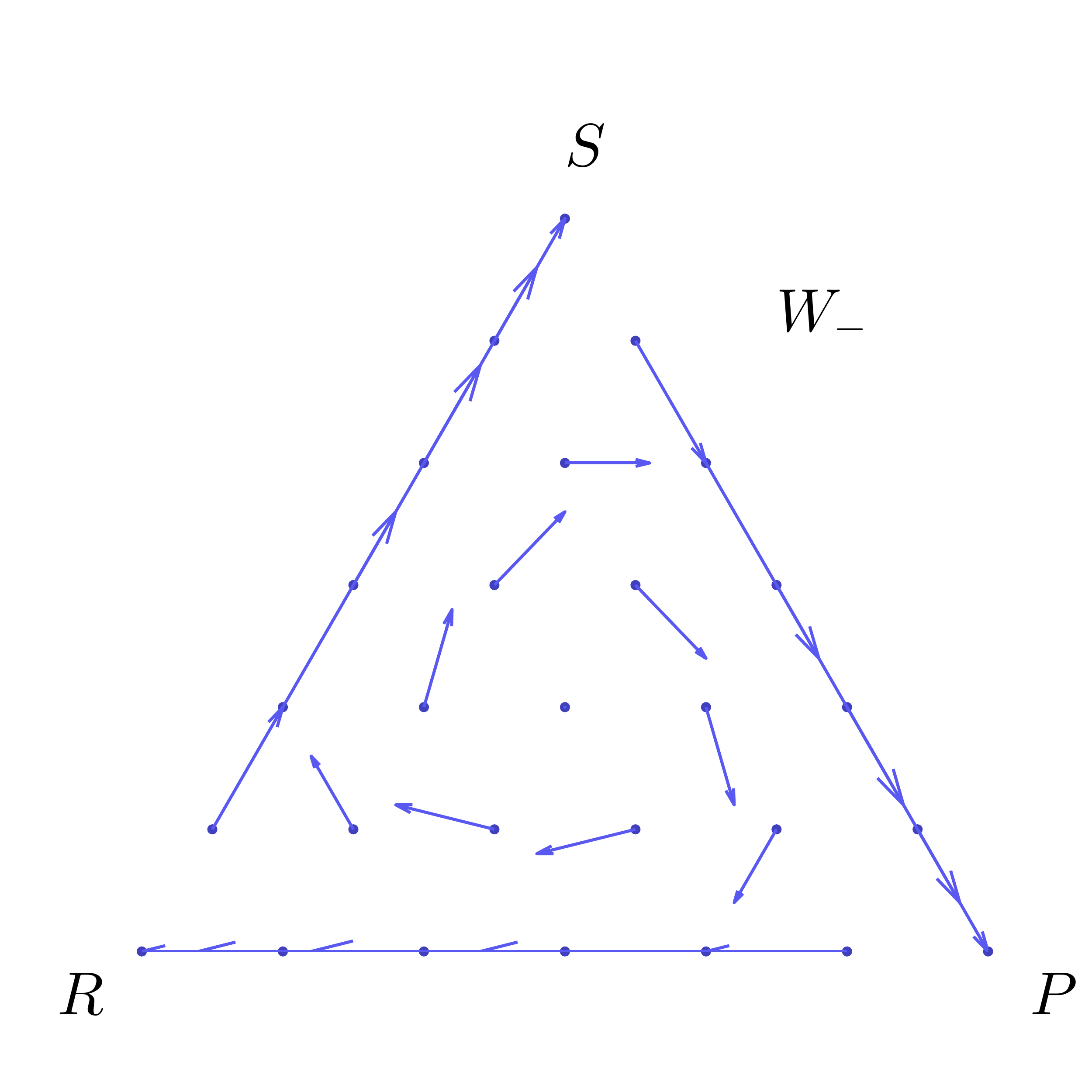}
       \includegraphics[width=1in]{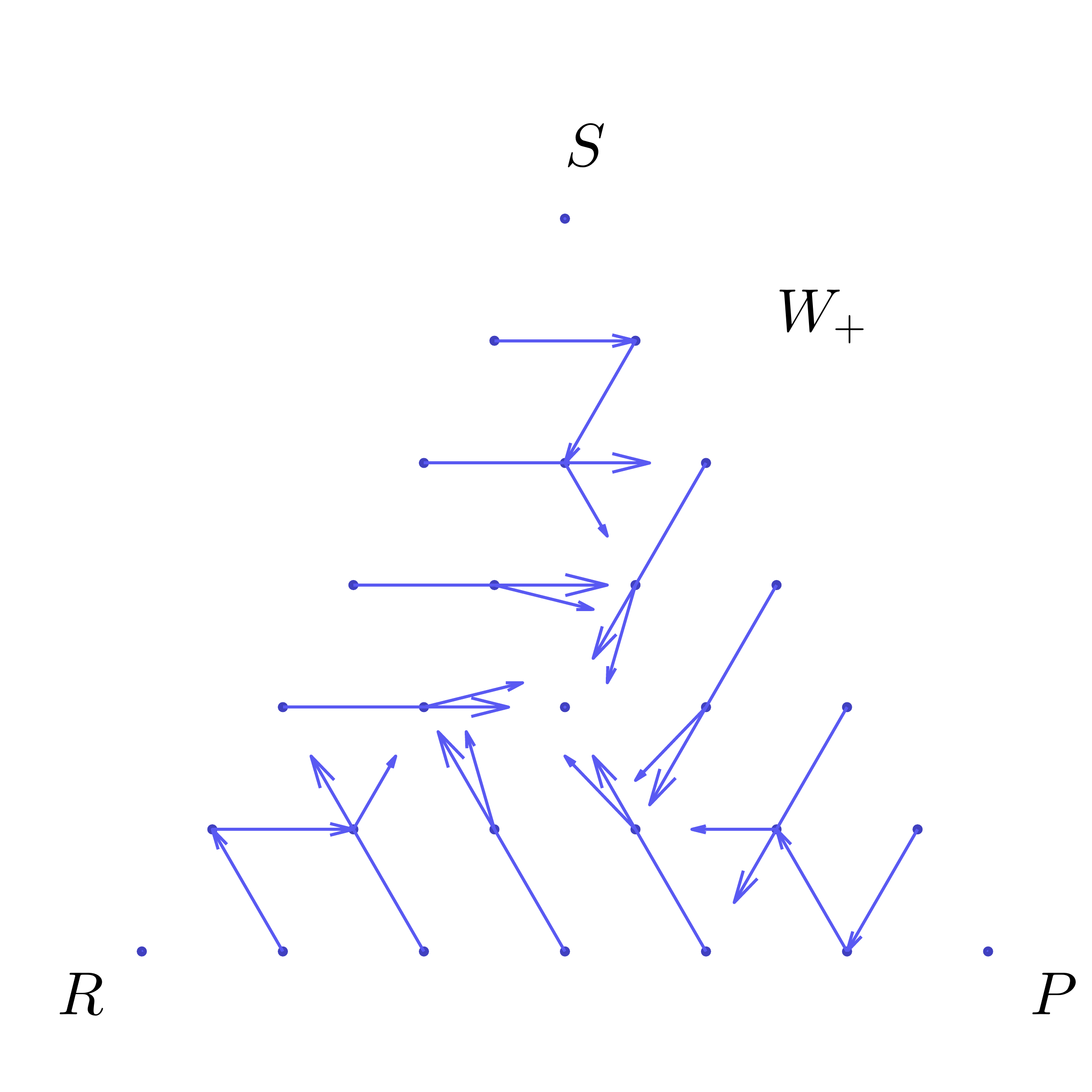}
      \end{center}
  \caption{
    \textbf{Theoretical spire patterns in extreme cases}. From left to right, $L_-=T_0=W_0=1$, $L_+=T_0=W_0=1$, $L_0=T_+=W_0=1$, $L_0=T_-=W_0=1$, $L_0=T_0=W_-=1$, $L_0=T_0=W_+=1$.}
   \label{fig:sixelements}
\end{figure}

Fig.~\ref{fig:sixelements} graph the six type of evolve patterns according to Eq.~\ref{eq:conditional}, given six combinational conditions (CC).
\begin{itemize}
  \item CC$_1$: $L_-=T_0=W_0=1$, $L_0=L_+=T_-=T_+=W_-=W_+=0$,
  \item CC$_2$: $L_0=T_-=W_0=1$, $L_-=L_+=T_0=T_+=W_-=W_+=0$,
  \item CC$_3$: $L_0=T_0=W_-=1$, $L_-=L_+=T_-=T_+=W_0=W_+=0$,
  \item CC$_4$: $L_+=T_0=W_0=1$, $L_-=L_0=T_-=T_+=W_-=W_+=0$,
  \item CC$_5$: $L_0=T_+=W_0=1$, $L_-=L_+=T_-=T_0=W_-=W_+=0$,
  \item CC$_6$: $L_0=T_0=W_+=1$, $L_-=L_+=T_-=T_+=W_-=W_0=0$.
\end{itemize}
The outcomes show that half of six motions cause counter-clockwise revolve, i.e., lose-left-shift, $L_-$, lose-right-shift, $L_+$, and tie-right-shift, $T_+$. On contrary, the remanent three motions cause clockwise revolve, i.e., tie-left-shift, $T_-$, win-left-shift, $W_-$, and win-right-shift, $W_+$. We notice that the lose-shift behaviors cause counter-clockwise collective revolve, no matter the individual shift direction is counter-clockwise or clockwise. On contrary, the win-shift behaviors cause clockwise collective revolve, no matter the individual shift direction is counter-clockwise or clockwise. And if people tie, the individual behavior cause same direction collective revolve, i.e., if people tie-right-shift, then the population revolve in the direction of counter-clockwise, and if people tie-left-shift, then the population revolve in the direction of clockwise.

\section{Discussion}
\subsection{Best response behavior}
There are many kind of behaviors, among them, there are two famous kinds of behaviors. One is the best response (BR) behaviors ~\cite{Porath1990,Matsui1992,Blume1995,Camerer1998,Hopkins1999,Juliusson2000,Crawford2001,Littman2001,Morris2004,Kukushkin2004,Conitzer2007,Rey-Biel2009} and the other is win-stay lose-shift (WSLS) behavior~\cite{nowak1993,Posch1999,Nowak2006,Imhof2007,Worthy2014}. First, we discuss the best response behaviors in different RPS games.

\begin{figure}[!ht]
\begin{center}
     \includegraphics[width=2in]{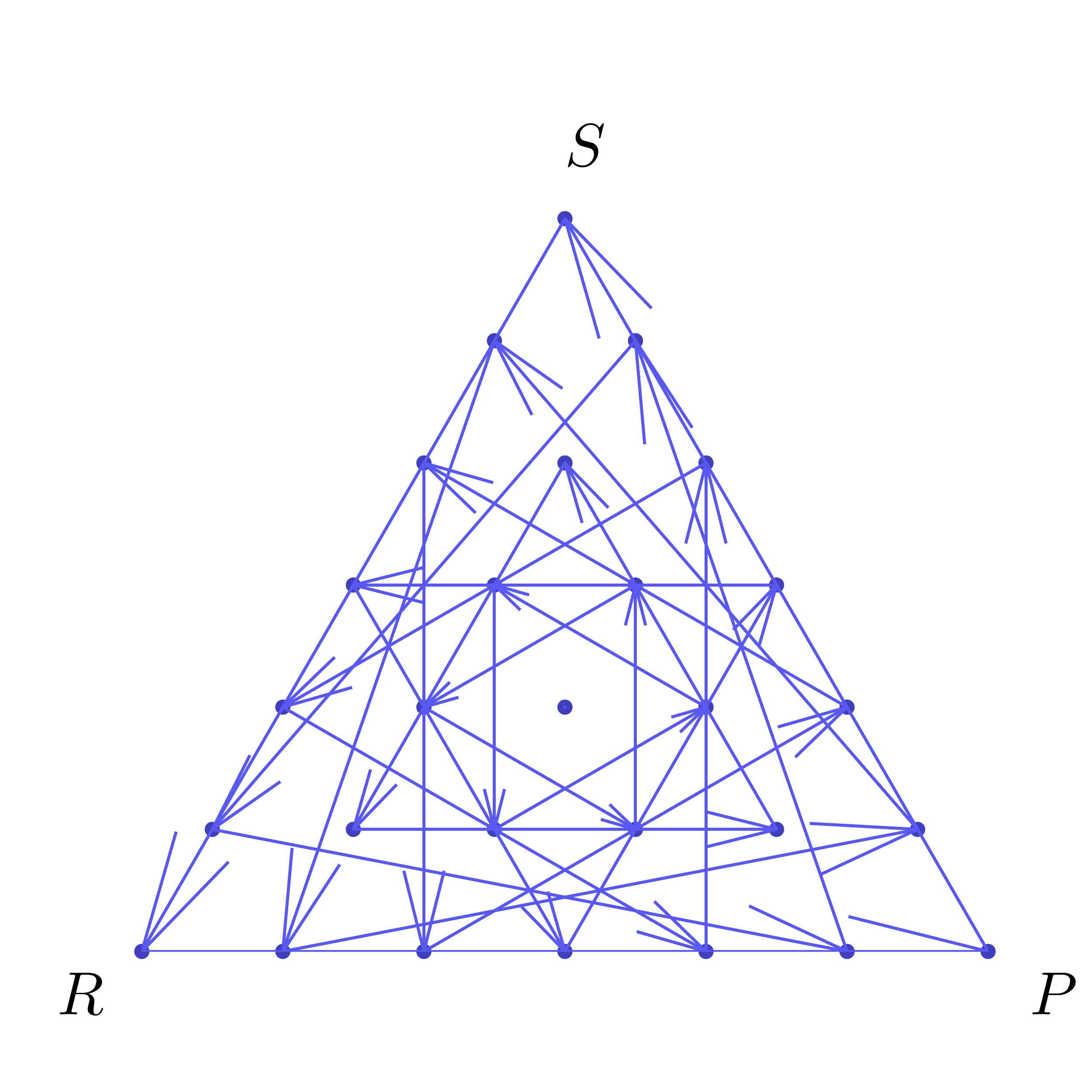}
     \includegraphics[width=3in]{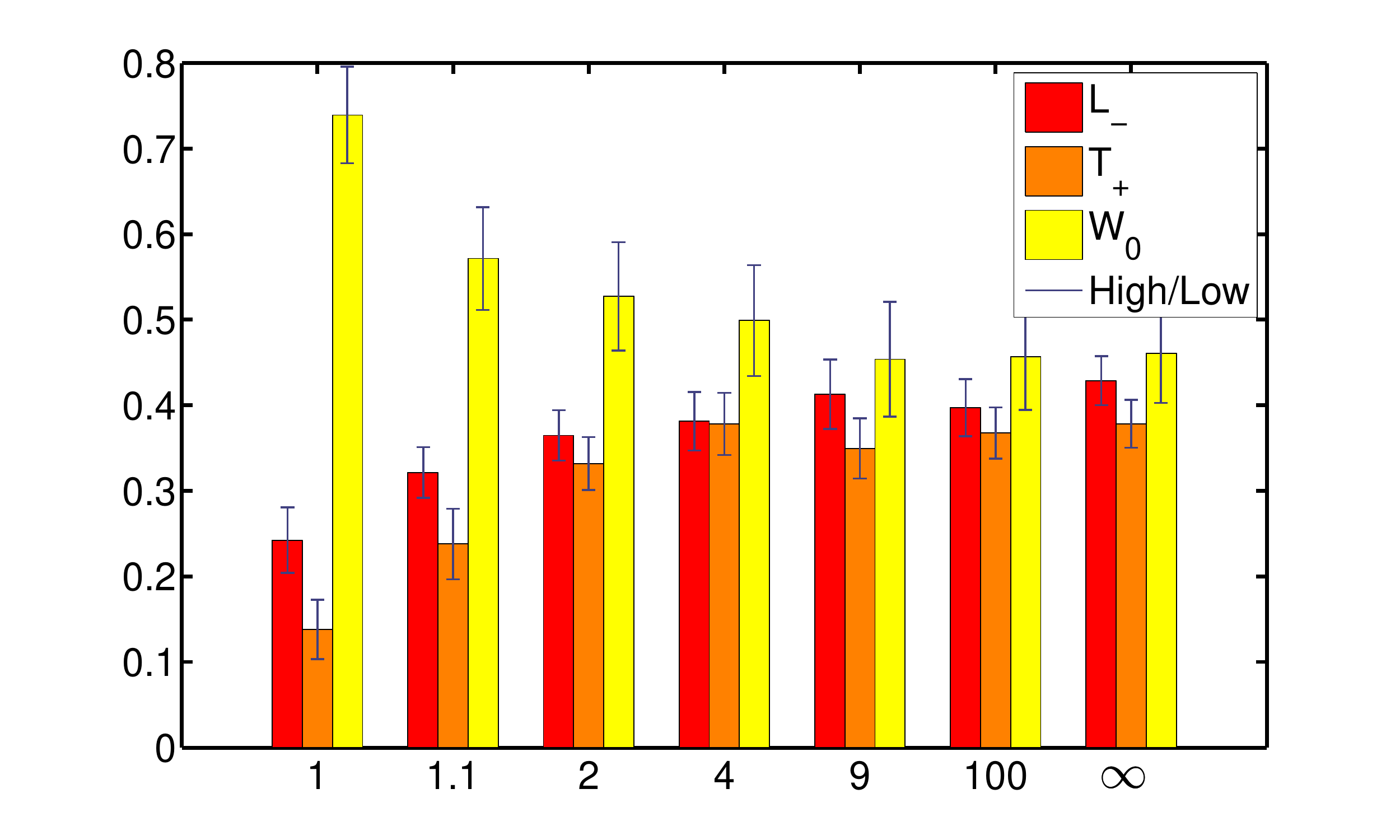}
      \end{center}
  \caption{
    \textbf{Best response}. Left, global pattern of pure best response behavior. Right, empirical best response proportion.}
   \label{fig:bestresponse}
\end{figure}

In the studied case of RPS game here, there are three kind of best response behaviors, i.e., $L_-$, $T_+$, and $W_0$. If people use pure best response strategy, then $L_-=T_+=W_0=1$, and the others all equal 0. Fig.~\ref{fig:bestresponse} left graphically represents the collective outcome of the individual pure best response behavior. It's clear that all jump out vectors form counter-clockwise revolve. So we can understand that the counter-clockwise population jump out vectors can be formed if people use best response strategy.

On the other side, the pure best response behavior hasn't provided the information about the relationship between behavior and different winning payoff $a$. So we turn back to the experimental data to investigate how best response behavior change with winning payoff $a$ changing. Fig.~\ref{fig:bestresponse} (right) illustrates the three best response behaviors' changing along the increasing of winning payoff $a$. As the winning payoff $a$ goes up, one of the best response behavior decline, i.e., win-stay (Spearman's rho =-0.2917, Prob $> |t|$ =       0.0000, obs=504), while the two other best response behavior ascend (for lose-left-shift, Spearman's rho =       0.3547, Prob $> |t|$ = 0.0000, for tie-right-shift, Spearman's rho =0.4317, Prob $> |t|$ = 0.0000, obs=504). As a whole, the best response behavior increasing as $a$ goes up (Spearman's rho =0.1010, Prob $> |t|$ =       0.0233, obs=504). This result indicate that different outcome of play, i.e., win or lose or tie, may influence people's choice in different directions.

\subsection{Win-stay lose-shift behavior}
\begin{figure}[!ht]
\begin{center}
    \includegraphics[width=2in]{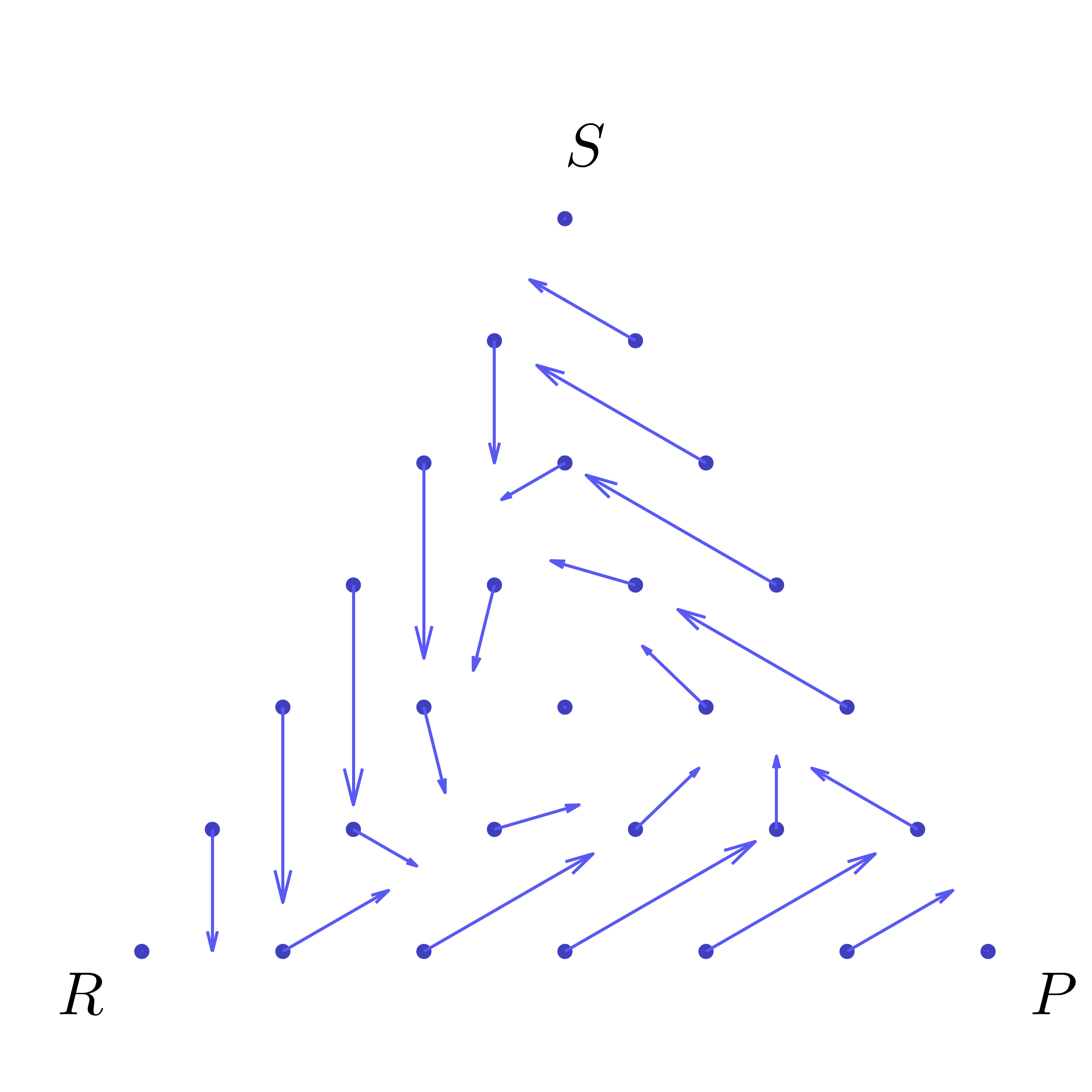}
          \includegraphics[width=3in]{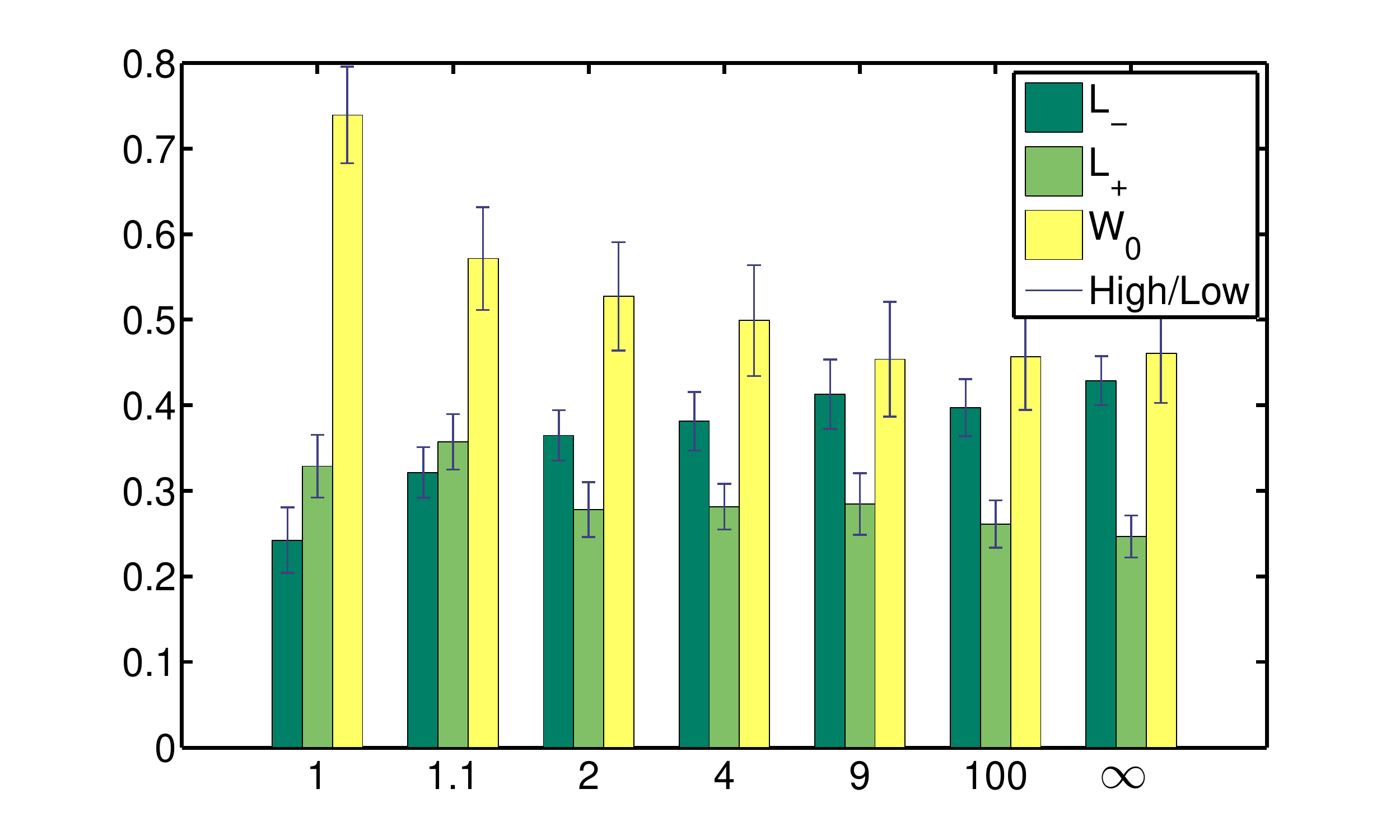}
           \end{center}
  \caption{
    \textbf{Win-stay lose-shift}. Left, global pattern of pure win-stay lose-shift behavior. Right, empirical win-stay lose-shift proportion.}
   \label{fig:winstayloseshift}
\end{figure}

In the studied case of RPS game here, there are three kind of WSLS behaviors, i.e., $L_-$, $L_+$, and $W_0$. If people use pure WSLS strategy, then $L_-+L_+=W_0=1$, and the others all equal 0. Fig.~\ref{fig:winstayloseshift} left graphically represents the collective outcome of the individual pure WSLS behavior. It's clear that all jump out vectors form counter-clockwise revolve. So we can understand that the counter-clockwise population jump out vectors also can be formed if people use WSLS strategy.

Fig.~\ref{fig:winstayloseshift} right graph the empirical three kind of win-stay lose-shift behaviors, i.e., $L_-$, $L_+$, and $W_0$. The win-stay behavior is decline when $a$ goes up (Spearman's rho =-0.2917, Prob $> |t|$ =0.0000, obs=504). The lose-left-shift behavior is increasing as $a$ goes up (Spearman's rho =0.3547, Prob $> |t|$ =0.0000, obs=504). But the lose-right-shift behaviors decline as $a$ goes up(Spearman's rho =-0.2249, Prob $> |t|$ =0.0000, obs=504). As a whole, the WSLS behavior decline as $a$ goes up (Spearman's rho =-0.2183, rob $> |t|$ =0.0000, obs=504).

\subsection*{Final remark}
The generalized RPS game is a representative model to show the dynamical strategy behavior appearing wildly in textbook of evolutionary game theory~\cite{Maynard1982evolution} and  behavior economics~\cite{Camerer2003}. In this paper, firstly and systematically, we provide the empirical evidences on how actual motions in these generalized RPS games depending on the controlled incentive $a$ (the winning payoff), both in individual and social levels. We hope that, these results can be a helpful message to promote the understanding our human social behaviors.
%

\section*{Acknowledgments}
We thanks Charles Plott for helful comments. We thank Hai-Jun Zhou for a recent collaboration on the finite-population Rock-Paper-Scissors game  [arxiv:1404.5199], especially the work [arxiv:1406.3668], which promote the present manuscript  greatly.  We thank Zunfeng Wang and Anping Sun for excellent assistant. This work was supported by the Fundamental Research Funds for the Central Universities (SSEYI2014Z), the State Key Laboratory for Theoretical Physics (Y3KF261CJ1), and the Philosophy and Social Sciences Planning Project of Zhejiang Province (13NDJC095YB).


\clearpage

\appendix{Appendix}

\begin{figure}[!ht]
\begin{center}
\includegraphics[width=1in]{PayoffMatrix0.pdf}
\includegraphics[width=1in]{azj0nor.pdf}
\includegraphics[width=1in]{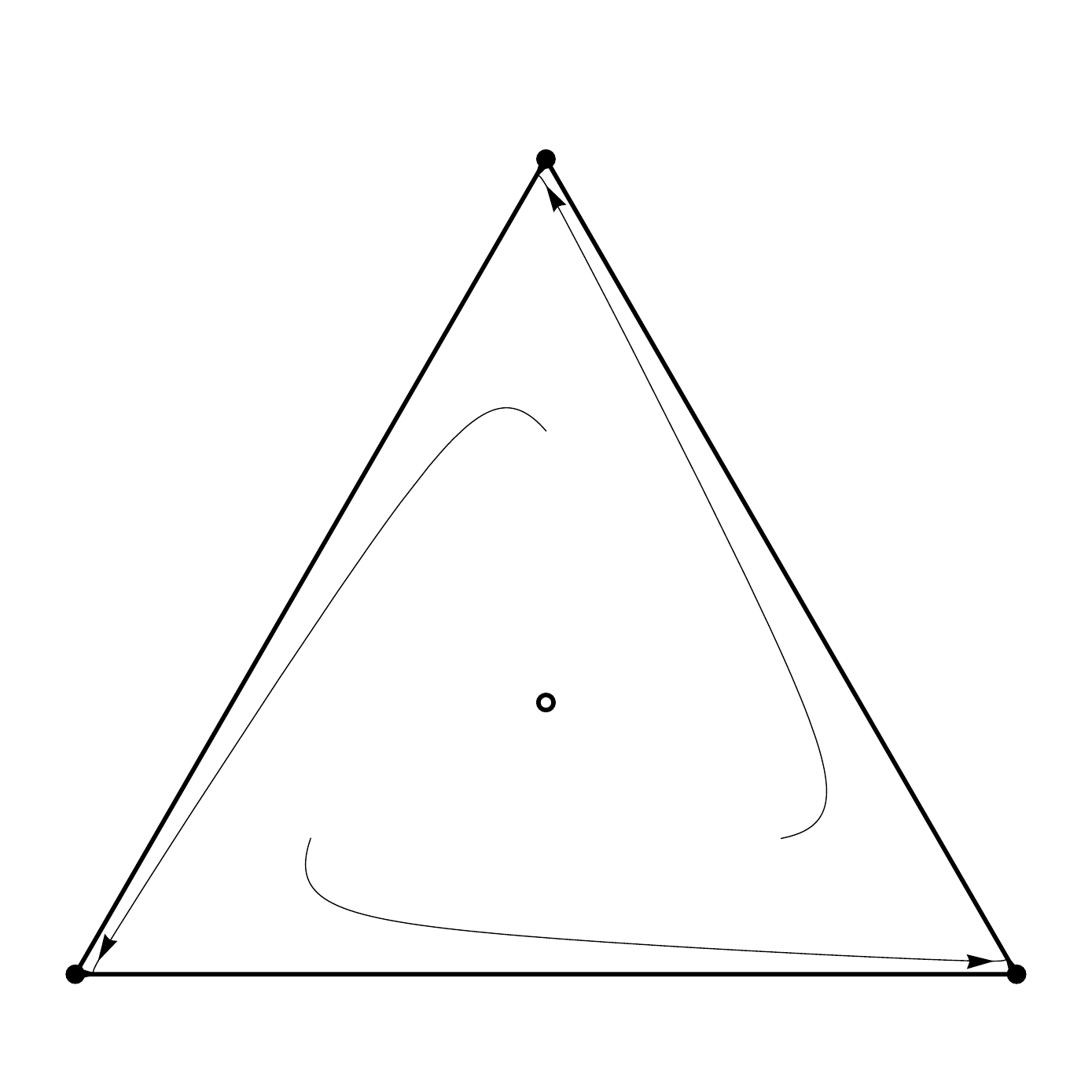}
\includegraphics[width=1in]{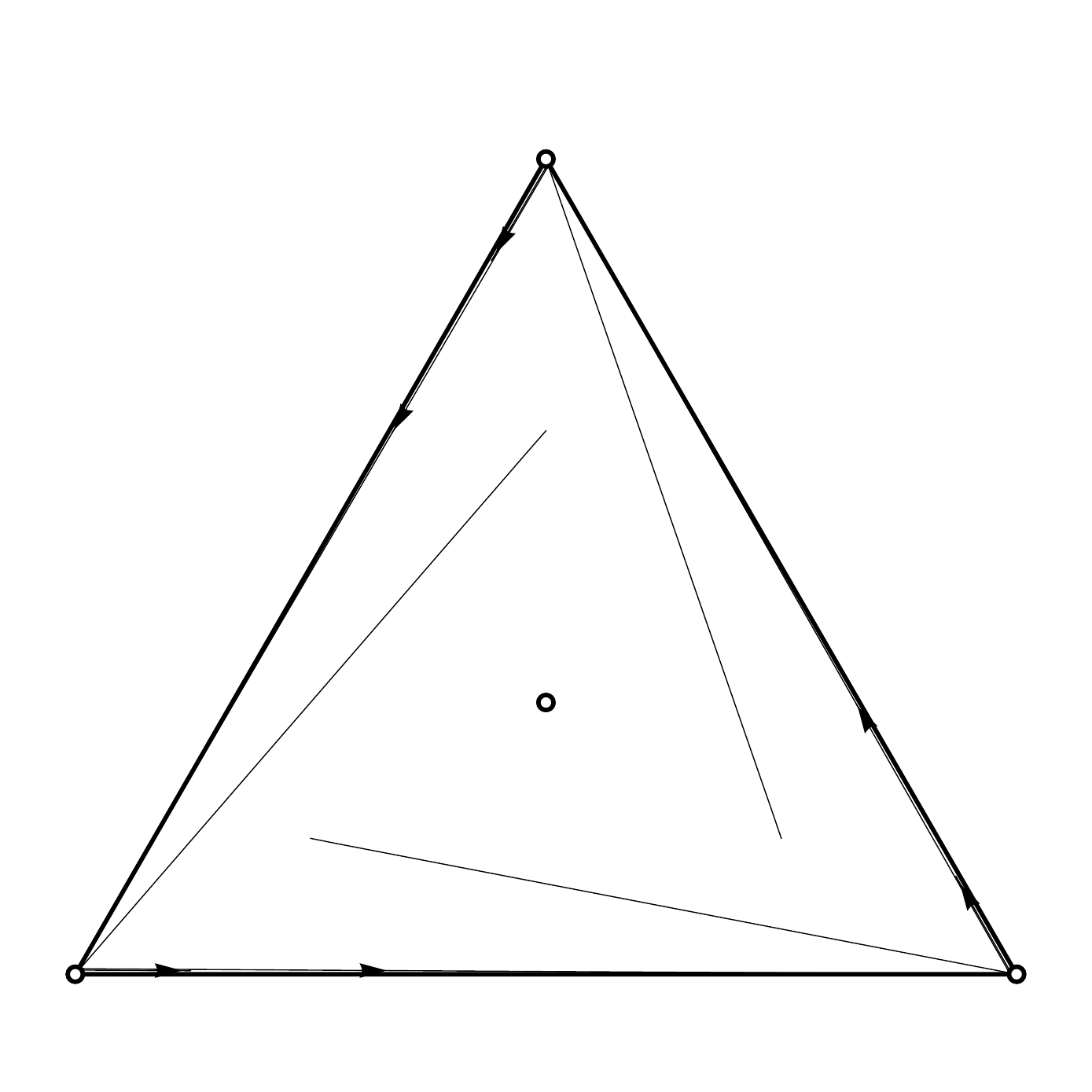}
       \includegraphics[width=1in]{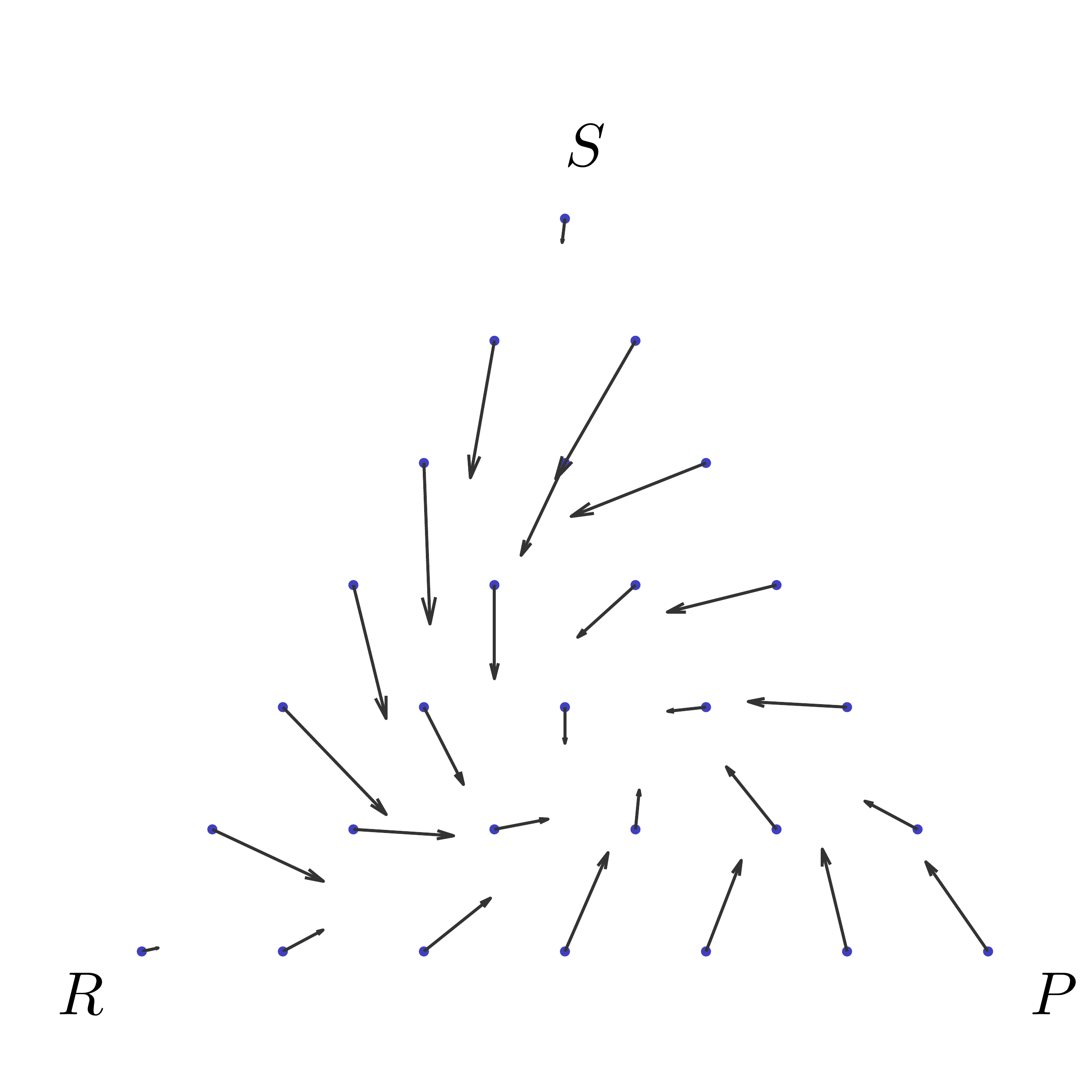}
       \includegraphics[width=1in]{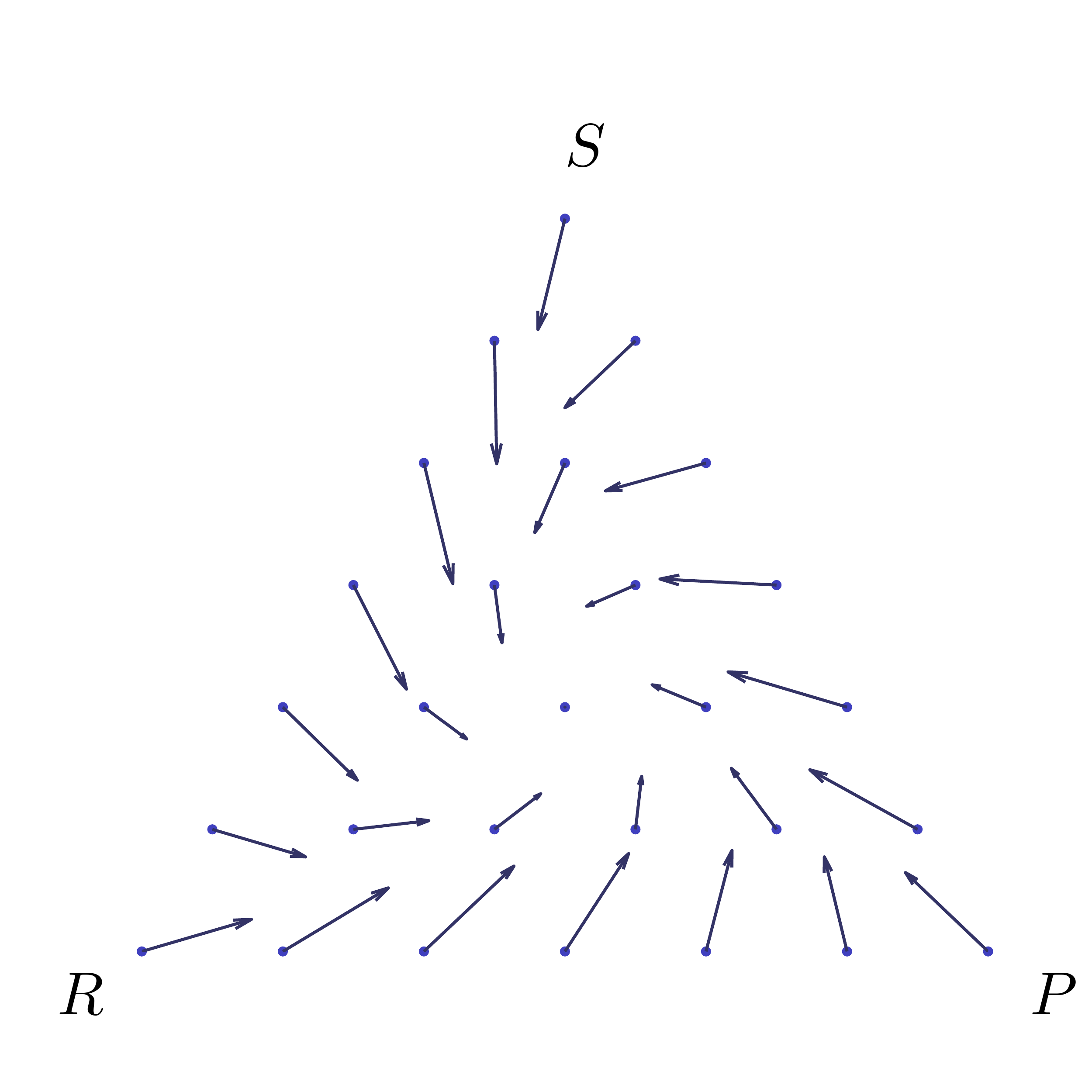}
\includegraphics[width=1in]{PayoffMatrix1.pdf}
\includegraphics[width=1in]{azj1nor.pdf}
\includegraphics[width=1in]{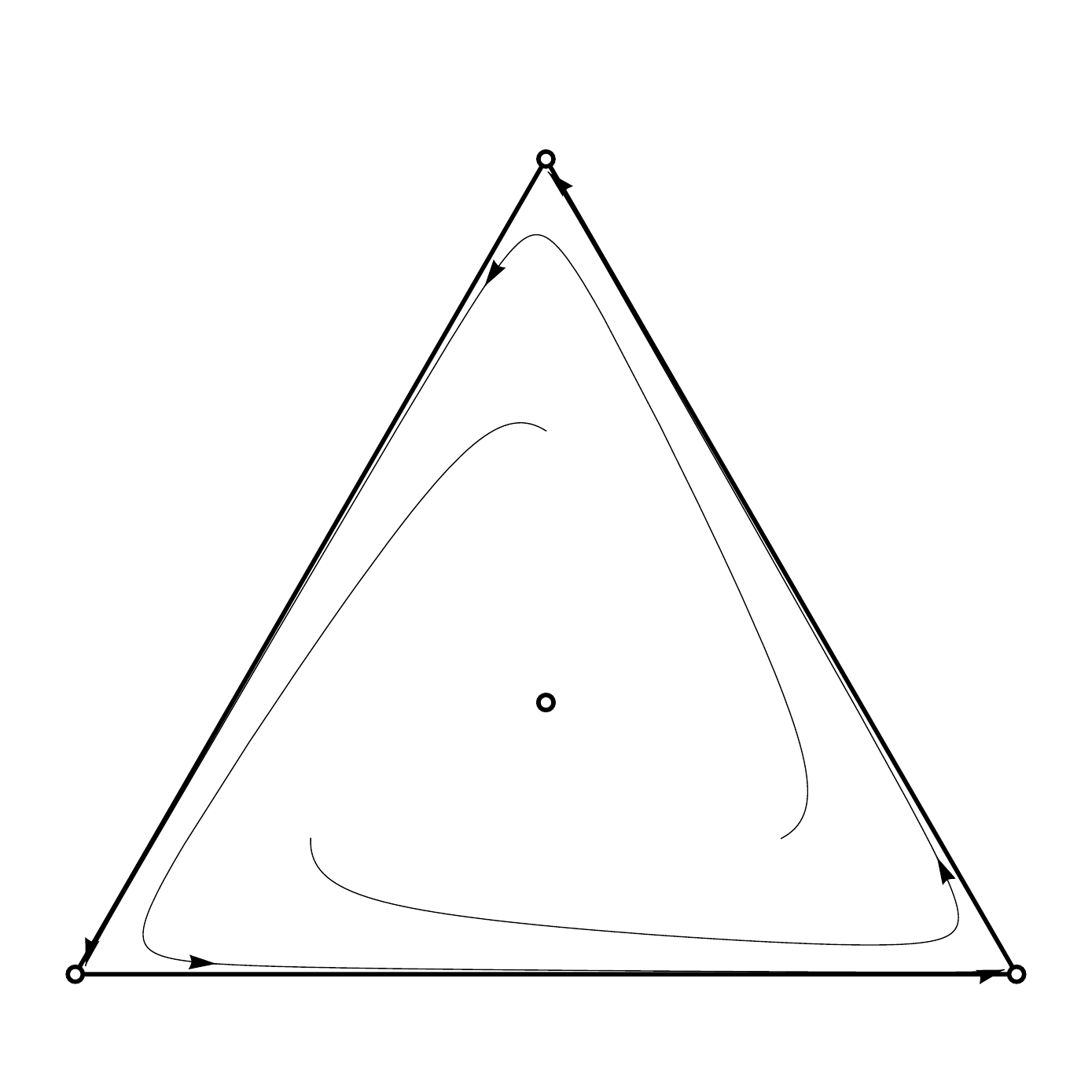}
\includegraphics[width=1in]{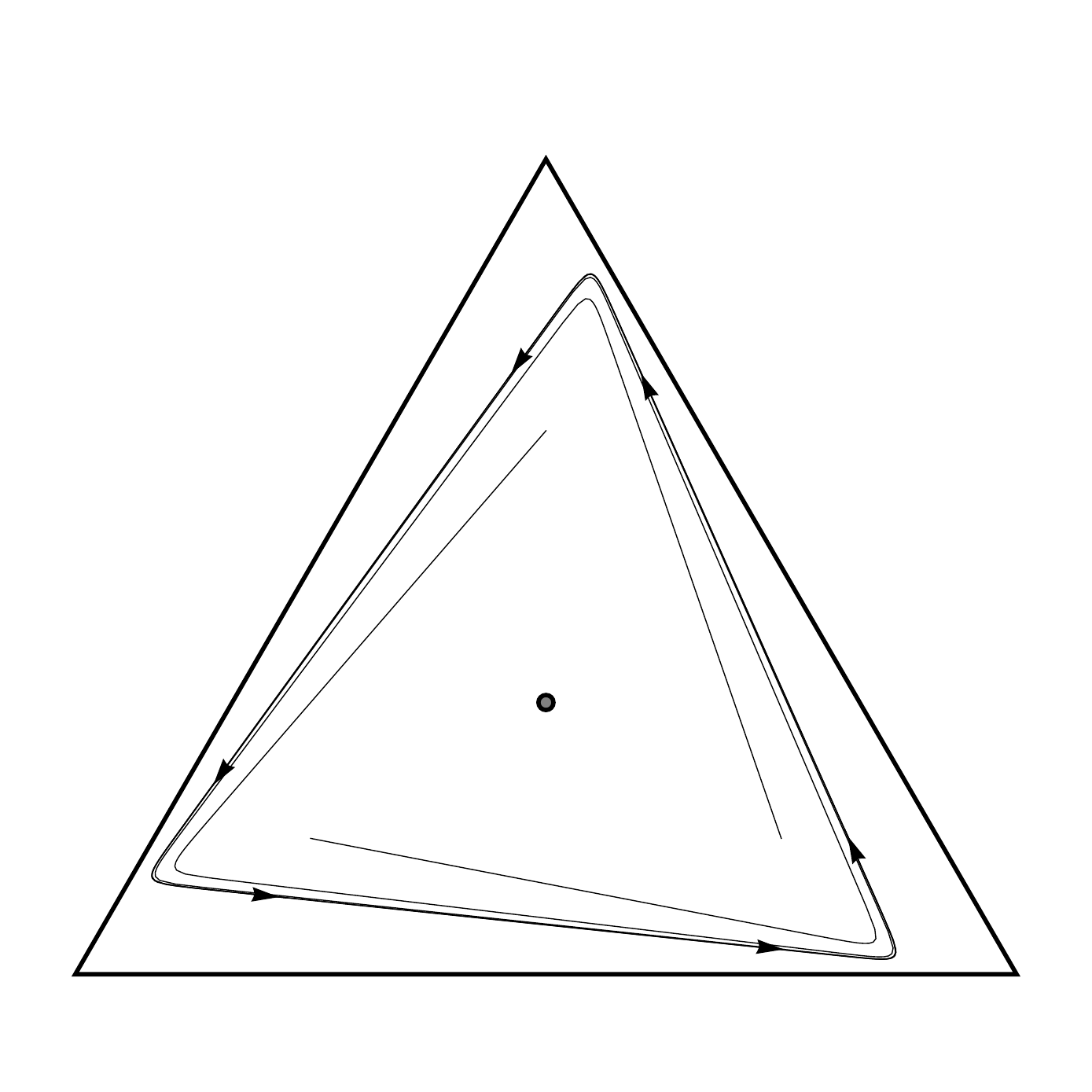}
       \includegraphics[width=1in]{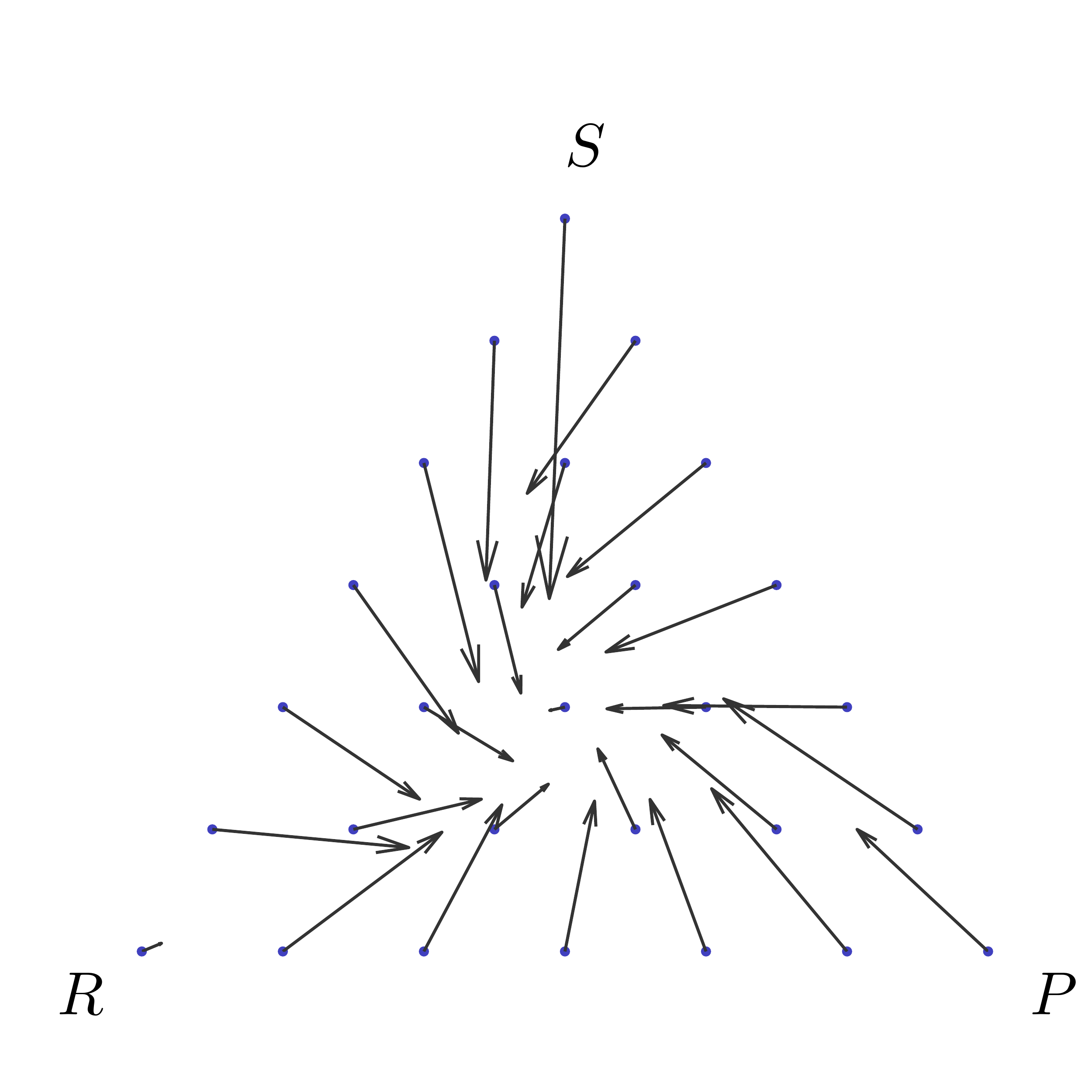}
       \includegraphics[width=1in]{RPSa1xb.pdf}
\includegraphics[width=1in]{PayoffMatrix2.pdf}
\includegraphics[width=1in]{azj2nor.pdf}
\includegraphics[width=1in]{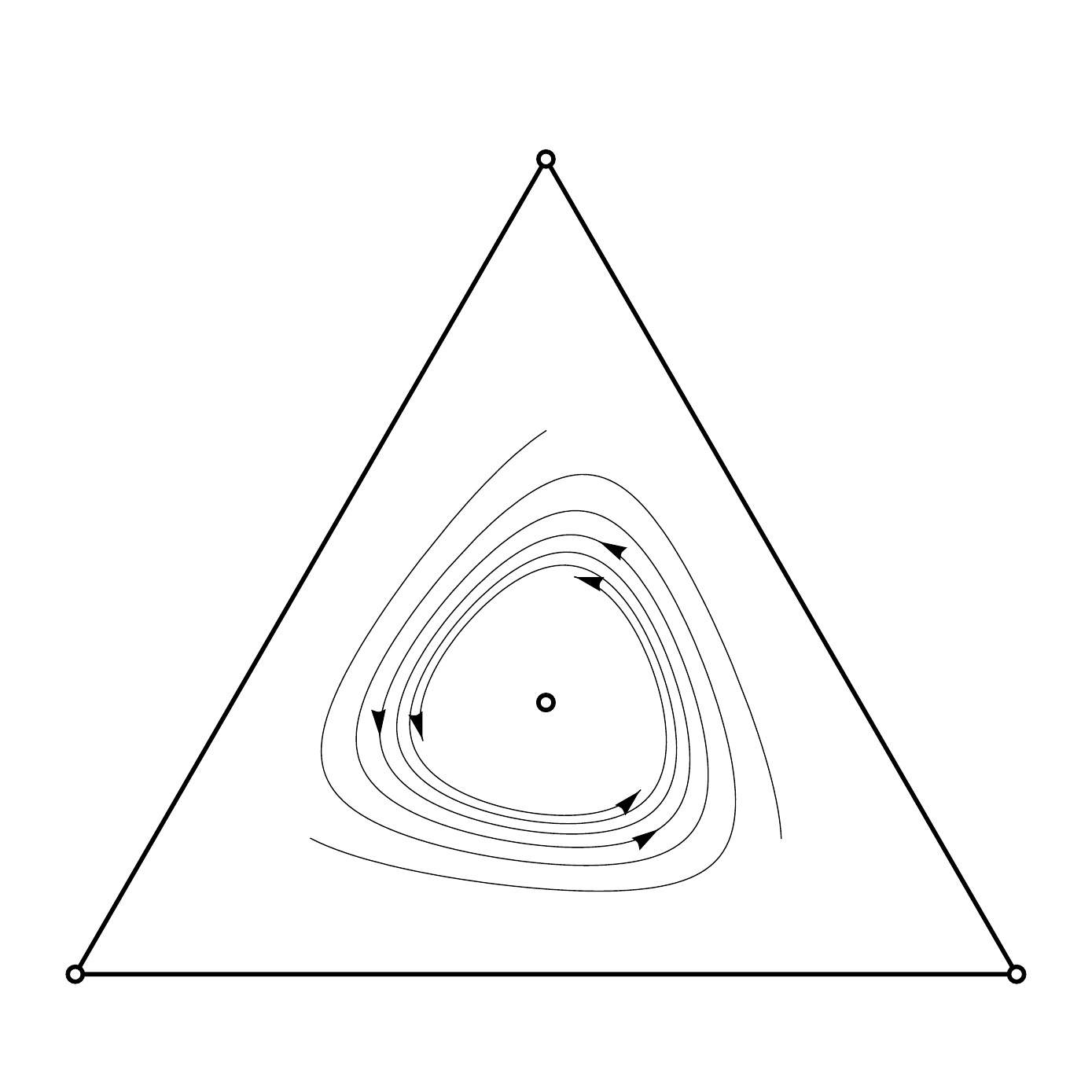}
\includegraphics[width=1in]{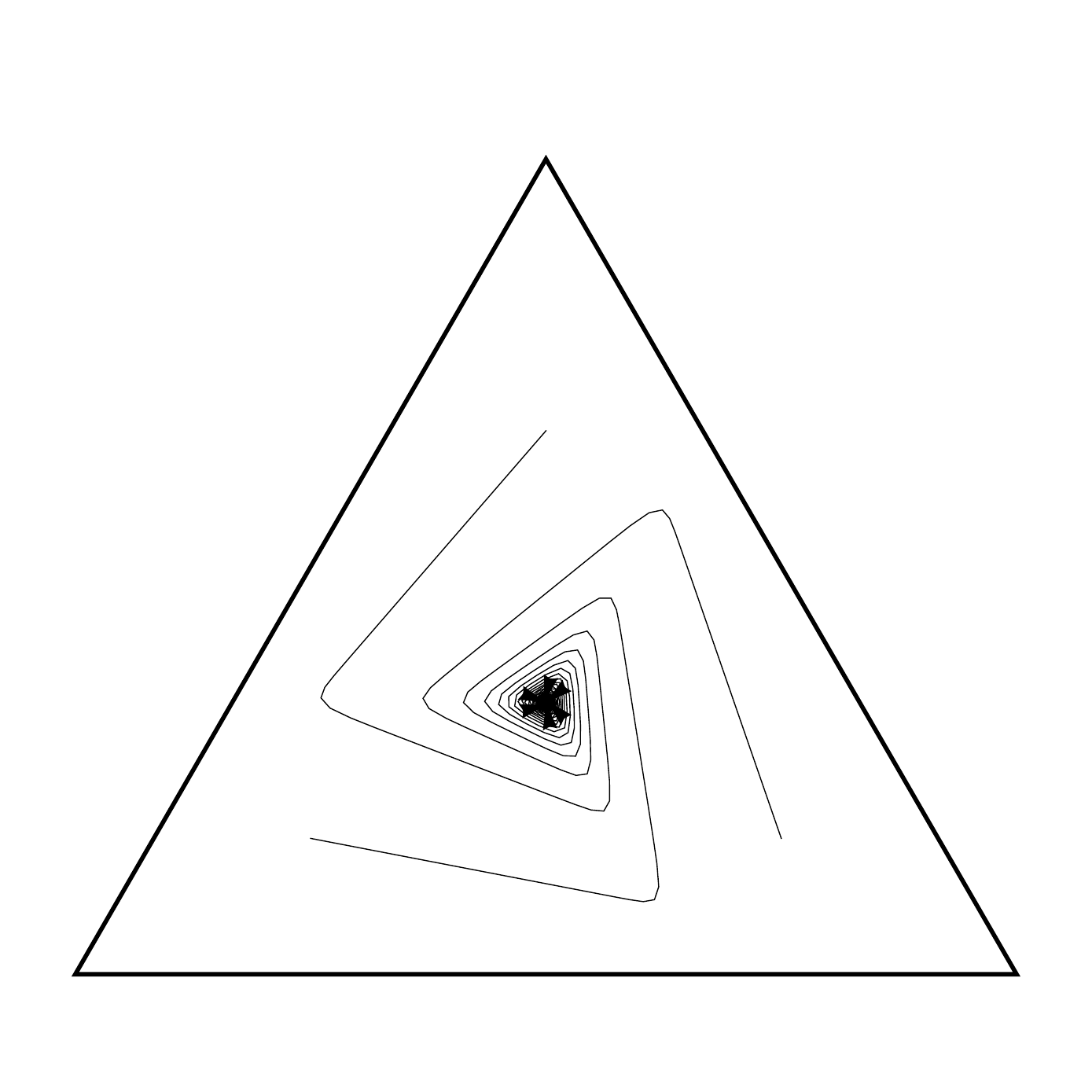}
       \includegraphics[width=1in]{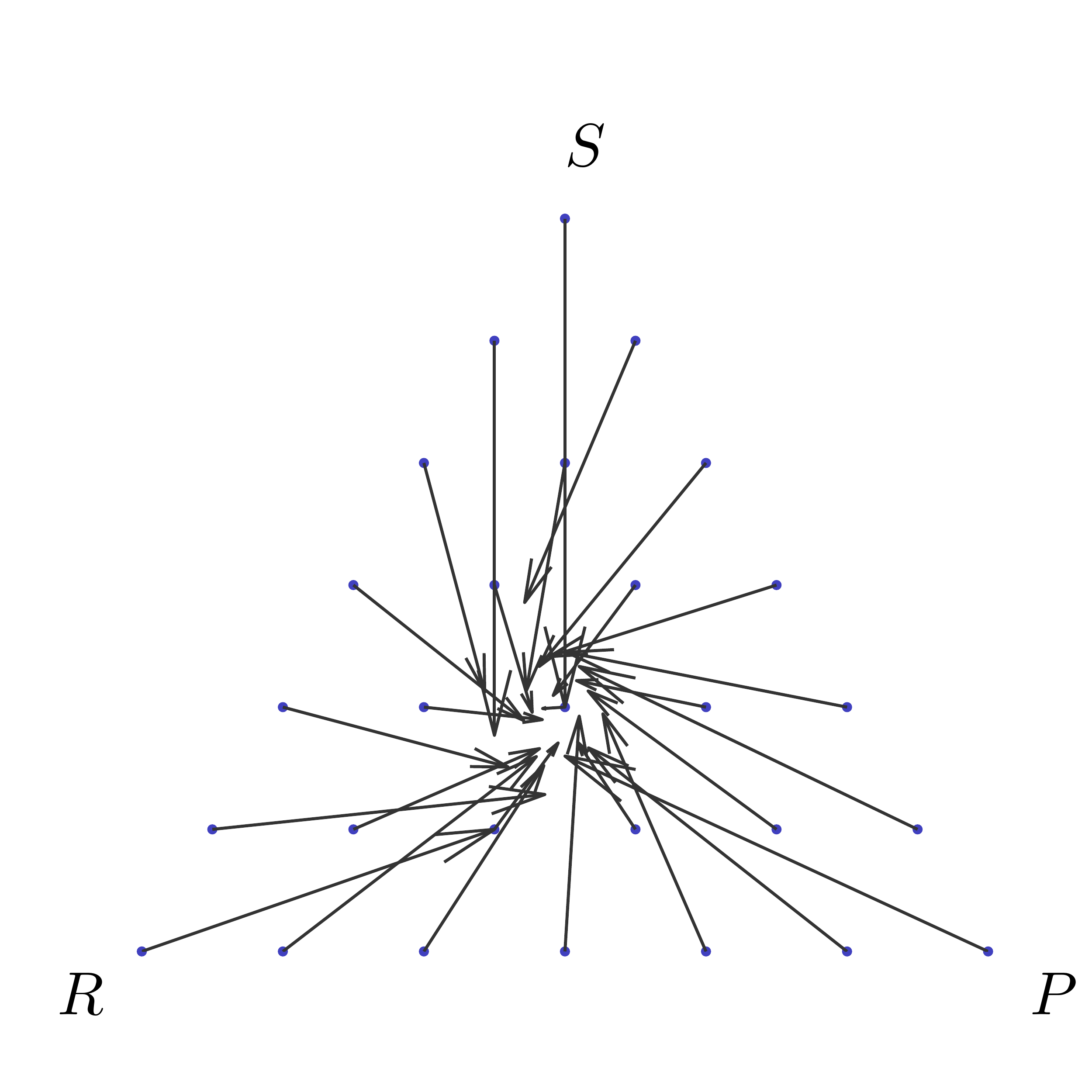}
       \includegraphics[width=1in]{RPSa2xb.pdf}
\includegraphics[width=1in]{PayoffMatrix4.pdf}
\includegraphics[width=1in]{azj4nor.pdf}
\includegraphics[width=1in]{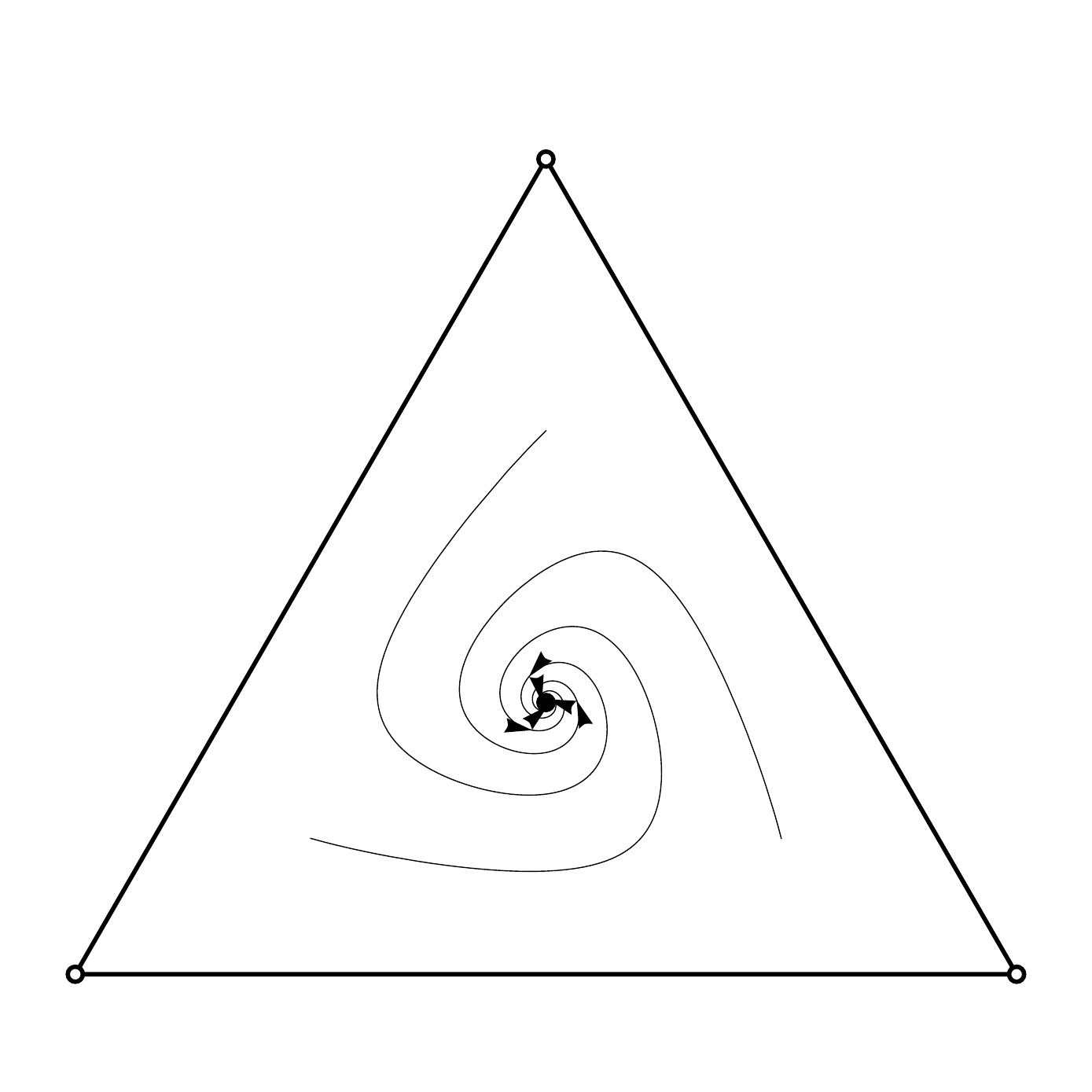}
\includegraphics[width=1in]{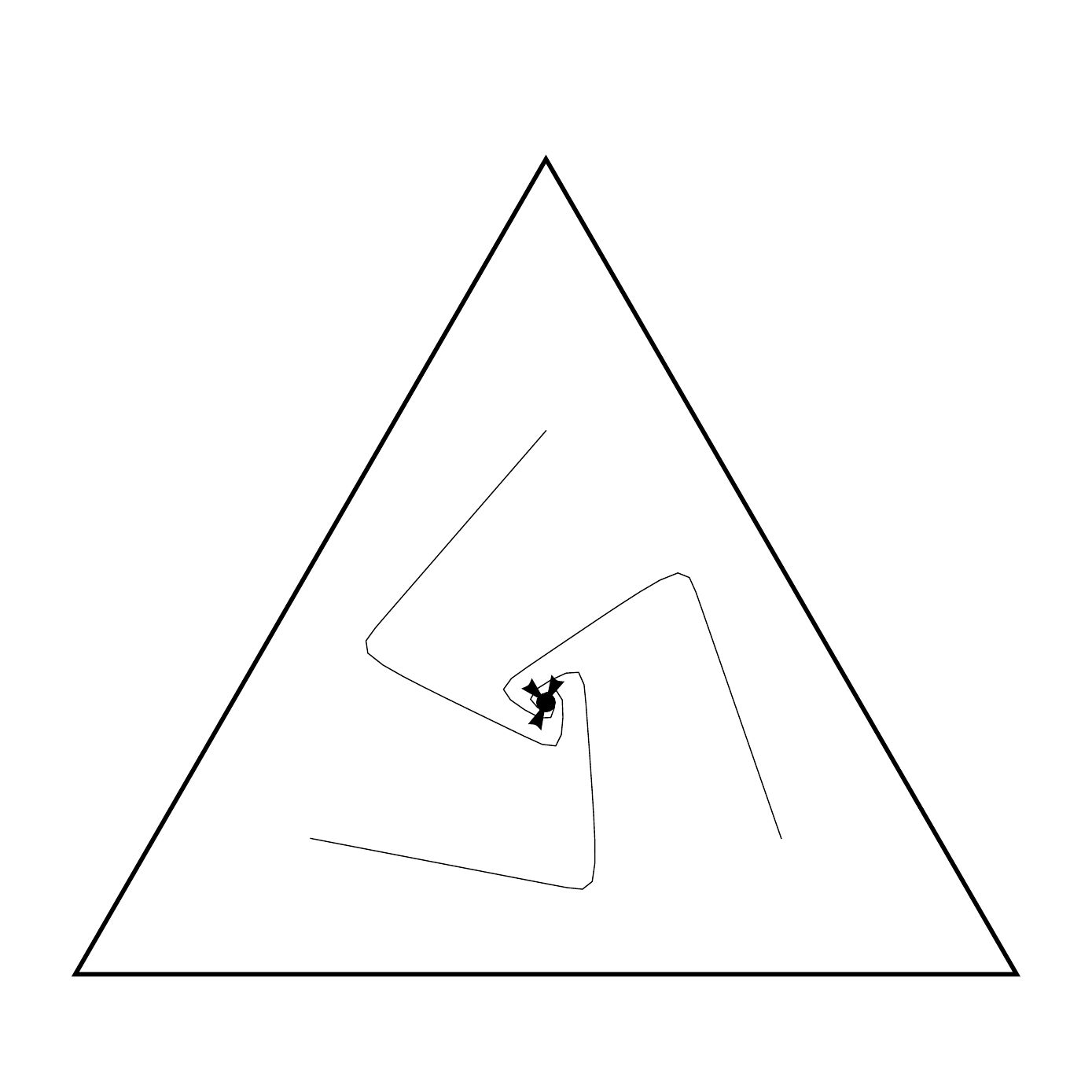}
       \includegraphics[width=1in]{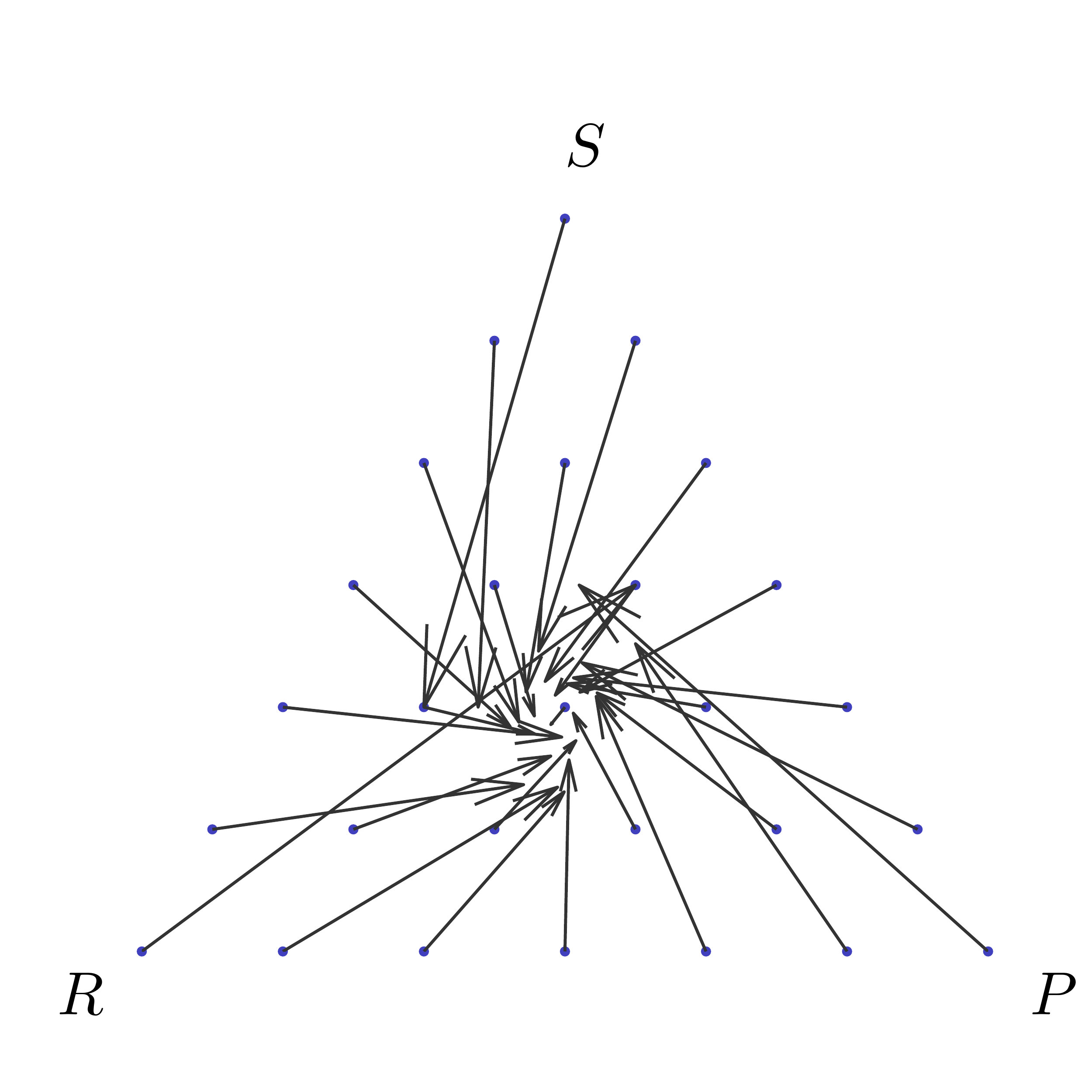}
       \includegraphics[width=1in]{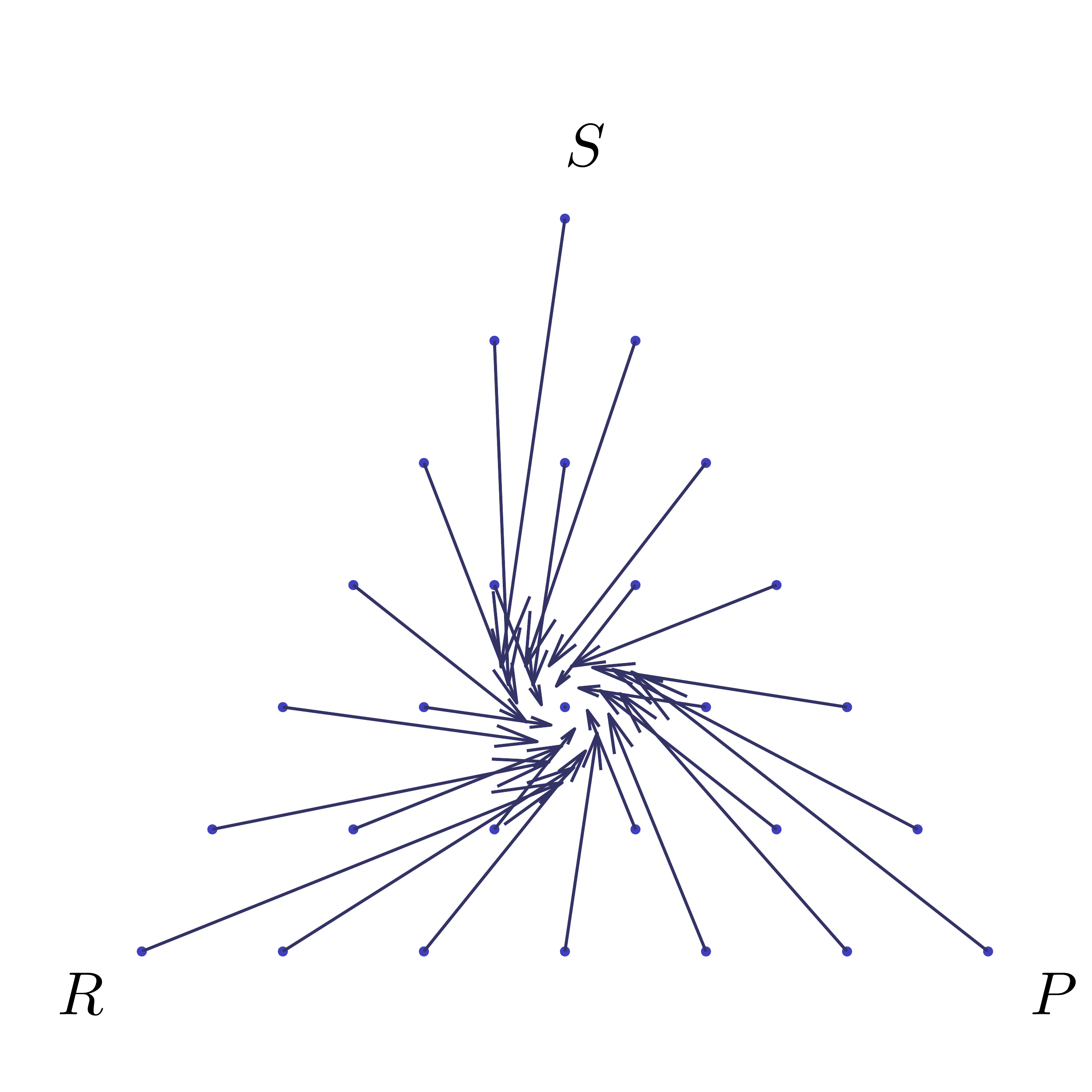}
\includegraphics[width=1in]{PayoffMatrix9.pdf}
\includegraphics[width=1in]{azj9nor.pdf}
\includegraphics[width=1in]{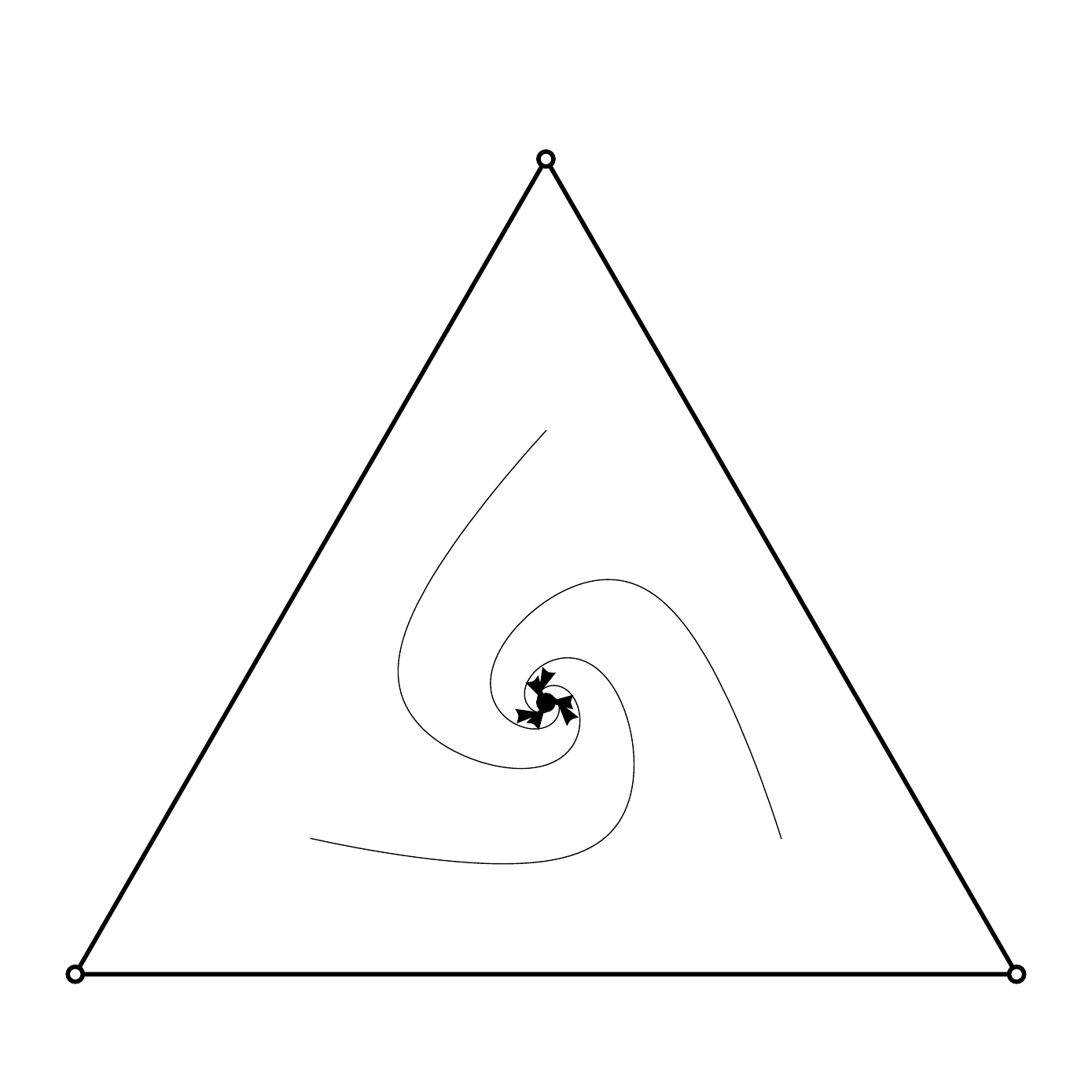}
\includegraphics[width=1in]{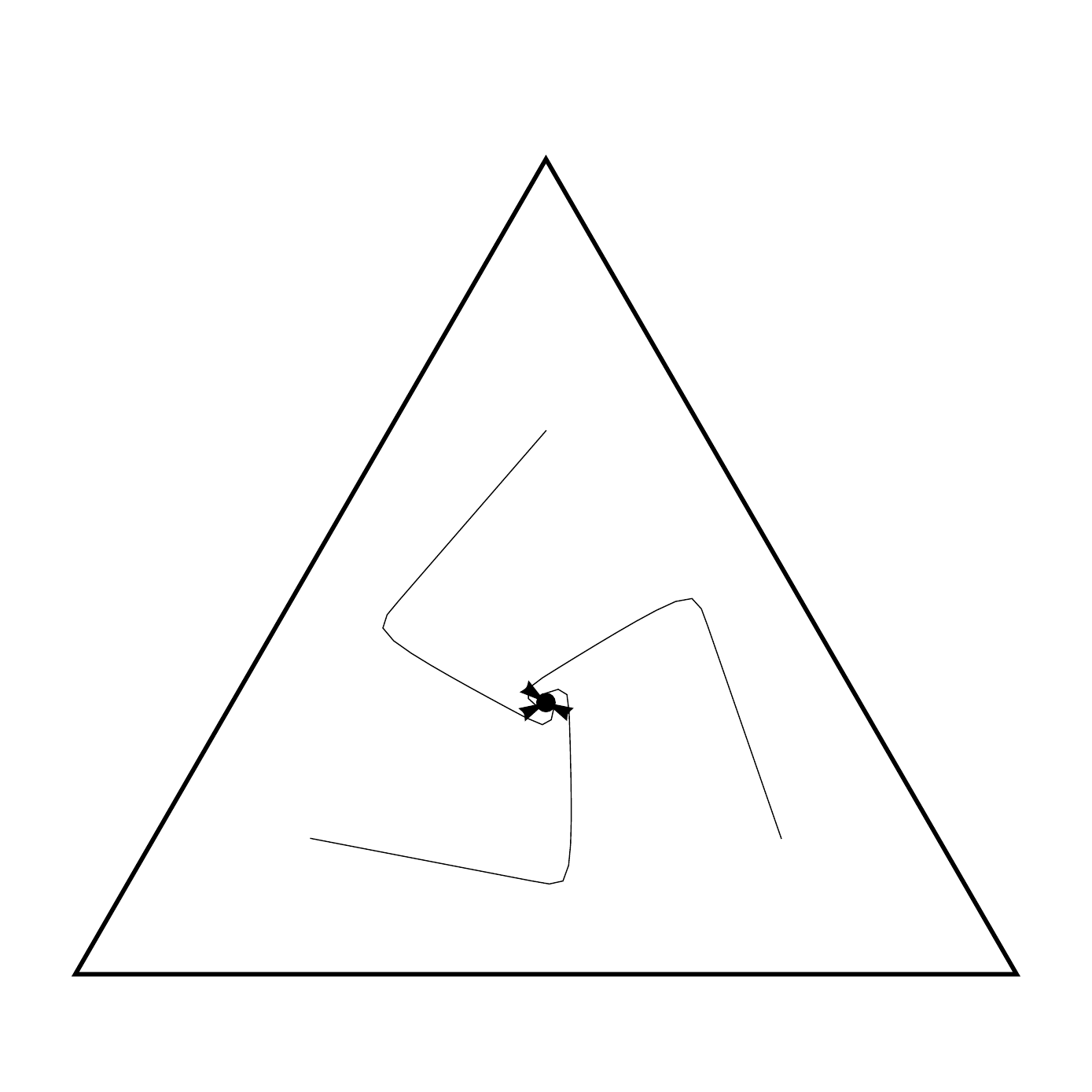}
       \includegraphics[width=1in]{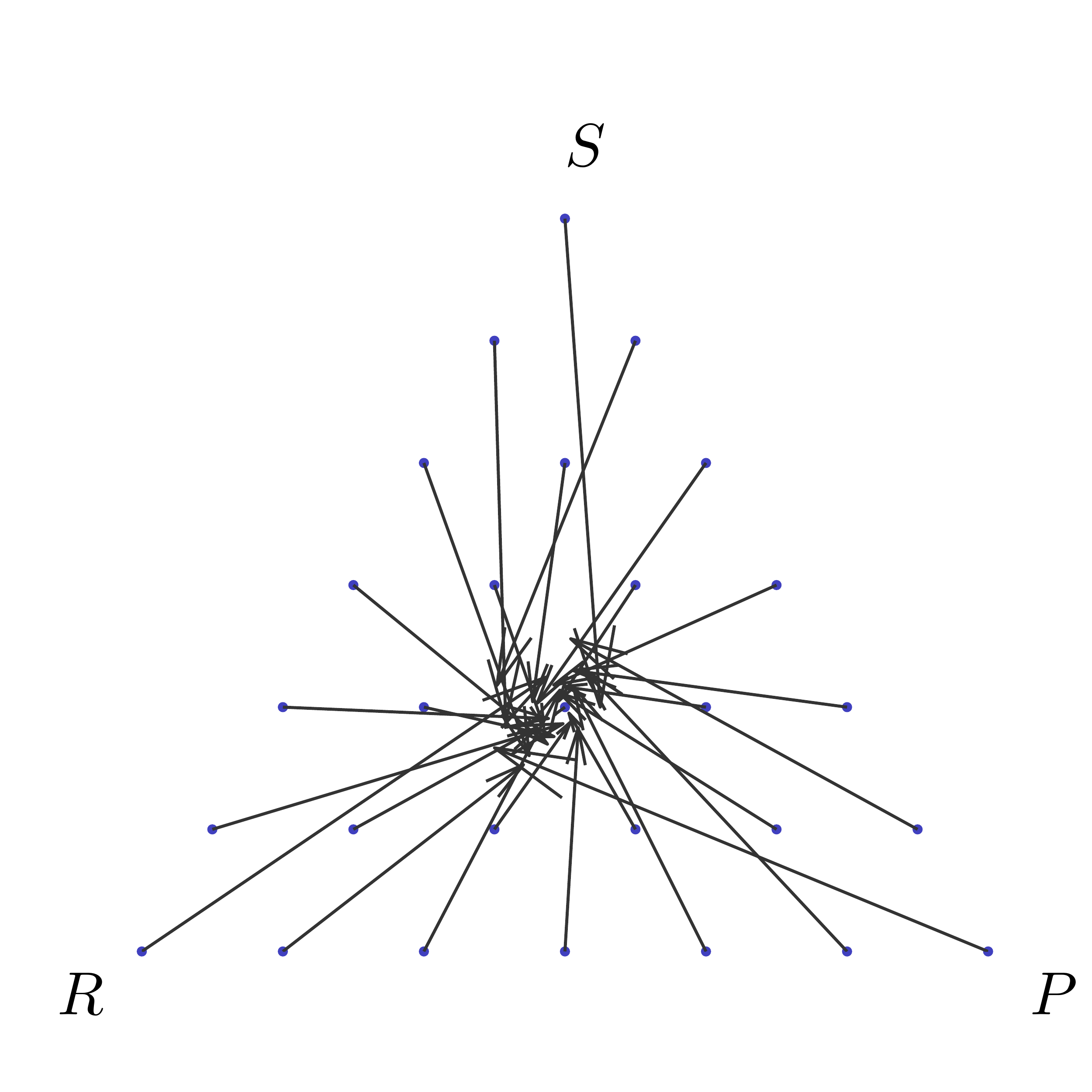}
       \includegraphics[width=1in]{RPSa9xb.pdf}
\includegraphics[width=1in]{PayoffMatrix100.pdf}
\includegraphics[width=1in]{azj100nor.pdf}
\includegraphics[width=1in]{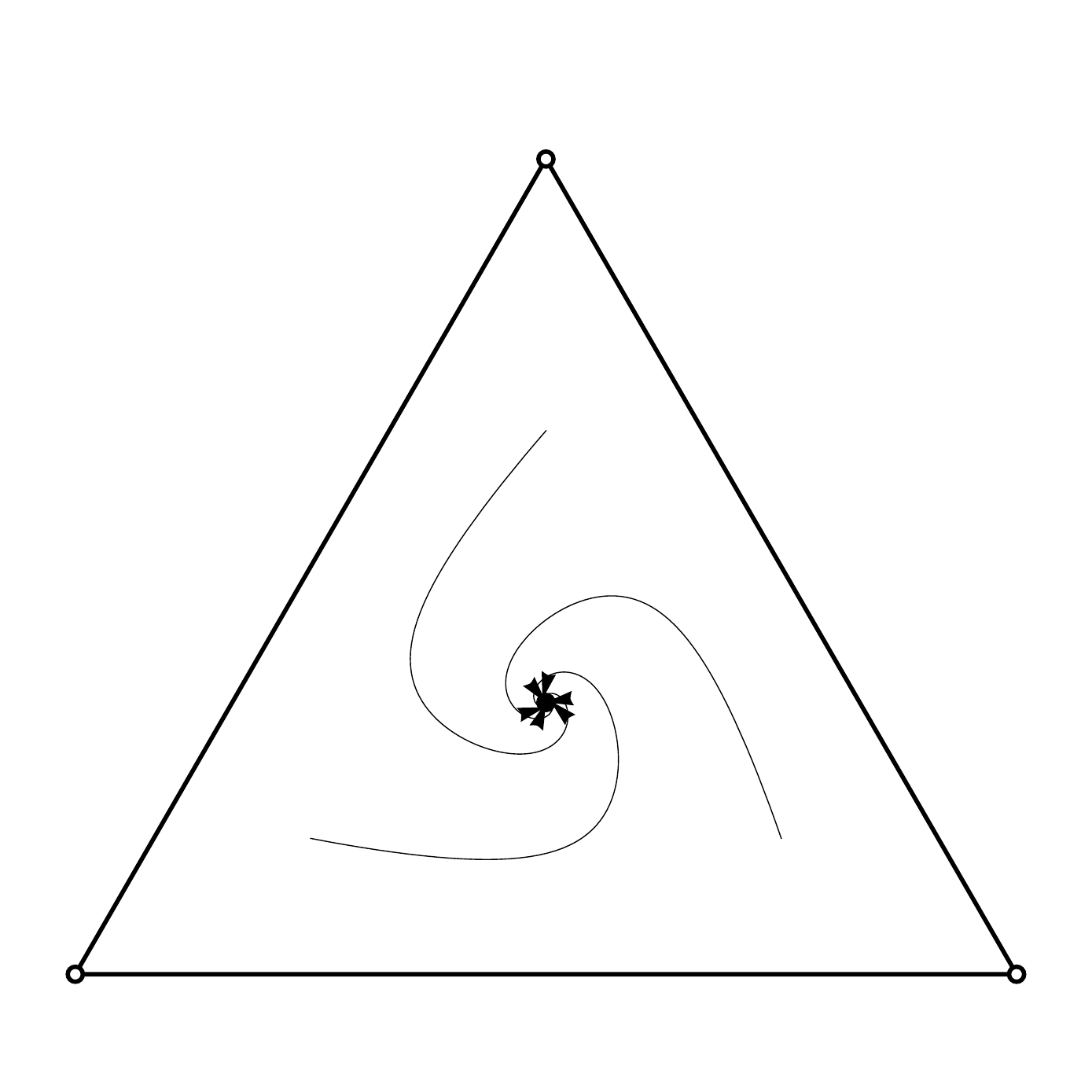}
\includegraphics[width=1in]{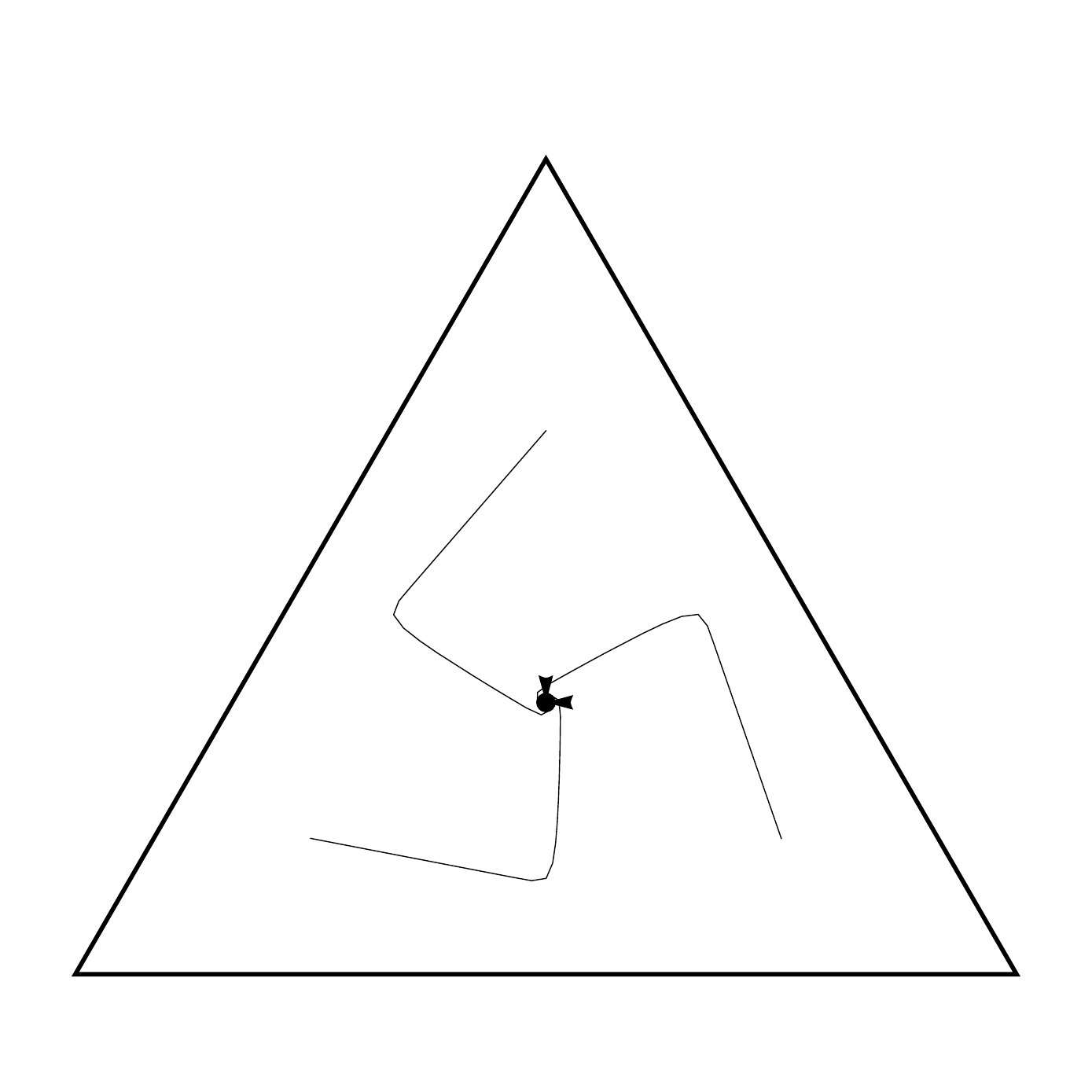}
       \includegraphics[width=1in]{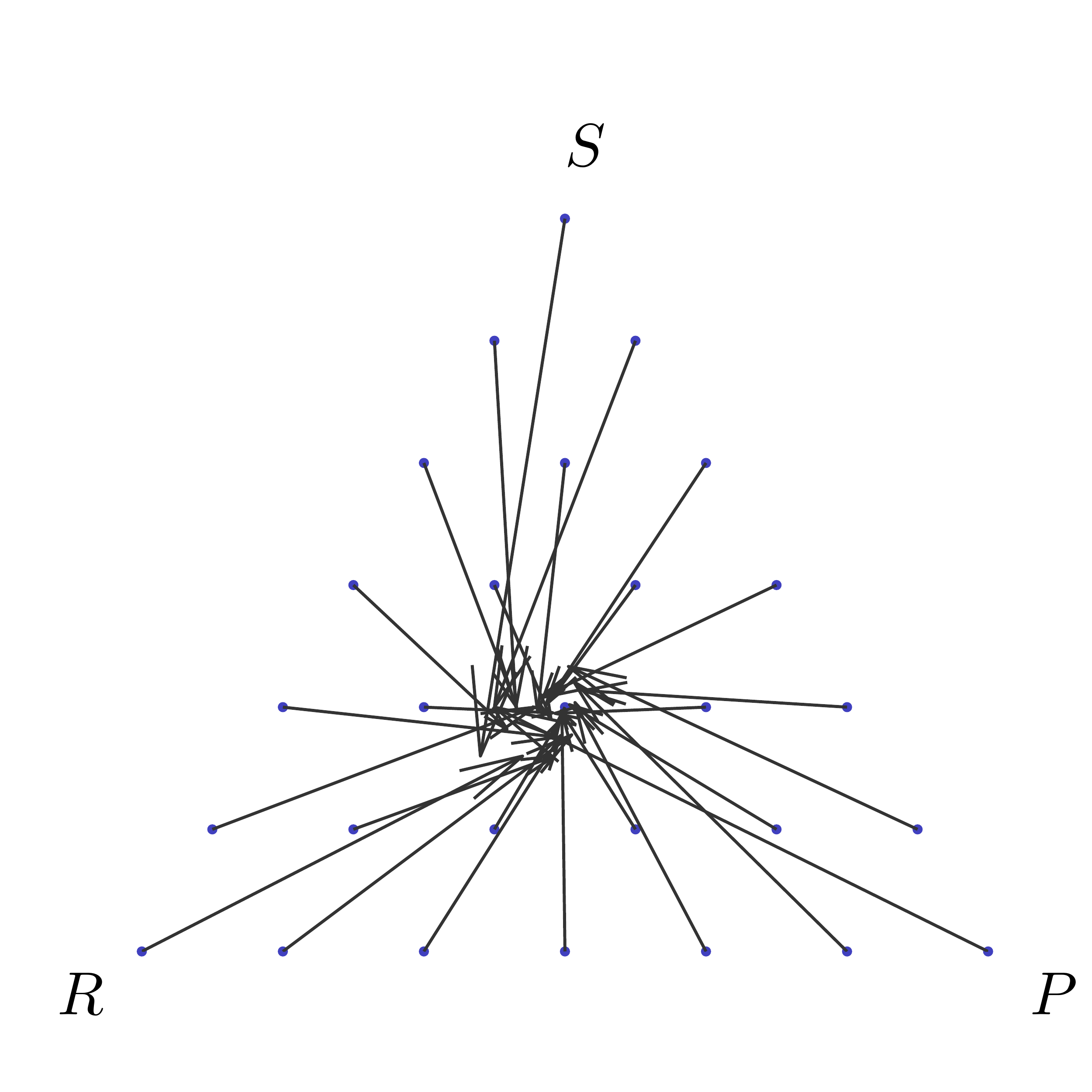}
       \includegraphics[width=1in]{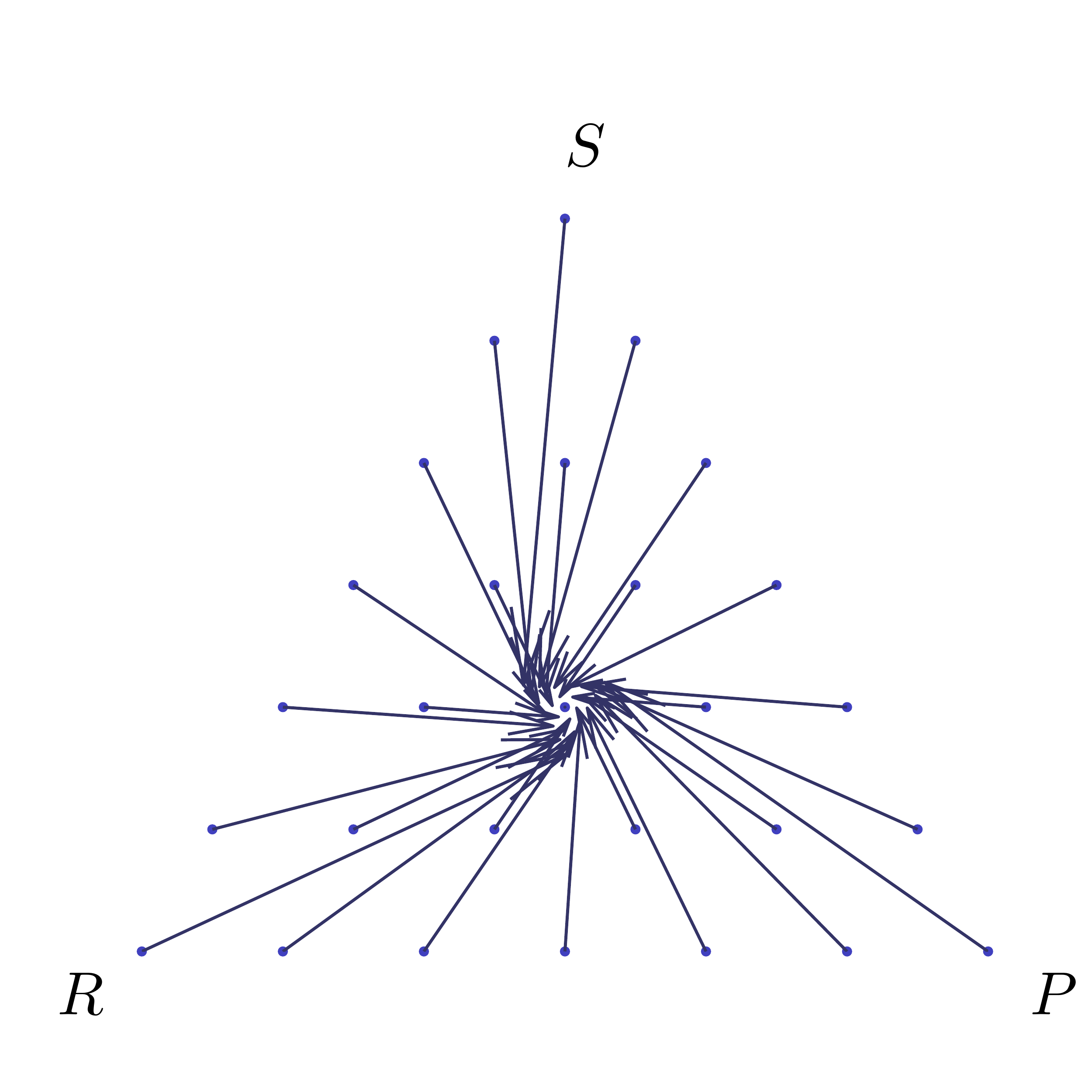}
\includegraphics[width=1in]{PayoffMatrix9999.pdf}
\includegraphics[width=1in]{azj9999nor.pdf}
\includegraphics[width=1in]{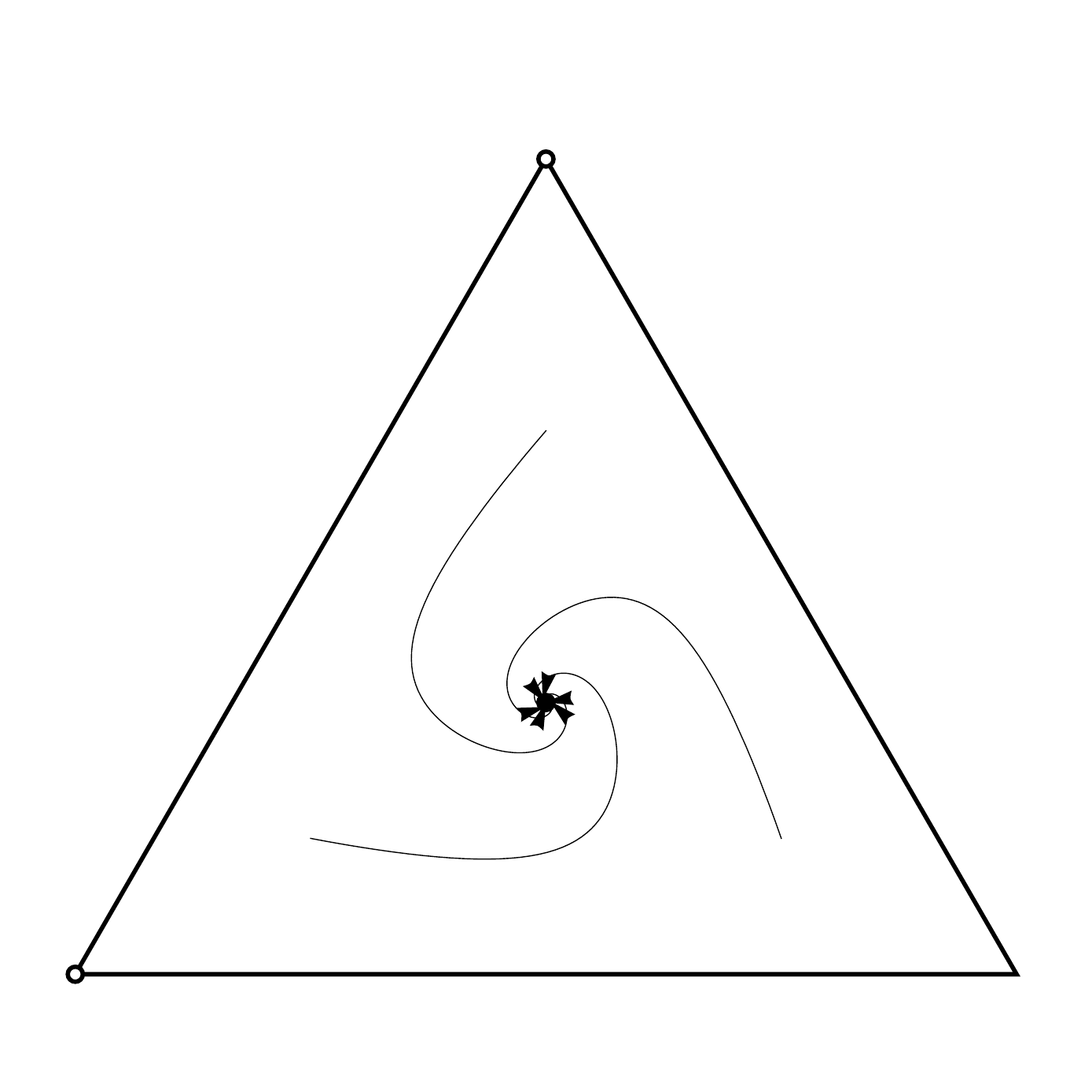}
\includegraphics[width=1in]{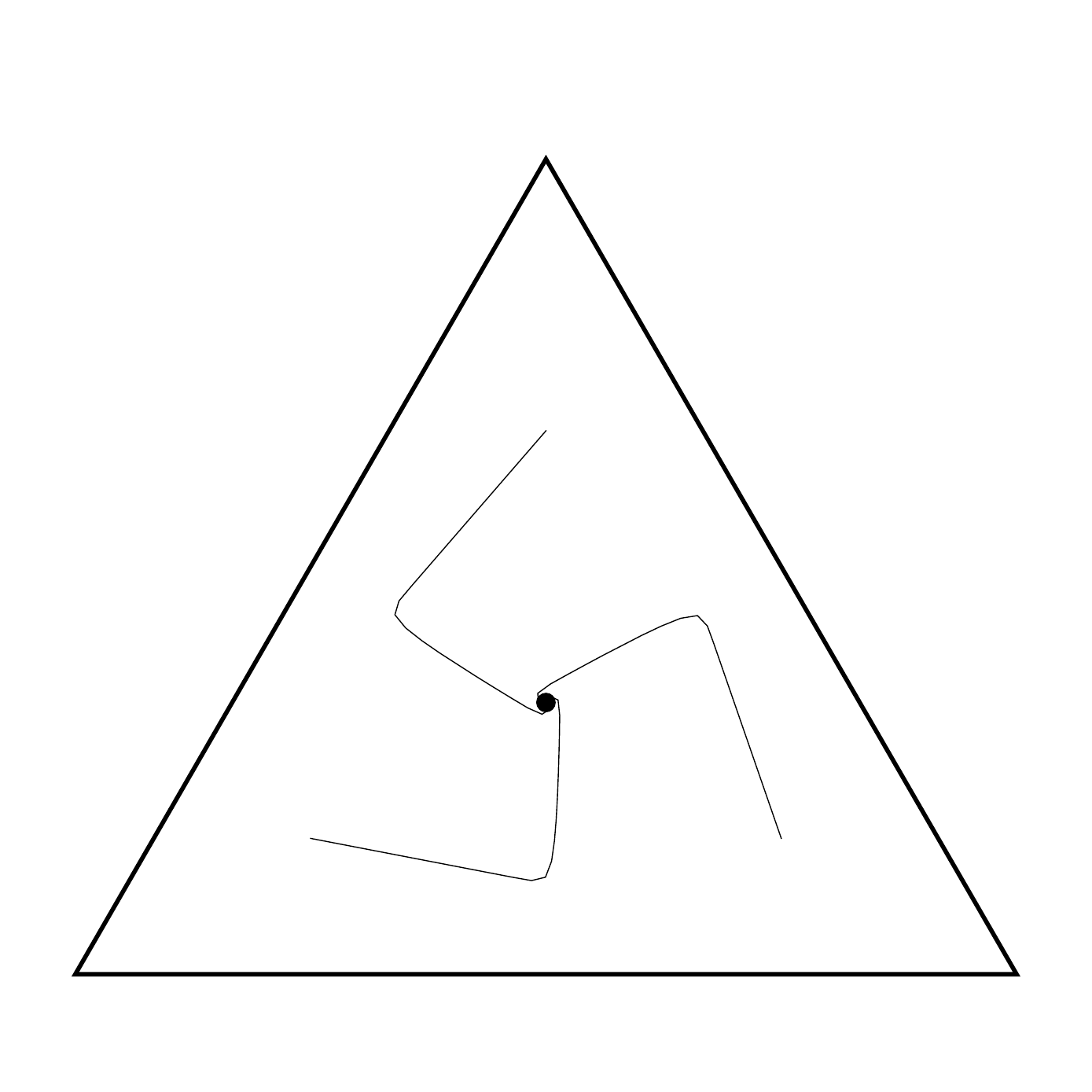}
       \includegraphics[width=1in]{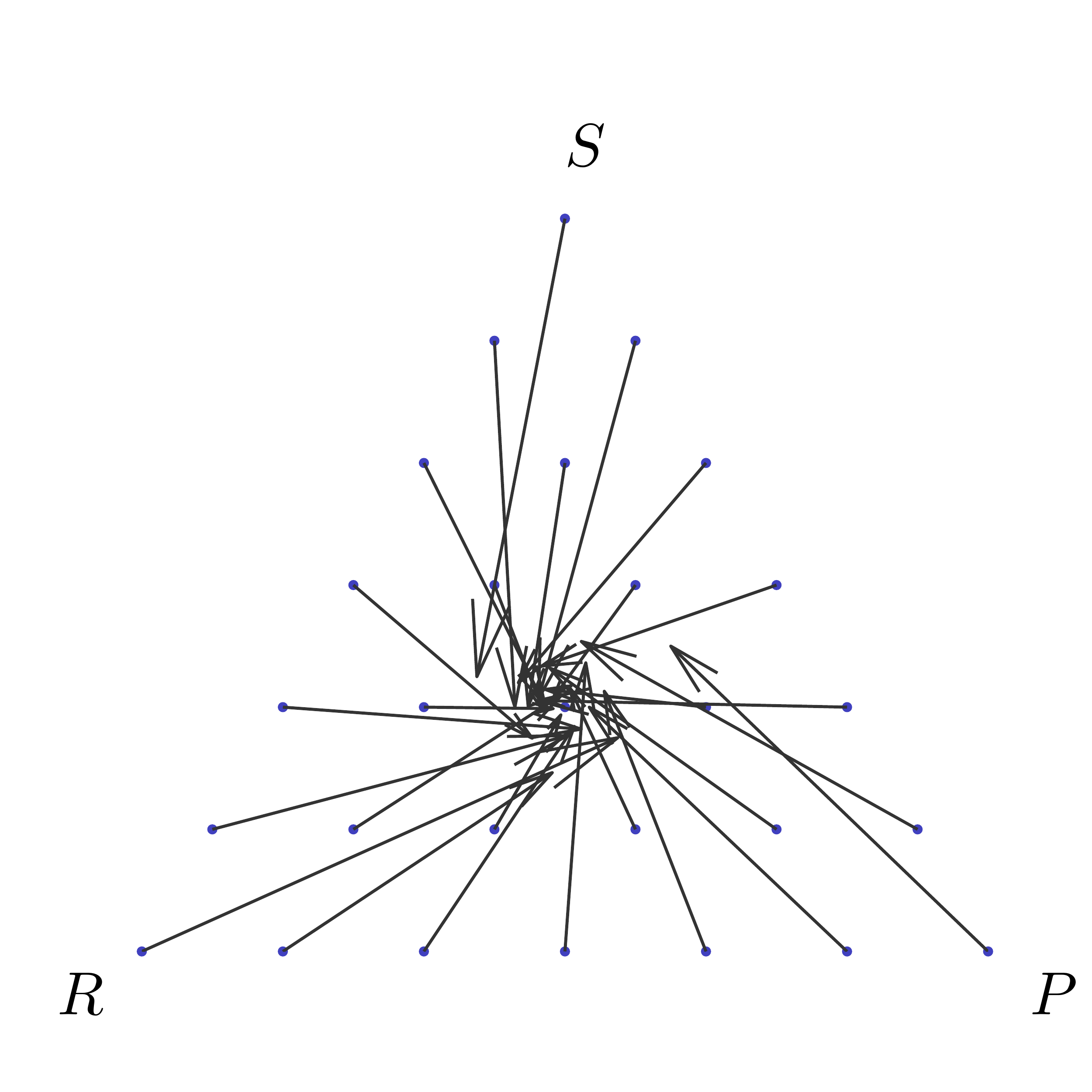}
       \includegraphics[width=1in]{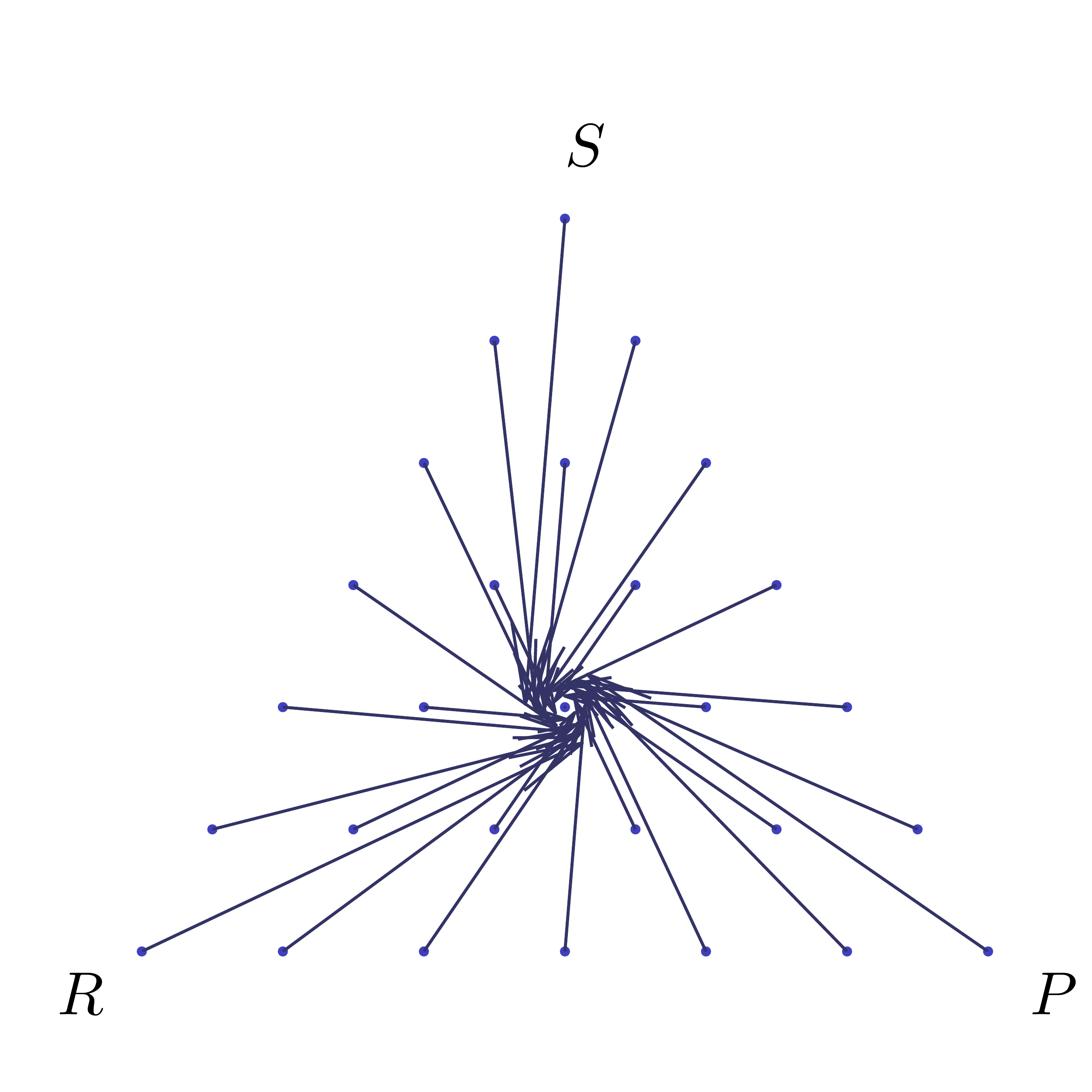}
             \end{center}
  \caption{
    \textbf{Experimental setting, theoretical prediction and experimental result}. The 1st column is the payoff matrix designed. The 2nd column comes from replicate dynamics, the 3rd column comes from logit dynamics, the 4th column comes from best response dynamics with continuous approximation in evolutionary game theory. The 5th column is the empirical results of the forward transition vector field. The 6th column is the forward transition vector field reconstruct by the equation \ref{eq:conditional} with the individual empirical WLT parameter shown in Fig.~\ref{fig:losecondition}~\ref{fig:tiecondition}~\ref{fig:wincondition}. In the case of payoff parameter (incentive) $a >$ 2, a $=$ 2 and a $<$ 2, according to the continuous replicator dynamics \cite{Sandholm2011}, the real part of the eigenvalue surrounding NE is negative, zero and positive, respectively, then the systems being expected to converge to, null and diverge from NE. }
   \label{fig:payoffandall}
\end{figure}

\clearpage

\end{document}